\documentclass[longauth]{aa}

\usepackage{graphicx}
\usepackage{txfonts}
\usepackage{hyperref}
\usepackage{lscape}
\usepackage{subcaption}

\usepackage{natbib}
\bibpunct{(}{)}{;}{a}{}{,} 

\usepackage{textgreek}
\usepackage{xargs}
\usepackage{xcolor}
\usepackage{xspace}

\let\oldtextsigma\textsigma
\renewcommand{\textsigma}{\oldtextsigma\xspace}
\let\oldAA\AA
\renewcommand{\AA}{\text{\oldAA}\xspace}
\def\w80{\ensuremath{w_{80}}\xspace}

\newcommandx{\fluxdcgs}[1][1=-20]{$\times 10^{[#1]}$~erg~s$^{-1}$~cm$^{-2}$~\AA$^{-1}$\xspace}

\newcommand{\Halpha}{\text{H\textalpha}\xspace}
\newcommand{\Hbeta}{\text{H\textbeta}\xspace}
\newcommand{\Hgamma}{\text{H\textgamma}\xspace}
\newcommand{\Hdelta}{\text{H\textdelta}\xspace}

\newcommandx{\permittedEL}[6][1=O,2=III,3=,4=,5=,6=]{\text{{#1}\,{\sc{#2}}{#3}{#4}{#5}{#6}}\xspace}
\newcommandx{\semiforbiddenEL}[6][1=O,2=III,3=,4=,5=,6=]{\text{{#1}\,{\sc{#2}}]{#3}{#4}{#5}{#6}}\xspace}
\newcommandx{\forbiddenEL}[6][1=O,2=III,3=,4=,5=,6=]{\text{[{#1}\,{\sc{#2}}]{#3}{#4}{#5}{#6}}\xspace}

\newcommandx{\HI}{\permittedEL[H][i]}
\newcommandx{\HIL}[1][1=3885]{\permittedEL[H][i][\,\textlambda][#1]}
\newcommandx{\HII}{\permittedEL[H][ii]}
\newcommandx{\HeI}{\permittedEL[He][i]}
\newcommandx{\HeII}{\permittedEL[He][ii]}
\newcommandx{\HeIL}[1][1=3889]{\permittedEL[He][i][\,\textlambda][#1]}
\newcommandx{\HeIIL}[1][1=4686]{\permittedEL[He][ii][\,\textlambda][#1]}
\newcommand{\OIII}{\forbiddenEL[O][iii]}
\newcommandx{\OIIIL}[1][1=5007]{\forbiddenEL[O][iii][\textlambda][#1]}
\newcommandx{\SIIIL}[1][1=5007]{\forbiddenEL[S][iii][\textlambda][#1]}
\newcommandx{\NeIIIL}[1][1=5007]{\forbiddenEL[Ne][iii][\textlambda][#1]}

\newcommandx{\NIL}[1]{\forbiddenEL[N][i][\textlambda][5200]}

\newcommandx{\OIL}[1][1=8446]{\permittedEL[O][i][\textlambda][#1]}
\newcommandx{\OILs}[1][1=8446]{\text{\textlambda {#1}}\xspace}

\newcommandx{\OIIL}[1][1=3727]{\forbiddenEL[O][ii][\textlambda][#1]}
\newcommandx{\SIIL}[1][1=6717]{\forbiddenEL[S][ii][\textlambda][#1]}

\newcommand{\NIV}{\semiforbiddenEL[N][iv]}

\newcommand{\NII}{\forbiddenEL[N][ii]}

\newcommandx{\NeVL}[1][1=3426]{\forbiddenEL[Ne][v][\textlambda][#1]}
\newcommandx{\NIIL}[1][1=6583]{\forbiddenEL[N][ii][\textlambda][#1]}

\newcommandx{\CIV}{\permittedEL[C][iv]}

\newcommandx{\CaT}{\permittedEL[Ca][ii][\textlambda][][8498,8542,8662]}
\newcommandx{\CaL}[1][1=8498]{\permittedEL[Ca][ii][\textlambda][#1]}
\newcommandx{\Ca}{\permittedEL[Ca][ii]}

\newcommandx{\Fef}{\forbiddenEL[Fe][ii]}

\newcommandx{\Fe}{\permittedEL[Fe][ii]}
\newcommandx{\FeL}{\permittedEL[Fe][ii][\textlambda]}

\newcommandx{\Feopt}{\permittedEL[Fe][ii][\textlambda][\textlambda][5190,][5320]}

\newcommandx{\CIII}{\semiforbiddenEL[C][iii]}
\newcommandx{\CIIIL}[1][1=1907]{\semiforbiddenEL[C][iii][\textlambda][#1]}
\newcommandx{\sOIIIL}[1][1=1665]{\semiforbiddenEL[O][iii][\textlambda][#1]}
\newcommandx{\NIIIL}[1][1=1750]{\semiforbiddenEL[N][iii][\textlambda][#1]}
\newcommandx{\NIII}{\semiforbiddenEL[N][iii]}
\newcommandx{\MgII}{\permittedEL[Mg][ii]}

\newcommand{\jwst}{\textit{JWST}\xspace}

\begin{document}

	\title{PRISMS. GHZ1. A standard tale of galaxy evolution with atypical ionizing conditions at $z$=9.878}
	\titlerunning{Galaxy evolution and ionizing conditions in GHZ1}

	\authorrunning{B. P\'{e}rez-D\'{\i}az et al.}

\author{Borja~P\'{e}rez-D\'{\i}az\inst{\ref{inst:INAF-ROME}}\thanks{\email{borja.perezdiaz@inaf.it}}
    \and Marco Castellano\inst{\ref{inst:INAF-ROME}}
    \and Javier \'{A}lvarez-M\'{a}rquez\inst{\ref{inst:CAB}}
    \and Luis Colina\inst{\ref{inst:CAB}}
    \and Lorenzo Napolitano\inst{\ref{inst:INAF-ROME}}
    \and Sarah Kendrew\inst{\ref{inst:ESA-BALTIMORE}}
    \and Alejandro Crespo G\'{o}mez\inst{\ref{inst:Baltimore}}
    \and Stefano Carniani\inst{\ref{inst:PISA}}
    \and Shouyi Wang\inst{\ref{inst:Beijing1}, \ref{inst:Beijing2}}
    \and Fan Zou\inst{\ref{inst:Michigan}}
    \and Abdurro'uf\inst{\ref{inst:Indiana}}
    \and Arjan Bik\inst{\ref{inst:Stockholm}}
    \and Karina Caputi\inst{\ref{inst:Groningen}} 
    \and Rom\'{a}n Fern\'{a}ndez-Aranda\inst{\ref{inst:CAB}}
    \and Adriano Fontana\inst{\ref{inst:INAF-ROME}}
    \and Yoshinobu Fudamoto\inst{\ref{inst:CHIBA}}
    \and Macarena Garc\'{\i}a-Mar\'{\i}n\inst{\ref{inst:ESA-BALTIMORE}}
    \and Mahmoud Hamed\inst{\ref{inst:CAB}}
    \and Yuichi Harikane\inst{\ref{inst:Tokyo}}
    \and Takuya Hashimoto\inst{\ref{inst:Tsukuba},\ref{inst:TCHoU}}
    \and Jakob M. Helton\inst{\ref{inst:Park}}
    \and Tiger Y.-Y. Hsiao\inst{\ref{inst:Austin}}
    \and Danial Langeroodi\inst{\ref{inst:Kavli},\ref{inst:Cavendish},\ref{inst:DARK}}
    \and Ruqiu Lin\inst{\ref{inst:Amherst}}
    \and Rui Marques-Chaves\inst{\ref{inst:Geneva}}
    \and G${\rm \ddot{o}}$ran ${\rm \ddot{O}}$stlin\inst{\ref{inst:Stockholm}}
    \and Pablo G. P\'{e}rez-Gonz\'{a}lez\inst{\ref{inst:CAB}}
    \and Enrico Piconcelli\inst{\ref{inst:INAF-ROME}}
    \and Carlota Prieto-Jiménez\inst{\ref{inst:CAB}, \ref{inst:UCM}}
    \and Pierluigi Rinaldi\inst{\ref{inst:Austin}, \ref{inst:CFC}}
    \and Guido Roberts-Borsani\inst{\ref{inst:London}}
    \and Bruno Rodr\'{\i}guez del Pino\inst{\ref{inst:CAB}}
    \and Paola Santini\inst{\ref{inst:INAF-ROME}}
    \and Tommaso Treu\inst{\ref{inst:California}}
    \and Ana Varo-O'Ferrall\inst{\ref{inst:CAB}, \ref{inst:UCM}}
    \and Jorge Zavala\inst{\ref{inst:Amherst}}}

   \institute{INAF – Osservatorio Astronomico di Roma, via Frascati 33, I-00078, Monteporzio Catone, Italy\label{inst:INAF-ROME}  
   \and Centro de Astrobiolog\'{\i}a (CAB), CSIC-INTA, Ctra. de Ajalvir km 4, Torrej\'on de Ardoz, E-28850, Madrid, Spain\label{inst:CAB}
   \and European Space Agency (ESA), ESA Office, Space Telescope Science Institute, 3700 San Martin Drive, Baltimore, MD 21218, USA\label{inst:ESA-BALTIMORE}
   \and Space Telescope Science Institute (STScI), 3700 San martin Drive, Baltimore, MD 21218, USA\label{inst:Baltimore}
   \and Scuola Normale Superiore, Piazza dei Cavalieri 7, I-56126 Pisa, Italy\label{inst:PISA}
   \and Department of Astronomy, School of Physics, Peking University, Beijing 100871, People’s Republic of China\label{inst:Beijing1}
   \and Kavli Institute for Astronomy and Astrophysics, Peking University, Beijing 100871, People’s Republic of China\label{inst:Beijing2}
   \and Department of Astronomy, University of Michigan, 1085 S University, Ann Arbor, MI 48109, USA\label{inst:Michigan}
   \and Department of Astronomy, Indiana University,727 East Third Street, Bloomington, IN 47405, USA\label{inst:Indiana}
   \and Department of Astronomy, Stockholm University, Oscar Klein Centre, AlbaNova University Centre, 106 91 Stockholm, Sweden\label{inst:Stockholm}
   \and Kapteyn Astronomical Institute, University of Groningen, P.O. Box 800, 9700AV Groningen, The Netherlands\label{inst:Groningen}
   \and Center for Frontier Science, Chiba University, 1-33 Yayoi-cho, Inage-ku, Chiba 263-8522, Japan\label{inst:CHIBA}
   \and Institute for Cosmic Ray Research, The University of Tokyo, 5-1-5 Kashiwanoha, Kashiwa, Chiba 277-8582, Japan\label{inst:Tokyo}
   \and Division of Physics, Faculty of Pure and Applied Sciences, University of Tsukuba, Tsukuba, Ibaraki 305-8571, Japan\label{inst:Tsukuba}
   \and Tomonaga Center for the History of the Universe (TCHoU), Faculty of Pure and Applied Sciences, University of Tsukuba, Tsukuba, Ibaraki 305-8571, Japan\label{inst:TCHoU}
   \and Department of Astronomy and Astrophysics, The Pennsylvania State University, University Park, PA 16802, USA\label{inst:Park}
   \and Department of Astronomy, University of Texas, Austin, TX 78712, USA\label{inst:Austin}
   \and Kavli Institute for Cosmology, University of Cambridge, Madingley Road, Cambridge CB3 0HA, UK\label{inst:Kavli}
   \and Cavendish Laboratory, University of Cambridge, 19 JJ Thomson Avenue, Cambridge CB3 0HE, UK\label{inst:Cavendish}
   \and DARK, Niels Bohr Institute, University of Copenhagen, Jagtvej 155A, 2200 Copenhagen, Denmark\label{inst:DARK}
   \and University of Massachusetts Amherst, 710 North Pleasant Street, Amherst, MA 01003-9305, USA\label{inst:Amherst}
   \and Geneva Observatory, Department of Astronomy, University of Geneva, Chemin Pegasi 51, CH-1290 Versoix, Switzerland \label{inst:Geneva}
   \and Departamento de F\'{i}sica de la Tierra y Astrof\'{i}sica, Facultad de Ciencias F\'{i}sicas, Universidad Complutense de Madrid, E-28040, Madrid, Spain\label{inst:UCM}
   \and Cosmic Frontier Center, The University of Texas at Austin, Austin, TX 78712, USA \label{inst:CFC}
   \and Department of Physics \& Astronomy, University College London, London, WC1E 6BT, UK\label{inst:London}
   \and Physics and Astronomy Department University of California Los Angeles CA 90095\label{inst:California}
   }

	\date{Received MONTH DAY, YEAR; accepted MONTH DAY, YEAR}



\abstract
{The first operating years of the \textit{James Webb Space Telescope} (\jwst) have revealed a surprising number of $z \gtrsim 10$ UV-bright objects, many of them characterized by their compactness, anomalous chemical abundance patterns and complex ISM physics (e.g. N-enhanced systems). To assess whether these objects are a phase of early galaxy evolution or a peculiar population, it is essential to understand the physics of systems without anomalous abundance patterns in this epoch.}
{We present a detailed analysis of GHZ1 ($z=9.878$), a moderately ($\mu = 1.715$) lensed UV-bright ($M_{UV} = -20.07$) galaxy in the Abell2744 field, characterized by an extended size ($R_{e} \sim 410$ pc). These properties, which contrast with the properties of many of the N-enhanced galaxies, make GHZ1 an excellent laboratory to assess what are the major drivers of early galaxy evolution and the origin of the differences.}
{By means of \jwst spectroscopic observations from Near Infrared Spectrograph (NIRSpec, PRISM-CLEAR configuration) and the Mid-Infrared Instrument Low Resolution Spectrograph (MIRI/LRS), we analyzed the full UV and optical emission from GHZ1. We used a spectro-photometric SED fitting to characterize the stellar evolution and the emission line spectra to constrain its chemical enrichment.}
{We find that GHZ1 is among the most massive systems at $z\sim 10$ (log(M$_{\star}$ [M$_{\odot}$]) = 9.17$_{-0.26}^{+0.27}$), with a moderate amount of dust attenuation ($A_{V} = 0.21_{-0.12}^{+0.12}$~mag) and ongoing star formation (with a median SFR $= 8.75_{-1.72}^{+1.71}$ M$_{\odot}\cdot$yr$^{-1}$ over the last 10 Myr). Its chemical enrichment patterns (12+log(O/H) = $7.81_{-0.11}^{+0.12}$, log(C/O) = $-0.66_{-0.13}^{+0.13}$, log(N/O)$< -1.04$) are consistent with a standard evolution dominated by primary production from massive stars. We also derive an \Halpha/\Hbeta Balmer ratio below Case B predictions. We speculate that this scenario can be caused by: \textit{i)} a gas of neutral H that makes Balmer lines optically thick; \textit{ii)} intrinsic line variability as \Halpha and the rest of Balmer lines were observed at different epochs ($\sim 45$ days rest frame); or, iii) complex outflow-driven kinematics affecting these lines. Overall, the derived properties of GHZ1 meet the expectations for a massive star-forming dominated system, although its atypical ionizing conditions require further investigation in order to anchor the differences in galaxy evolution at high $z$.}
{}

\maketitle


\section{Introduction}
\label{intro}
Our understanding of early galaxy evolution has been revolutionized by the first years of observations with \jwst. The discovery of a high abundance of UV-bright galaxies in the first 500 Myr of the Universe has been a remarkable result, unexpected on the basis of pre-JWST predictions \citep[e.g.][]{Castellano_2022, Naidu_2022, Harikane_2023, Perez-Gonzalez_2023, Donnan_2024}. As a result, ongoing efforts are devoted to understanding which physical processes might be responsible for such excess, both from theoretical \citep[e.g.][]{McGaugh_2024, Menci_2024, Ferrara_2025, Mauerhofer_2025, Burgarella_2026} and observational \citep[e.g.][]{Whitler_2023, Dressler_2024, Kokorev_2025, Santini_2026} perspectives.

Whereas large and deep photometric surveys are necessary to create a census of these galaxies and to expand their search to higher redshifts \citep[e.g.][]{Rodighiero_2023, Perez-Gonzalez_2025, Castellano_2025, Gandolfi_2026}, spectroscopic follow-up campaigns are essential to confirm \citep[e.g.][]{Arrabal-Haro_2023, Napolitano_2025a,Zavala_2025, Tang_2026, Alvarez-Marquez_2026} and to understand which physical processes are governing galaxy evolution at such early times. The ionized gas-phase of the interstellar medium (ISM) offers a unique window not only to assess the physical conditions and complement the information from photometric observations, but also to explore the chemical footprints of these processes \citep{Maiolino_2019}.

Indeed, spectroscopic follow-ups of these $z \gtrsim 10$ systems revealed that they harbor an ISM in extreme conditions \citep[e.g.][]{Calabro_2024, Hariakane_2025, Roberts-Borsani_2026, Castellano_2026}, likely connected to their unusual UV-brightness, and in many cases showing a compact structure \citep[e.g.][]{Cameron_2023, Senchyna_2024, Castellano_2026}. The analysis of the ISM chemical composition of many of these sources yielded another peculiarity: many of these compact objects are characterized by a high N abundance (log(N/O) > -1.0) for their O abundances (12+log(O/H) < 8.1), in contrast to standard models of chemical evolution that predict an almost flat abundance (log(N/O)$\sim$-1.4) due to the primary production by massive stars \citep[e.g.][]{Henry_2000, Nicholls_2017}. Moreover, some of them show strong evidence for early ignition of Active Galactic Nuclei (AGN), as it is the case for GHZ9 \citep{Napolitano_2025b}, not only characterized by their compactness, but also for their X-ray emission \citep{Kovacs_2024}.

A great variety of scenarios have been proposed to explain these anomalous log(N/O) ratios. \citet{Berg_2026} found direct evidence for Wolf-Rayet (WR) stars in RXCJ2248 ($z\sim
6$). These stars are found after $~3-6$ Myr of bursts of star formation and they are located within regions that display an enhancement of N and He due to their mass loss through stellar winds \citep{Esteban_1995}. Very Massive Stars (VMS, $> 100$ M$_{\odot}$) can produce elevated N/O ratios due to a rapid mass-loss \citep[e.g.][]{Vink_2023}. Medium-resolution observations of CEERS-1019 ($z \sim 8.67$) have revealed the presence of VMS signatures in their spectrum \citep{Marques-Chaves_2026b}. Other scenarios include differential galactic winds driven by core-collapse supernovae \citep{Rizzuti_2025} or formation of stars over a pre-enriched gas \citep{McClymon_2026}, or even more speculative ones invoking supermassive stars (SMS, $> 10^{4}$ M$_{\odot}$, e.g.\citealp{Charbonnel_2023}).

However, available NIRSpec-based constraints suffer from significant limitations. In fact, beyond $z \gtrsim 7$, NIRSpec cannot cover the \NIIL[6548,6584] lines, so the bulk of studies on N-enhanced objects at high-$z$ rely on the measured UV lines, which are enhanced in the most extreme conditions (high densities, stellar atmospheres contamination, ...), with some exceptions measured from their optical counterparts \citep{Stiavelli_2025, Cameron_2026, Tripodi_2026}. Indeed, \citet{Zhu_2025} showed that mainly N-enhanced systems exhibit detectable \NIII and \NIV UV lines. The follow-up with \jwst/MIRI of the well established N-enhanced system GN-z11 showed no detections of \NIIL[6548,6584] \citep{Alvarez-Marquez_2025}, favoring again that the major drivers of such conditions are the regions traced by the UV lines. Whereas the UV emission has been interpreted as dominated by AGN activity \citep{Maiolino_2024}, the complete coverage of its UV-optical rest-frame emission shows increasing evidence that massive stars are the dominant source of ionization \citep{Alvarez-Marquez_2025, Chen_2026, Nakane_2026}. 

The PRImordial galaxy Survey with MIRI Spectroscopy (PRISMS) program (ID 8051, PIs: Javier Álvarez-Márquez and Luis Colina; see also \citealt{Alvarez-Marquez_2026, Marques-Chaves_2026}) aims precisely at analyzing early galaxy evolution ($9.7 < z < 10.4$) by observing Intermediate-UV ($-20.5 \leq M_{UV} \leq -17.6$) galaxies, by means of a coherent analysis of their UV and optical emission, which necessarily requires NIRSpec and MIRI spectroscopy. In this work, we present a detailed analysis of GHZ1 (RA=3.5119, Dec=-30.3719), initially identified as a $z>10$ galaxy candidate based on the \jwst/NIRCam photometry in the GLASS field \citep{Treu_2022} by \citet{Castellano_2022}. The follow-up analysis presented by \citet{Napolitano_2025a}, based on NIRSpec/PRISM data, confirmed it at a slightly lower spectroscopic redshift ($z = 9.875$). GHZ1 is notable for its extended size ($R_{e}\sim$ 410 pc, \citealt{Yang_2022}) and stellar mass ($\log(\mathrm{M}_{\star}[\mathrm{M}_{\odot}]) \sim 8.9$, \citealt{Santini_2023}), making\footnote{We have corrected the original values from magnification, as they were published before a lensing model was available ($\mu = 1.715$, \citealt{Bergamini_2023}). We also correct them to the spectroscopic redshift $z = 9.878$ derived in this work.} it one of the most massive and extended objects at Cosmic Dawn. 

\begin{figure}[h!]
    \centering
    \includegraphics[width=0.99\linewidth]{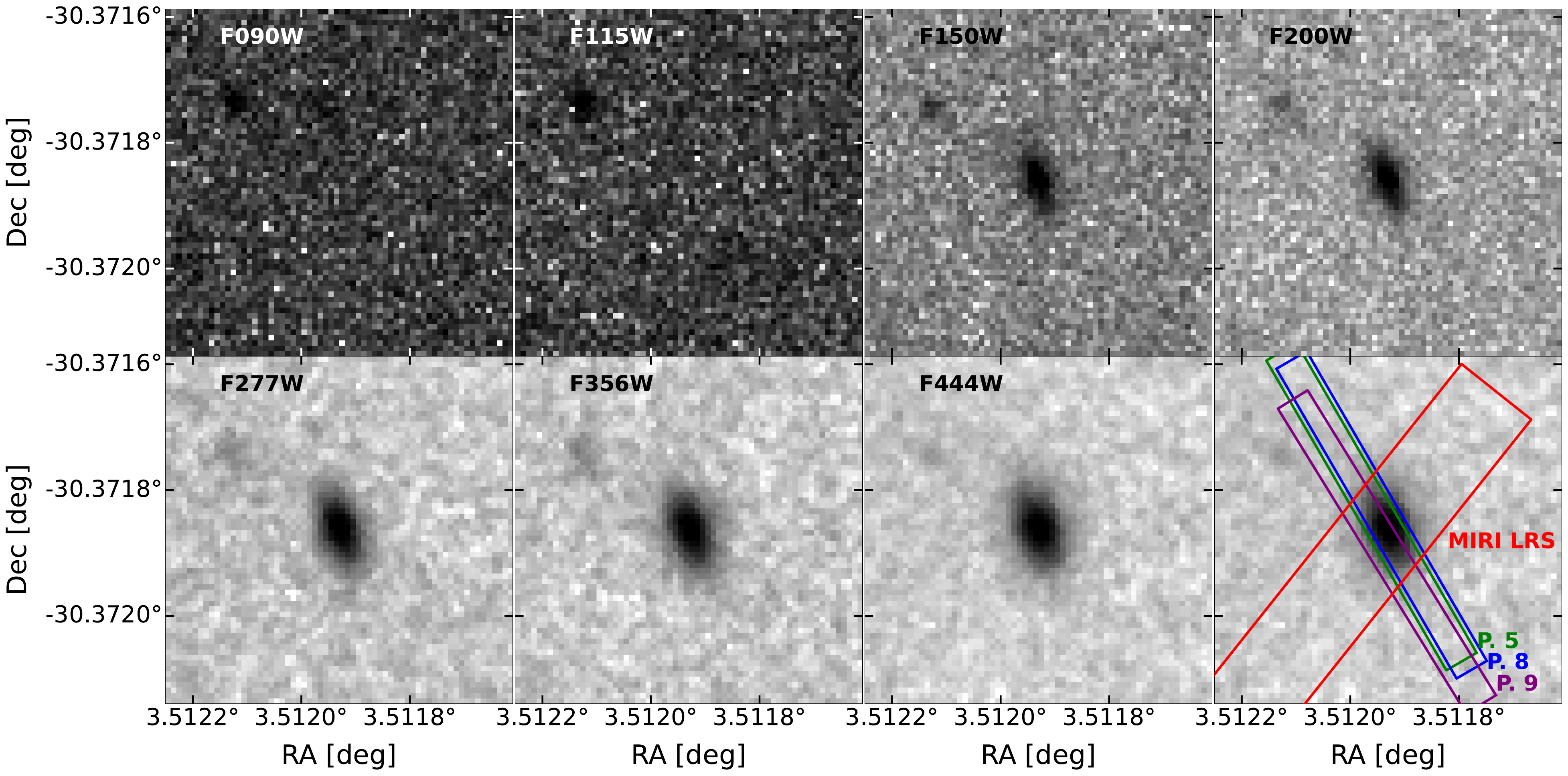}
    \caption{Mosaic of \jwst/NIRCam broadband images for GHZ1. From top to bottom, left to right, we show the cutouts (2 arcsec by side) centered in GHZ1 for F090W and F115W (non-detections) and F150W, F200W, F277W, F356W and F444W \citep[ASTRODEEP][]{Merlin_2024}. The last image shows the F444W image with the slits positions for the NIRSpec/PRISM (green, blue and purple for each pointing) and MIRI/LRS (red) observations of the target}. 
    \label{Mosaic_nircam}
\end{figure}

The analysis of these systems is essential to achieve a complete picture of early galaxy evolution. First, to complete the picture of the mass-metallicity relation, we need to constrain the high-mass end of these high-$z$ systems, where such relations are poorly sampled \citep[e.g.][]{Nakajima_2023, Curti_2024, Koller_2026, Pollock_2026}. Second, the properties of massive extended systems at $z \sim 10$ can yield tight constraints on the ubiquity and/or duration of extreme phases that might alter their chemical enrichment, and enable quantifying the role of compactness in the evolutionary pathways of high-$z$ sources \citep[e.g.][]{Charbonnel_2023, Schaerer_2024}. Third, the prevalence of dust at high-$z$ needs to be explored in massive systems to constrain theoretical scenarios that explain the high abundance of ultraviolet-bright galaxies \citep[e.g.][]{Ferrara_2025, Burgarella_2026}.

This paper presents a detailed analysis of the chemical and physical properties of GHZ1, a continuation of a series of works of the PRISMS project \citep{Alvarez-Marquez_2026, Marques-Chaves_2026}. The manuscript is organized as follows. Sec. \ref{s: data} introduces the MIRI and NIRSpec observations and outlines the key elements of the data reduction processes. Sec. \ref{s: methods} presents the analysis of the main spectroscopic features (emission lines, continuum,...) and the spectral energy distribution (SED). Sec. \ref{s: characteristic} provides a detailed analysis of the characteristic physical and chemical properties of the source. Sec. \ref{s: discussion} offers a motivated discussion of GHZ1 properties in the context of high-$z$ sources. Finally, Sec. \ref{s: conclusions} summarizes the main findings of this work.

Throughout this work, we have assumed as solar abundances the photosphere measurements reported for the Sun \citep{Asplund_2009}, the AB magnitude system \citep{Oke_1983} and a flat \ensuremath{\Lambda}CDM cosmological model with $\Omega_{M} = 0.31$ and $H_{0} = 67.7$ km$\cdot$s$^{-1}\cdot$Mpc$^{-1}$ \citep{Planck_2020}.

\section{Data}\label{s: data}
In this section, we describe the new data obtained with MIRI/LRS for GHZ1, along with ancillary data from NIRSpec/PRISM observations. We provide a quick summary of the most important steps regarding the data reduction processes, while further details can be found in the works presented by \citet{Alvarez-Marquez_2026} (MIRI/LRS) and \citet{Napolitano_2025a} (NIRSpec/PRISM).

\subsection{MIRI/LRS}\label{ss: miri}
MIRI/LRS \citep{Kendrew_2015} observations for GHZ1 were carried out as part of Cycle 4 program ID 8051 (PIs: Javier Álvarez-Márquez and Luis Colina). The configuration of the observations (visit 17) relied on a two-column mosaic and a two-point dither pattern, which gives four equidistant positions along the slit (each of them using 149 groups and 14 integrations in FASTR1 mode). The total exposure time amounts to 23,299 s ($\sim 6.5$ h).

The 2D and 1D calibrated spectra are retrieved following the methodology described by \citet{Alvarez-Marquez_2026}. In short, data were processed with pipeline v1.20.2 \citep{Bushouse_2025} and context 1464 of the Calibration Reference Data System. Stages one and two were performed following the standard procedure, although for the \texttt{jump} step the \texttt{find\_showers} keyword  was activated to search and correct the cosmic ray showers event. Before stage three, we run the same three customized steps outlined in \citet{Alvarez-Marquez_2026}: {\it i)} a wavelength masking to constrain the final spectra in the [4.85$\mu$m, 8.05$\mu$m] range; {\it ii)} a master background correction, individually subtracted from each of the images; and, {\it iii)} a residual background subtraction to correct for temporal background variations and gradients from the mean correction.

During stage three, the final 1D spectrum is obtained using an aperture equal to 0.44$^{\prime\prime}$, which accounts for $~75\%$ of the encircled energy for an unresolved source at 5$\mu$m. Given the MIRI PSF size at 5$\mu$m ($\sim0.27^{\prime\prime}$), the apparent size of GHZ1 \citep[$\sim0.1^{\prime\prime}$,][]{Yang_2022} makes the source fully compatible with the point-like assumption. The noise level for the final 1D spectrum is estimated from two background extracted spectra over the full exposure of the observations \citep[see][for further details]{Alvarez-Marquez_2026}.

\subsection{NIRSpec/PRISM}\label{ss: nirspec}
GHZ1 was observed with \jwst/NIRSpec in the PRISM-CLEAR configuration during the Cycle 2 program GO-3073 (PI: Marco Castellano). The observing configuration was designed based on a three-shutter slit with a three-point nodding pattern, yielding a total exposure time of 19,701 s ($\sim 5.5$ h) for the three visits. To obtain calibrated 2D and 1D spectra we used the same data reduction process described in detail by \citet{Napolitano_2025a} and \citet{Castellano_2026}. In short, data were processed with the standard calibration pipeline v1.13.4, using the Calibration Reference Data System mapping 1197, and replicating the methodology outlined in \citet{Arrabal-Haro_2023}.

As GHZ1 stands out for being an extended source (see Fig. \ref{Mosaic_nircam}), we account for any potential slit loss in the spectroscopic observations. Hence, we compare the observed NIRCam photometry \citep[Bands F150W, F200W, F277W, F356W, F444W from the ASTRODEEP catalog,][]{Merlin_2024} to the synthetic photometry obtained by multiplying the reduced NIRSpec/PRISM spectrum by the corresponding throughput curves\footnote{The official datasets are available at \url{https://jwst-docs.stsci.edu/jwst-near-infrared-camera/nircam-instrumentation/nircam-filters}.}.

On average, we find that the observed photometry is consistent ($0.97\pm0.05$) with the synthetic photometry retrieved from the NIRSpec/PRISM spectrum. Hence, we do not apply any slit loss correction factor.

\subsection{Cross-calibration and combined NIRSpec and MIRI spectra}\label{ss: combined}

\begin{figure*}[h!]
    \centering
    \includegraphics[width=1.0\linewidth]{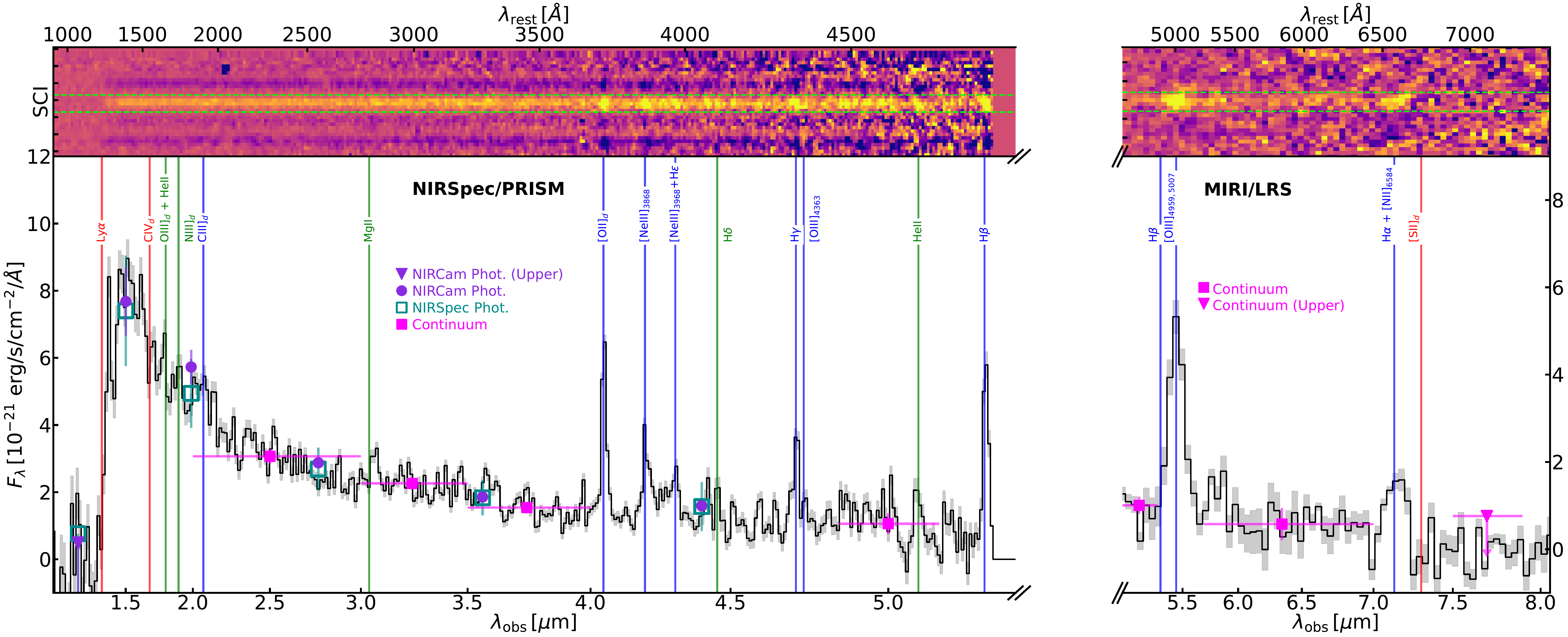}
    \caption{The combined NIRSpec/PRISM (left) and MIRI/LRS spectrum (right). The two upper panels show the corresponding 2D spectra for each instrument, showing the extraction window for the 1D spectrum with lime horizontal lines. The lower panels show the extracted 1D spectrum, and the errors are shown as gray-shaded regions for each bin. JWST/NIRCam photometric measurements from ASTRODEEP \citep{Merlin_2024} are shown as purple circles, or triangles for upper limits, while NIRSpec synthetic photometry derivations are shown as darkcyan open squares. Similarly, continuum detections are shown as magenta squares, or triangles for upper limits, with horizontal error bars showing the range for stacking. Vertical lines identify key features: red for undetected lines, green for tentative features (2$<$S/N$<$3) and blue for detected emission lines (S/N$\geq$3).} 
    \label{Spec_nirmir}
\end{figure*}

We show in Fig. \ref{Spec_nirmir} the individual MIRI/LRS (right plot) and NIRSpec/PRISM (left plot) spectra for GHZ1. The NIRCam coverage (purple points) shows the agreement between the observed photometry and the spectrum. We also include the detections of free-line continuum emission (magenta squares) at 2.5, 3.3, 4.05, 5.05 and 6.35 $\mu$m, and a $3\sigma$ upper limit (magenta triangle) for the reddest continuum at 7.7 $\mu$m observed with MIRI/LRS.

The detection of free-line continuum emission at $\lambda = 4.9-5.2$~$\mu$m in both NIRSpec/PRISM and MIRI/LRS, allows us to perform a cross-calibration between the two instruments. The continuum flux derived from NIRSpec/PRISM yields 1.06($\pm$0.30)$\times$10$^{-21}$erg$\cdot$s$^{-1}\cdot$cm$^{-2}\AA^{-1}$, while from MIRI/LRS we derive a median flux of 1.01($\pm$0.19)$\times$10$^{-21}$ erg$\cdot$s$^{-1}\cdot$cm$^{-2}\AA^{-1}$, demonstrating consistency between the two instruments within the uncertainties. Hence, no scaling factor is applied to cross-calibrate them.

We combine NIRSpec/PRISM and MIRI/LRS observations by masking the \Hbeta emission in the former, and including the MIRI/LRS emission from the overlapping region ($> 5.0\ \mu$m) between both instruments. We use this combined spectrum (\texttt{GHZ1-spec}), together with the observed photometry, as inputs for the SED analysis of the main physical properties of GHZ1 (see Sec. \ref{ss: SED_fitting} for further details).

\section{Methods}\label{s: methods}

\subsection{Emission line fitting}\label{ss: lines}
Emission lines were fitted following a similar approach to that presented by \citet{Napolitano_2025a}. Below we summarize our Markov chain Monte Carlo (MCMC) method, implemented with the \textsc{EMCEE} package \citep{emcee}.

Following \citet{Napolitano_2025a}, we identify any tentative emission line feature over a range marked by four times the Gaussian standard deviation ($\sigma_{\lambda}$) given by the spectral resolution of each instrument. In the case of partially blended lines (e.g. \Hgamma and \OIIIL[4363]), the window is cropped to 3$\sigma_{\lambda}$. The local continuum is estimated from two windows, each $5\sigma_{\lambda}$ wide, located immediately bluewards and redwards of the emission-line window. Using the resulting fitting region, spanning $14\sigma_{\lambda}$ in total, we then simultaneously fit the continuum (linear fit), a single Gaussian fit for individual emission lines and unresolved doublets/multiplets or multiple Gaussians if they can be resolved.

As for any MCMC approach, the initial \textit{walkers} (i.e. the independent initial values) are critical for the correct implementation of the MCMC fit. As initial points for the continuum fit, the flux in the bluest and reddest-most wavelength ranges is used to infer the linear slope. These values are then varied within 50\% of uncertainty, but adding as a prior the non-negative value of the intercept. As the instrumental resolution for both NIRSpec/PRISM and MIRI/LRS dominates any broadening of the lines, the Gaussian standard deviation is limited to that given by the resolution of the corresponding instrument ($\sigma_{\lambda}$), allowing a variation of 5\% (20\%) of uncertainty for NIRSpec (MIRI) observations. The centroids of the Gaussians are also limited to the maximum values observed in the corresponding ranges, allowing for a 5\% uncertainty as well. The best-fit model is then taken from the median of the posterior distributions, and the uncertainties from the corresponding 68th percentile.

The flux of each emission line is calculated in two different ways depending on the nature of the lines. For unresolved doublet/multiplets or single lines, the continuum-subtracted flux is computed from a direct integration (by means of the Simpson integration method) of the observed flux over the 4$\sigma_{\lambda}$ range, and the corresponding uncertainty from the observed flux error and the propagated error from the integrated method. In partially blended emission lines, as the direct integration method cannot distinguish each contribution, the standard flux is given by the fitted Gaussian values, and the error comes from the propagation of such parameters. 

The rest-frame equivalent width (EW) is then computed from the ratio between the total flux (from the above procedure) and the expected continuum flux under the centroid ($\mu_{i}$) of the line. The uncertainty in the EW estimation is derived following Eq. 7 from \citet{Vollmann_2006}.

We show in Appendix~\ref{ap: lines} the best-fit models for all emission lines. We present in Table~\ref{Table_lines} the measurements for all detected emission lines. We also provide 3$\sigma$ upper limits for tentative (2 $<$ S/N $<$3) features. 

We measure the redshift of the source from the flux-weighted centroids of the brightest lines, namely \OIIL[3727], \NeIIIL[3868], \Hgamma, \Hbeta and \Halpha, in comparison to their corresponding vacuum wavelengths. The average redshift obtained for GHZ1 is $z = 9.878\pm0.007$, consistent with the value reported by \citet{Napolitano_2025a} ($z = 9.875\pm0.008$). We find consistent measurements for the fluxes and EWs of those lines reported by \citet{Napolitano_2025a}.

\begin{table}[h!]
	\caption{Detected (S/N $>$ 2) emission features.}
	\label{Table_lines}
	\centering
	\begin{tabular}{c | c  c  } 
		\hline\hline
		\textbf{Line} & \textbf{Flux} & \textbf{Equivalent Width} \\ 
		\textbf{(1)} & \textbf{(2)} & \textbf{(3)} \\ \hline
        \HeII+\sOIIIL & $<$ 3.8 & -\\
        \NIIIL[1746.1753] & $<$ 3.5 & -\\ 
		\CIIIL[1907,1909] & 3.9$\pm$1.1 & 8.7$\pm$4.9\\
        $^{\dagger}$\MgII & 3.3$\pm$0.6 & 13.7$\pm$4.1\\
        \OIIL[3726,3728] & 11.5$\pm$0.7 & 68.1$\pm$3.9\\
        $^{\dagger\dagger}$\NeIIIL[3868] & 5.1$\pm$0.6 & 25.4$\pm$3.9\\
        $^{\dagger\dagger}$\HIL[3885] & $<$ 2.2 &  - \\
        \NeIIIL[3967] & 3.3$\pm$0.4 & 19.9$\pm$3.9 \\
        \Hdelta & $<2.0$ & - \\
        \Hgamma & 5.1$\pm$0.8 & 37.5$\pm$13.5 \\
        \OIIIL[4363] & 1.4$\pm$0.4 & 11.6$\pm$5.4 \\
        \HeIIL[4686] & $< 3.5$ & - \\
        $^{\ddagger}$\Hbeta & 10.3$\pm$0.5 & 174.0$\pm$18.8 \\
        $^{\ddagger}$\Hbeta+\OIII & 79.4$\pm$8.8 & 785$\pm$97 \\
        \Halpha+\NII & 19.1$\pm$3.9 & $897\pm468$ \\        
	\end{tabular}
	\tablefoot{Column (1) shows the identified feature. Column (2) gives the observed flux and error not corrected from magnification in 10$^{-19}$ erg$\cdot$s$^{-1}\cdot$cm$^{2}$. Column (3) gives the equivalent width and error in rest-frame $\AA$.\\ $^{\dagger}$ We do not consider the \MgII feature as real, since its velocity shift and appearance in some of the dithers suggest that the feature might not be real. \\$^{\dagger\dagger}$ The \HIL[3885] is listed here as being the result from the fully blended multi-gaussian fit to the \NeIIIL[3868] line. \\ $^{\ddagger}$ We list both the measurement of \Hbeta from NIRSpec and \Hbeta+\OIIIL[4959,5007] from MIRI (see Appendix \ref{ap: lines}).}
\end{table}

\subsection{The UV shape: slope and magnitude}
UV properties are essential to fully understand the ionizing mechanisms responsible for the observed emission. Two quantities are particularly useful due to their simplicity and accessibility from the data: the UV slope ($\beta_{UV}$) and UV magnitude ($M_{UV}$). 

We derive the UV slope by fitting a power-law continuum to the rest-frame UV emission (see Appendix \ref{ss: uv_slope} and Fig. \ref{fig_slope}), which leads to a robust measurement of $\beta_{UV} = -1.85\pm 0.05$. We compute the absolute UV magnitude by evaluating the observed spectrum at rest-frame $\lambda = 1500\AA$. From the redshift ($z = 9.878\pm0.007$) and magnification correction \citep[$\mu = 1.715$,][]{Bergamini_2023}, we find that the absolute UV magnitude is $M_{UV} = -20.07\pm 0.07$ mag. These values are consistent with those reported by \citet{Napolitano_2025a}.

\begin{figure}
\centering
\includegraphics[width=0.8\linewidth]{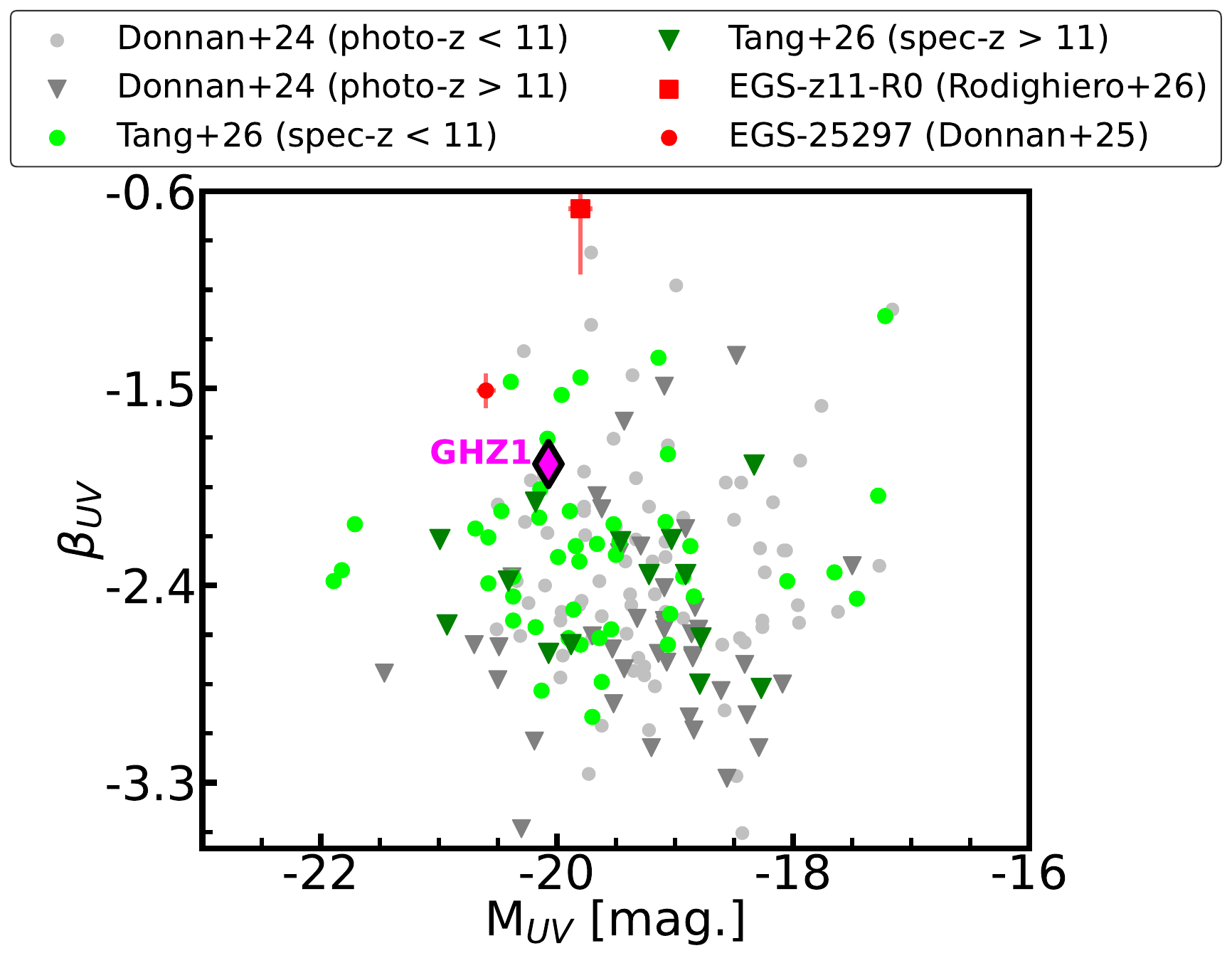}
\caption{Relation between the UV magnitude ($M_{UV}$) and the UV slope ($\beta_{UV}$) for GHZ1 compared to $z>9$ systems.}
\label{fig_uv}
\end{figure}

We show in Fig.~\ref{fig_uv} how GHZ1 compares to other high-$z$ ($>$9) sources in terms of these two UV properties. For comparison, we select the photometric catalog presented by \citet{Donnan_2024} and the spectroscopically-confirmed catalog from \citet{Tang_2026}. We also include two examples of the reddest galaxies spectroscopically confirmed: EGS-25297 \citep{Donnan_2025} and EGS-z11-R0 \citep{Rodighiero_2026}. We note that GHZ1 belongs to the population of bright UV galaxies, with a UV slope that is redder than the average of spectroscopic samples at $z\sim10$. Nonetheless, its properties are compatible with the sample within the scatter.

\subsection{SED fitting}\label{ss: SED_fitting}
We perform a spectro-photometric SED fitting using \textsc{bagpipes} v1.0.3 \citep{Carnall_2018, Carnall_2019}, for both the spectrum and photometric measurements of GHZ1. In particular, we introduced as input the broadband photometry from ASTRODEEP \citep{Merlin_2024} and \texttt{GHZ1-spec}, i.e., the combined emission of NIRSpec and MIRI after masking the \Hbeta emission in NIRSpec, for which we have a complete census of the emission lines from \CIIIL[1907,1909] to \Halpha and from the continuum up to $6500\AA$. All observables are introduced after magnification correction \citep[$\mu = 1.715$,][]{Bergamini_2023}.

We use as stellar templates the BPASS v2.1 models \citep{Eldrige_2017}, assuming an initial mass function based on \citet{Kroupa_1993} with an upper cut-off in stellar mass of 300 M$_{\odot}$. We assume the non-parametric star formation history (SFH) model given by \citet{Iyer_2019} with five lookback time bins and with a star formation rate (SFR) prior in the range 0.01-50 M$_{\odot}\cdot$yr$^{-1}$. We fit dust attenuation by assuming the \citet{Cardelli_1989} law, with a prior on the attenuation magnitude given by $A_{V} \in [0, 2.0]$. Nebular emission is modeled following the default prescriptions in \textsc{bagpipes} \citep{Carnall_2019}, but assuming higher density values log($n_{e}$ [cm$^{-3}$]) = 3, a tight metallicity\footnote{The ionizing stellar populations are characterized by the same metallicity as the gas.} Z $\in [0.1, 0.15]$ Z$_{\odot}$ and ionization priors log$U$ $\in [-2.6, -2.3]$. These constraints on the nebular conditions are motivated by the emission line analysis presented in Sec. \ref{ss: phys_chem}.

A proper characterization of the systematics involving the SED fitting requires an exploration of the impact of the most critical inputs assumed: the SFH model and the attenuation curve. For the former, we test different star formation history (SFH) models, namely: delayed exponentially declining, bursts with continuity prior, and double power-law. For the latter, we have also explored two other attenuation curves, namely \citet{Calzetti_2000} and \citet{Salim_2018}.

\begin{figure}
\centering
\includegraphics[width=0.9\linewidth]{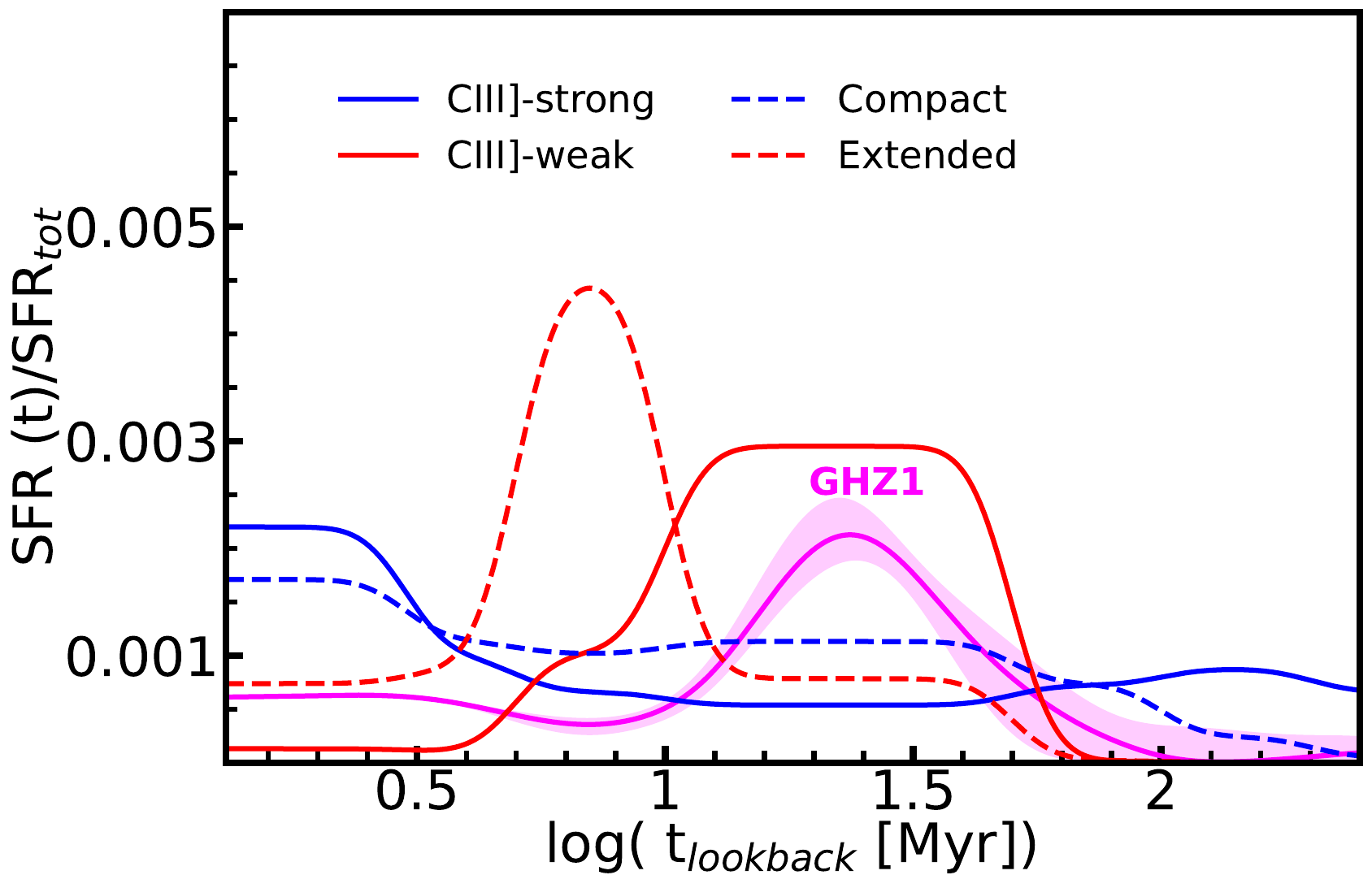}
\caption{Reconstructed SFH for GHZ1, normalized to the total star formation rate over its age (magenta line). The magenta shaded area represents the upper and lower limit uncertainties for the SFH. The average SFHs for \CIII-strong (continuous blue), \CIII-weak (continuous red), compact (dashed blue) and extended (dashed red) sources from \citet{Roberts-Borsani_2026} are shown as reference samples at $z \sim 10$. SFHs have been smoothed with a gaussian kernel for graphical representation.}
\label{Fig_sfh}
\end{figure}

The best fit model provided by \textsc{bagpipes} points\footnote{The uncertainties provided here are the result of the variations for each property under different SFH models and attenuation curves convolved with the intrinsic uncertainty for each of them.} towards an ongoing burst of star formation (SFR = 8.25$_{-1.71}^{+1.72}$ M$_{\odot}\cdot$yr$^{-1}$ over the last 10 Myr) and total stellar mass of log(M$_{\star}$ [M$_{\odot}$]) = 9.17$_{-0.26}^{+0.27}$, magnification corrected. This mass value is slightly higher than the one reported by \citet{Santini_2023}, which was derived assuming an over-simplified SFH, not appropriate for high redshift galaxies \citep[see][]{Santini_2026}. The attenuation is fit to $A_{V} = 0.21_{-0.12}^{+0.12}$. Based on the best fit values, GHZ1 is characterized by a specific star formation rate (sSFR) of $\sim10^{-8.2}$ yr$^{-1}$. Taking as reference value the cosmic sSFR = 0.2 t$_{H}^{-1}$($z$) \citep[being t$_{H}$($z$) the Hubble time at redshift $z$,][]{Chaikin_2026}, the threshold to consider star forming galaxies at $z\gtrsim10$ is $\sim 10^{-9.4}$ yr$^{-1}$, qualifying GHZ1 as a robust star forming dominated system. This sSFR is also below the threshold \citep[sSFR$^{\star} $= 10$^{-7.6}$ yr$^{-1}$,][]{Ferrara_2024} for these high-$z$ systems to become super-Eddington and develop outflows that might expel their dust content, consistent with the redder ($\beta_{UV} = -1.85$) UV slope measured.

The SFH retrieved also provides an important look at the growth history of GHZ1. For comparison, we include some of the results discussed by \citet{Roberts-Borsani_2026} for $z \gtrsim 10$ sources. In particular, we compare GHZ1 to two groups of opposite systems: \CIII weak (EW(\CIII) $<10\AA$) vs \CIII strong (EW(\CIII) $>10\AA$), and extended ($R_{e}>400$ pc) vs compact ($R_{e} < 100$ pc). We show in Fig. \ref{Fig_sfh} the reconstructed SFH for GHZ1 compared to these four groups of systems, assuming for consistency the same attenuation curve \citep{Salim_2018}. While the properties of GHZ1 (EW(\CIII) $=8\pm5\ \AA$; $R_{e}\sim410$ pc) qualify this system as \CIII-weak and extended, the direct comparison shown in Fig. \ref{Fig_sfh} reveals that GHZ1 actually shows a mixed behavior. Whereas it is characterized by a strong, old\footnote{We note that the age for this old stellar population is highly dependent on the dust attenuation curve assumed: for \citet{Salim_2018} we derive an age of $\sim 40$ Myr, whereas for \citet{Cardelli_1989} is $\sim 120$ Myr.} burst like \CIII-weak or extended sources, its shape and duration show a composite picture, being as old as for \CIII-weak sources, but narrower as extended sources. Additionally, GHZ1 shows a rising star formation, although not as strong as those observed in \CIII-strong or compact sources.

\subsection{Nebular dust-attenuation from Balmer ratios}\label{ss: reddening}
The combined NIRSpec/PRISM and MIRI/LRS data allow us to constrain dust reddening by measuring their Balmer decrements. In particular, we use the \Hgamma and \Hbeta detections from NIRSpec/PRISM and the \Halpha+\NIIL feature from MIRI/LRS. In this last case, we cannot deblend the \NII contribution, but according to photoionization modeling predictions, such contamination is of the same order as the observational error ($\sim$10$^{-19}$ erg$\cdot$s$^{-1}\cdot$cm$^{-2}$). This is consistent with the non-detections of \NII emission in MIRI/MRS observations, as it is the case for MACS0647{\ensuremath{-}}JD \citep{Hsiao_2024} or GN-z11 \citep{Alvarez-Marquez_2025}. We have also used the upper limit on \Hdelta to check the consistency in dust attenuation.

We compute the predicted Balmer decrements assuming Case B recombination conditions, the average electron temperature ($T_{e} = 1\cdot10^{4}$K), and under two different assumptions for the density conditions, namely $n_{e} = 10^{2}$ cm$^{-3}$ and $n_{e} = 10^{6}$ cm$^{-3}$. Table~\ref{Balmer_ratios} shows the relevant predictions and the observed ratios.

\begin{table}[h!]
	\caption{Theoretical (Case B) and observed Balmer ratios.}
	\label{Balmer_ratios}
	\centering
	\begin{tabular}{c | c  c  c} 
		\hline\hline
		\textbf{Ratio} & \boldmath{$n_{e} = 10^{2}$} & \boldmath{$n_{e} = 10^{6}$} & \textbf{Observed}\\ 
		\textbf{(1)} & \textbf{(2)} & \textbf{(3)} & \textbf{(4)}\\ \hline
        \Halpha/\Hbeta & 2.863 & 2.803 & 1.85$\pm$0.43 \\
        \Halpha/\Hgamma & 6.118 & 5.958 & 3.63$\pm$0.91 \\
        \Hbeta/\Hgamma & 2.137 & 2.123 & 1.97$\pm$0.21 \\
	\end{tabular}
    \tablefoot{Density is shown in units of cm$^{-3}$.}
\end{table}

As reported in Table \ref{Balmer_ratios}, the observed Balmer ratios do not exhibit the characteristic excess of the redder wavelengths due to dust attenuation. Focusing our attention on \Halpha, we observe that all Balmer ratios involving this line yield systematically lower values than those predicted by the Case B recombination scenario, which can be considered as a strong indication for the deviation from Case B (see Sec. \ref{ss: missing_ha} for further details). On the other hand, the \Hbeta/\Hgamma ratio is consistent with the theoretical predictions. When assuming no dust attenuation, the \Hdelta emission is predicted to be F(\Hdelta)$\sim 2.8\times10^{-19}$ erg$\cdot$s$^{-1}\cdot$cm$^{-2}$, which is above the upper limit measured in GHZ1.

We perform alternative fits to the \Halpha emission, testing different trends for the underlying continuum shape. We find that all fits lead to Balmer ratios below Case B predictions (\Halpha/\Hbeta $\in$ [1.5, 2.2]). Similarly, we have explored alternative reductions for the MIRI (Appendix \ref{ss: sarah}) and NIRSpec (JADES reduction, Appendix \ref{ss: stefano}; DJA reduction, Appendix \ref{ss: dja}) data. Overall, we conclude that all reductions lead to \Halpha/\Hbeta ratios below Case B ($\leq 2.3$), although the tensions are reduced in some cases.

We investigate in Appendix \ref{ss: 2d_spec}
any potential losses from the extraction window used in the 2D spectra, and we find no evidence of significant extended line emission as compared to the adjacent continuum emission that could have reconciled the observed ratio. We conclude that the \Halpha/\Hbeta ratio is systematically lower than Case B, despite the tension being partially reduced in some of our tests. We further explore the physical interpretations in Sec. \ref{ss: balmer_anom} and discuss the potential implications of these findings in Sec. \ref{ss: missing_ha}.
\section{Characterizing GHZ1: from ionizing conditions to chemical enrichment}\label{s: characteristic}
The flux of any emission line depends on several factors: the electron conditions within the ISM ($T_{e}$ and $n_{e}$), the ionizing structure (usually inferred from global quantities as the ionization parameter\footnote{The ionization parameter is defined as $U \equiv Q(H)/(4\pi R^{2}_{0}n_{e}c)$, being $Q(H)$ the number of ionizing photons per second, $R_{0}$ the distance of the inner face of the cloud to the ionizing source, and $c$ the speed of light. This parameter serves as a proxy of the coupling efficiency between the ionizing source and the surrounding ISM.} $\log U$) and the total amount of H, He and heavier elements (O, N, C, ...). In this section, we provide a quantitative and detailed analysis on such conditions.
\subsection{Ionizing mechanism}\label{ss: ionization}
The simultaneous detection of moderate- (e.g. \OIIIL[4959, 5007]) and low-ionized (e.g. \OIIL[3726,3729]) species allows us to fully explore the wide variety of conditions responsible for line emission in galaxies. Furthermore, the upper limits measured for other key high-ionized species (e.g. \HeIIL[4686]) will also provide meaningful constraints on such conditions.

\begin{figure}
\centering
\includegraphics[width=1.0\linewidth]{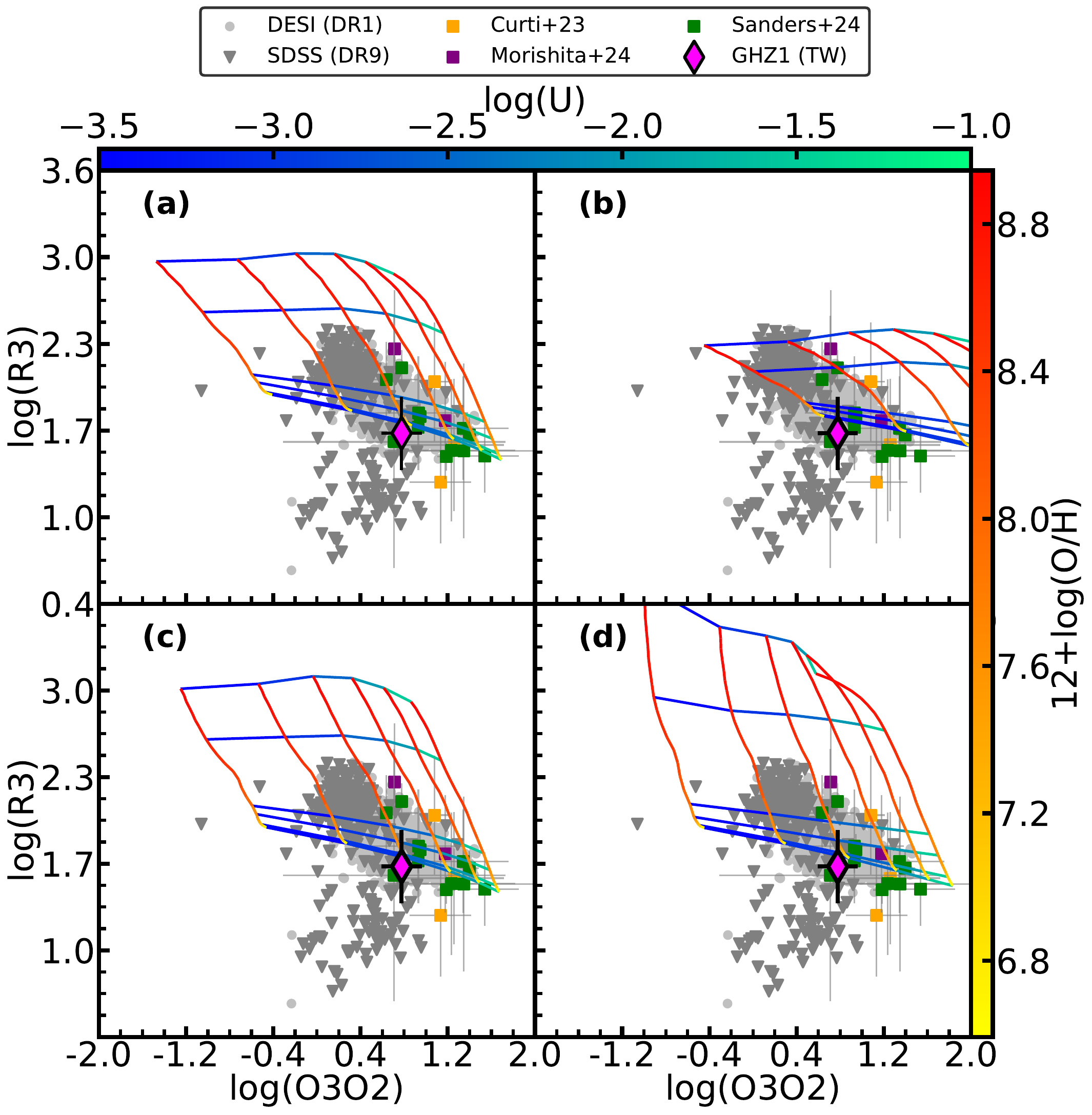}
\caption{Diagnostic diagram based on the R3 and O3O2 observables. The color grid shows the predictions of star forming dominated systems from photoionization models (see Appendix \ref{ap: photomodels}) for different scenarios. Panel (a): a burst of 1 Myr age and density $n_{e} = 10^{3}$ cm$^{-3}$. Panel (b): a burst of 1 Myr age and density $n_{e} = 10^{5}$ cm$^{-3}$. Panel (c): a burst of 3 Myr age and density $n_{e} = 10^{3}$ cm$^{-3}$. Panel (d): a burst of 3 Myr age, density $n_{e} = 10^{3}$ cm$^{-3}$ and without dust and depletion. References for the samples can be found in Table \ref{Tab_gal_ref}.}
\label{fig_r3o3o2}
\end{figure}

Given the great variety of O emission lines retrieved, we can perform a direct diagnostic that suppresses any uncertainty due to the metal abundances. In particular, we use the R3 ($\equiv$\OIIIL[4959, 5007]/\OIIIL[4363]) vs O3O2($\equiv$\OIIIL[4959, 5007]/\OIIL[3726,3728]) diagram \citep{Perez-Diaz_2026} which directly compares electron temperature (R3, i.e. metallicity) and ionization (O3O2). From Fig. \ref{fig_r3o3o2}, we can see that GHZ1 lies in the region populated by both local and high-$z$ star forming systems.

\begin{figure}
\centering
\includegraphics[width=1.0\linewidth]{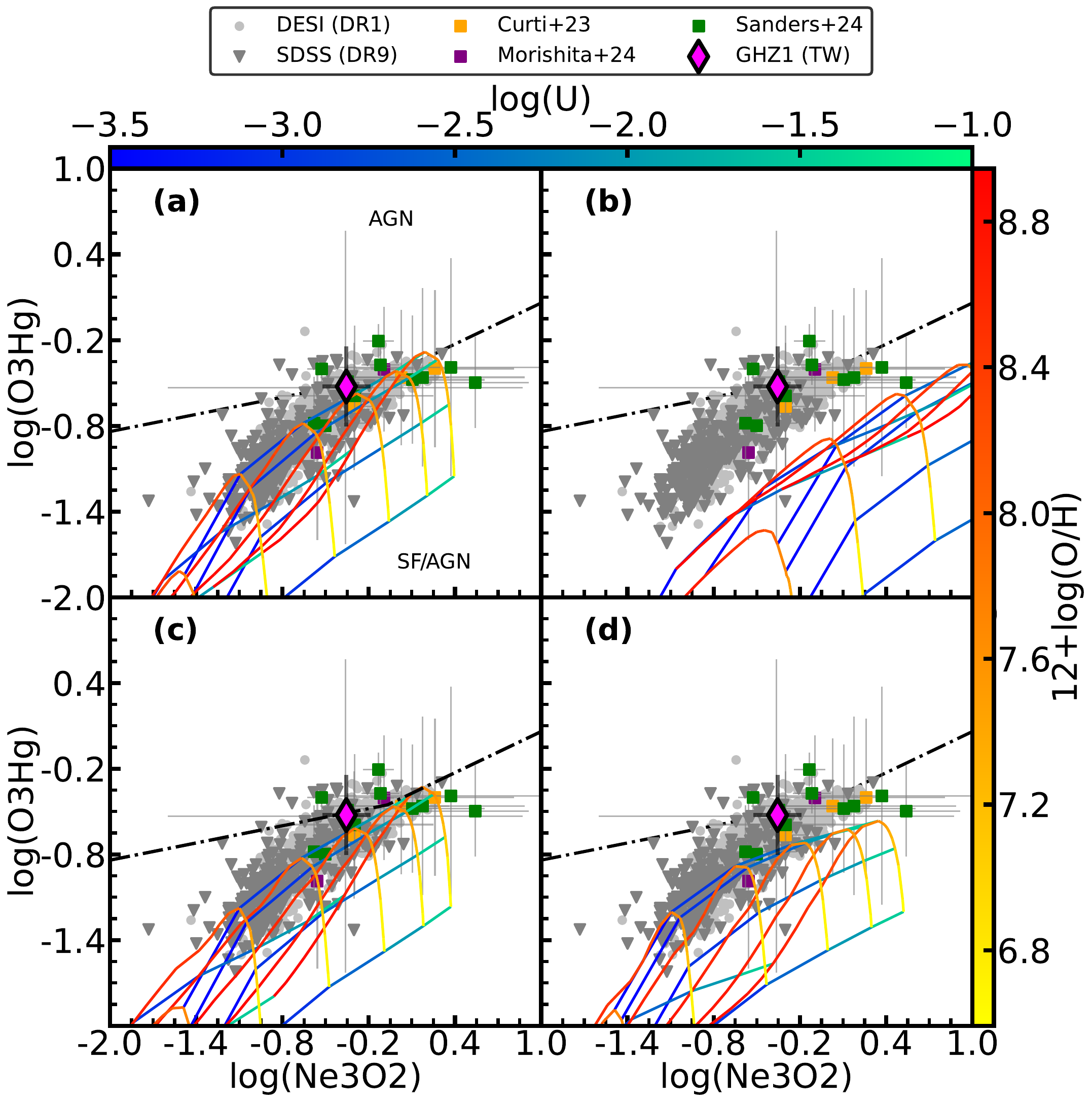}
\caption{Same as Fig. \ref{fig_r3o3o2} but for the O3Hg and Ne3O2 observables. The dashed-dotted line shows the demarcation proposed by \citet{Mazzolari_2024} to separate AGN-dominated systems.}
\label{fig_o3hgne3o2}
\end{figure}

We also analyze the O3Hg ($\equiv$\OIIIL[4363]/\Hgamma) vs Ne3O2 ($\equiv$\NeIIIL[3868]/\OIIL[3727]) diagram, recently proposed by \citet{Mazzolari_2024} as a powerful diagnostic for the AGN nature of a source. As AGN radiation fields produce a larger number of high-ionizing photons than those from young, massive stars, there is an increase in the upper levels of O$^{++}$ leading to brighter \OIIIL[4363] emission. From Fig. \ref{fig_o3hgne3o2}, we cannot disentangle whether there is some AGN contribution, but certainly this would be minor compared to that powered by star formation.

We inspect the profile of the Balmer lines for the presence of a broad component. From NIRSpec/PRISM, the width of \Hgamma ($\sigma = 501 \pm 35$ km$\cdot$s$^{-1}$) and \Hbeta ($\sigma = 417\pm 58$ km$\cdot$s$^{-1}$) is consistent with the spectral resolution (504 km$\cdot$s$^{-1}$ and 397 km$\cdot$s$^{-1}$ respectively). The \Halpha emission line width ($\sigma = 2011 \pm 453$ km$\cdot$s$^{-1}$) is slightly broader than the value predicted from the instrumental resolution curve ($\sim 1500$ km$\cdot$s$^{-1}$). Overall, the broadening of the Balmer lines is consistent within $\sim 1\sigma$ with that produced by the instrumental resolution. However, it must be noted that if there is any broader counterpart in \Hgamma or \Hbeta, considering the low S/N of the continuum for the current data, such component cannot be recovered. With the available low spectral resolution and S/N characteristics of the observations, the data do not support, neither discard, the presence of an AGN.

Additional hints for the AGN activity can be found from the detection of its X-ray emission and other indicators of hard ionizing fields. In particular, we found no X-ray signature coming from this source (see Appendix \ref{s:xray} for more details). The lack of highly-ionized species (see the upper limit on the \HeIIL[4686] feature) also disfavors the AGN scenario. From the upper limit on the X-ray luminosity in the 2-10 keV band (L$_{X} < 1.2\times10^{44}$ erg/s), based on the relations reported by \citet{Jin_2012}, the total AGN contribution to the optical emission (e.g. in terms of \Hbeta) is expected to be below $26\%$.

Overall, emission line diagnostics point towards either a partially or a non-contaminated SF dominated galaxy. Furthermore, its extended ($\sim 410$ pc) size contrasts to the expected compactness of $z \sim 10$ AGN-dominated sources \citep[e.g.][]{Napolitano_2025b}. In terms of the kinematic properties of Balmer lines, we only detect some slight broadening in the \Halpha feature, although it is mainly dominated by the instrumental broadening and relatively low S/N. While GHZ1 is consistent with a SF-dominated source, we cannot rule out the possibility that there might be some AGN contamination. 
\subsection{The physical origin of low Balmer ratios}\label{ss: balmer_anom}
Having assessed that the dominant source of ionization is star formation, we can deeply investigate the significantly low \Halpha/\Hbeta ratio reported in Sec. \ref{ss: reddening}. Deviations from Case B recombination predictions have been previously reported in the literature \citep[e.g.][]{Pirzkal_2024, Yanagisawa_2024, Scarlata_2024, McClymont_2025}, and interpreted under two main scenarios: 1) Balmer lines are not optically thin in the gas conditions for these objects \citep{Yanagisawa_2024}; or, 2) the emission lines mainly come from matter-bounded nebulae \citep{McClymont_2025}. 
\subsubsection{Radiation-bounded vs matter-bounded}\label{ss: matter-bounded}
\begin{figure}
\centering
\includegraphics[width=1\linewidth]{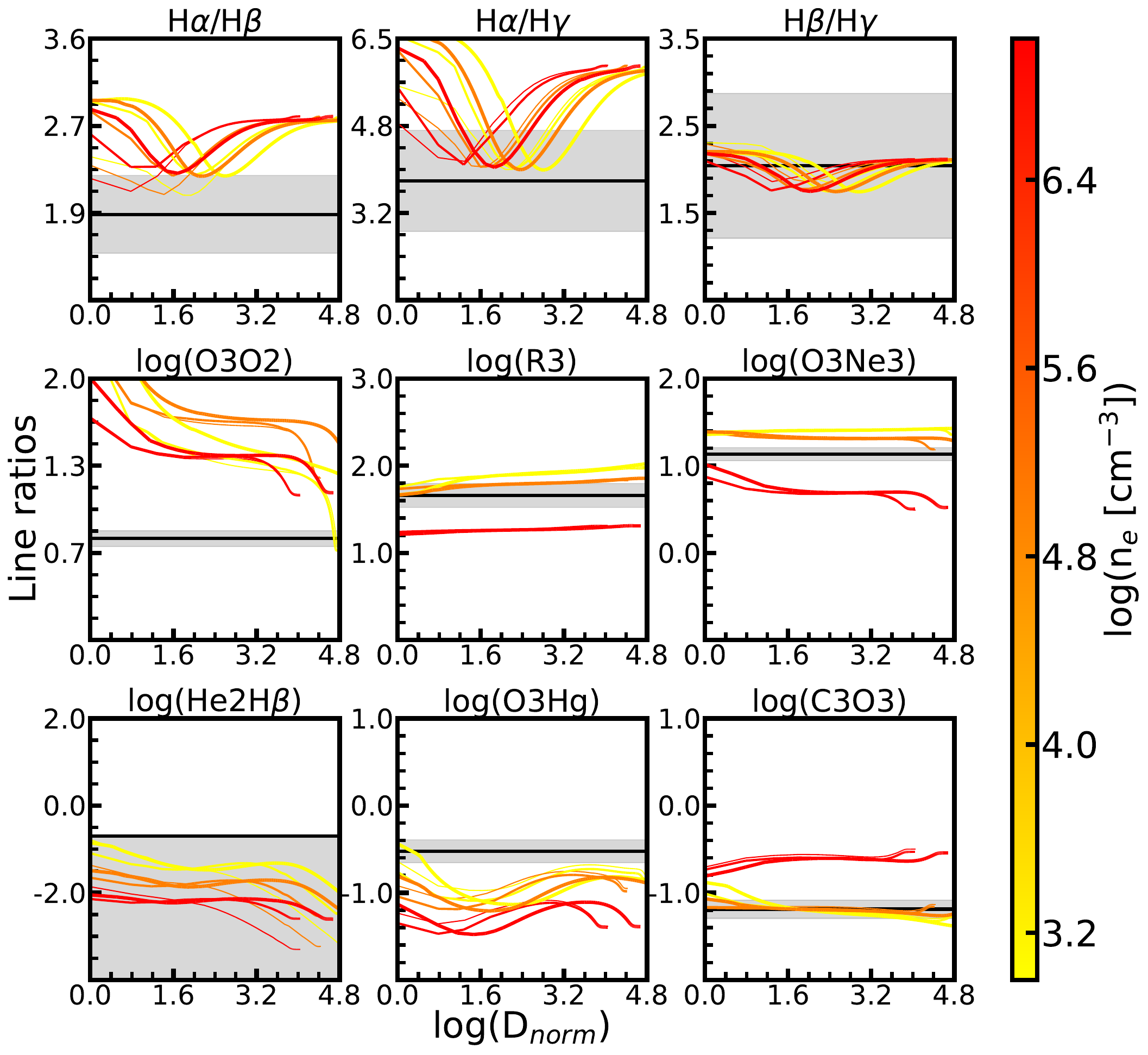}
\caption{Spatial resolution within photoionized models of the most relevant emission line ratios as a function of the normalized (based on the radii of the inner face) depth. Horizontal black lines and gray-shaded areas represent GHZ1 observations of the corresponding line ratios and associated uncertainties.}
\label{fig_ism}
\end{figure}

To analyze the possibility of a matter-bounded scenario, we used the photoionization grid of models from the NUVOLOSO project (see Appendix \ref{ap: photomodels}, P\'{e}rez-D\'{\i}az et al. in prep.) and analyzed the cumulative emission of the brightest UV and optical lines as a function of the depth within the nebula (with one being the illuminated face of the cloud). Results are shown in Fig. \ref{fig_ism}. Models have been computed for a fixed metallicity (12+log(O/H) = 7.9 and log(C/O) = -0.65) which mimic the expected conditions for GHZ1 (see Sec. \ref{ss: phys_chem} for more details), and the ionization parameter is set to -2.5, scaling as a function of density to preserve the total flux of ionizing photons.

We find that Balmer ratios are better reproduced by a matter-bounded nebula, although the extremely low \Halpha/\Hbeta and \Halpha/\Hgamma ratios put tight constraints on such conditions. However, we also observed that this would lead to a significantly higher O3O2 ratio, compared to that observed in GHZ1. Interestingly enough, the C3O3 ($\equiv$\CIIIL[1907,1909]/\OIIIL[4959,5007]) ratio, while behaving steadily across the nebula depth, has notably sensitivity to the high density conditions, as the ratio involves emission lines from transitions with different critical densities.

Under the matter-bounded scenario explored here, we also need to analyze the simultaneous effects on the entire set of observed emission lines. First, the ionization degree of the nebula needs to be low enough to compensate the higher O3O2 ratios characteristic of matter-bounded models. This first assumption naturally results in higher density conditions. Secondly, as for matter-bounded cases the C$^{2+}$/O$^{2+}$ shows a mostly flat profile, compensating the elevated C3O3 ratios requires a much lower log(C/O) assumption than log(C/O)$\sim$-0.65. However, such high-density conditions and lower abundances would necessarily lead to a decrease in the R3 ratio (e.g. the auroral line \OIIIL[4363] would be pumped, \citealt{Perez-Diaz_2026}), which is not observed in GHZ1. 

The matter-bounded scenario can reproduce the anomalous Balmer ratios observed in GHZ1, but the predictions for other line ratios (e.g. O3O2, R3) would be inconsistent with the current observations. Whereas we cannot discount the combined emission from matter-bounded and radiation-bounded clouds, the fact that the Balmer lines might be primarily affected by the matter-bounded clouds would eventually induce changes in other collisionally excited lines (CELs) observed. Hence, we need to explore other scenarios that might explain the anomalous Balmer ratios.

\subsubsection{Optical depths from Balmer lines}\label{ss: depths}
While in the previous section we discuss the possibility of a matter-bounded scenario and its implications, in this section we evaluate the possibility that Balmer line ratios are affected by the intrinsic optical depths associated to their transitions.

Following the procedure presented in \citet{Yanagisawa_2024}, we assumed that each observed Balmer line ($H_{n'n}$, being $n'$ and $n$ the upper and lower levels in the transition) intensity is given by
\begin{equation}
\label{eq: balmer depth} I \left( H_{n'n} \right)  = n \left( \mathrm{H}_{n'} \right) \epsilon_{n'n} \cdot 10^{-0.4\cdot A_{V} f_{\lambda_{n'n}}} \cdot \exp \left( -\tau_{n'n} \right).
\end{equation}
The first term is the overall intrinsic emission without post-processing, i.e. the result of the total abundance of H$^{+}$ ions transitioning from the $n'$- to $n$-level and the corresponding emissivity ($\epsilon_{n'n}$). The second term introduces the dust-attenuation correction, which in this case we assume the \citet{Cardelli_1989} extinction curve ($A_{\lambda}/A_{V}$). Finally, the third term introduces the self-attenuation due to the column of H atoms that might reabsorb the same photons, i.e. those already at the $n$-level. In particular, we model the optical depth $\tau_n$ as \citep{Yanagisawa_2024}
\begin{equation}
\label{depth} \tau_{n'n} = T_{n'n} \left( \frac{N(\mathrm{H}(n))}{2.9\cdot10^{13} \ \mathrm{cm^{-2}}} \right) \left( \frac{18 \ \mathrm{km}\cdot\mathrm{s}^{-1}}{v_{doppl} \left( T_{e} , n_{e}  \right)} \right),
\end{equation}
being $N(\mathrm{H}(n))$ the H column density for any Hydrogen series to the $n$-level and $v_{doppl}$ the half-width velocity of the ions under the assumption of Maxwellian distributions. We have applied the same normalization as \citet{Yanagisawa_2024} so for the Balmer series ($n=2$), specifically for \Halpha, \Hbeta and \Hgamma, the corresponding normalization factors $T_{n'}$ are 1.0, 0.14 and 0.05, respectively.

\begin{figure}
\centering
\includegraphics[width=1\linewidth]{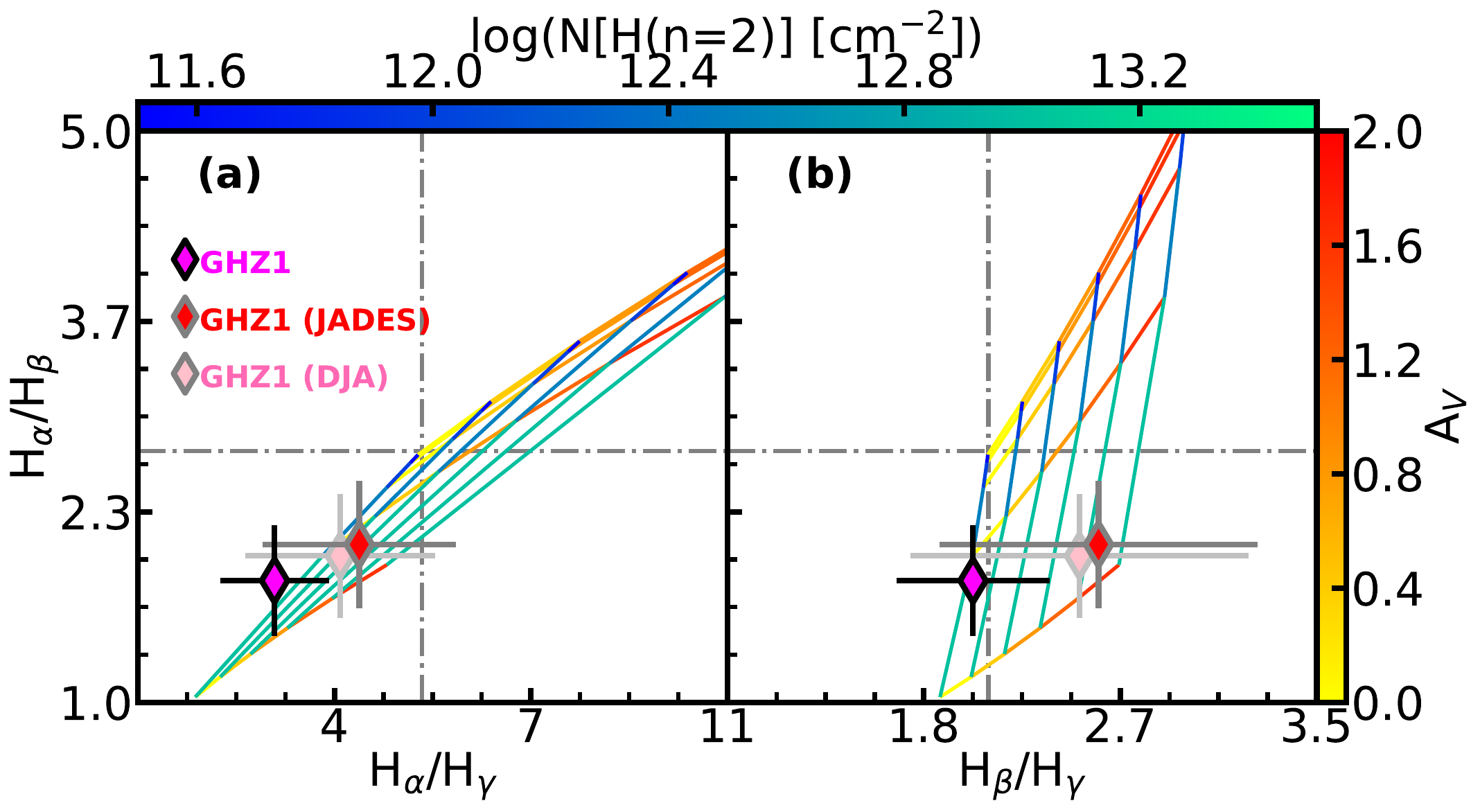}
\caption{Diagnostic diagram for Balmer optical depth. Dotted-dashed gray lines represent the theoretical Balmer ratios from Case B. We also show the Balmer decrements retrieved from alternative reductions to the data (see Appendix \ref{ap: alt_reductions}).}
\label{fig_balmer}
\end{figure}

In Fig. \ref{fig_balmer} we show the grid of models, for which $A_{V}$ and $N_{H}$ are set as free parameters. The theoretical Balmer ratios are computed with \textsc{pyneb} \citep{pyneb}, assuming standard conditions\footnote{Assuming other conditions, like low temperature ($T_{e} \sim 7,000$ K) and high density (log($n_{e}$ [cm$^{-3}$]) = 6) does not introduce significant changes in the grids.} (i.e. $T_{e} = 18,000$ K and $n_{e} = 1,000$ cm$^{-3}$). We find that some dust attenuation is needed ($A_{V} = 0.52^{+0.68}_{-0.38}$), but together with a moderately high column density ($\log(\mathrm{H}(n) \ [\mathrm{cm}^{-2}]) = 13.29^{+0.18}_{-0.21}$). Under this scenario, the predicted \Hdelta flux would be $\sim 2.03\times10^{-19}$ erg$\cdot$s$^{-1}\cdot$cm$^{-2}$, which is more compatible with the upper limit found in this work in contrast to the solely Case B predictions.

Overall, we conclude that the anomalous Balmer ratios might be caused by both a moderately-low dust attenuation and a significant suppression due to the column density of H atoms. This scenario is consistent with the $\beta_{UV} \sim -1.8$ found, which is redder than the expected value for dust-free stellar populations ($\beta_{UV} \sim [-2.6,-2.3]$, \citealt{Katz_2025}). Hence, to analyze the intrinsic emission line spectra, we correct all the retrieved emission lines for reddening ($A_{V} = 0.2_{-0.1}^{+0.1}$ mag), with an extra correction due to the optical depth that is only applied to the Balmer lines.
\subsubsection{Line variability}\label{sss: variablity}
As the anomalous Balmer ratios are those involving the \Halpha emission line (MIRI/LRS), with the ratio between \Hbeta and \Hgamma (NIRSpec/PRISM) behaving as expected, it is important to analyze the possibility of line emission variation between both observations. NIRSpec/PRISM observations were carried out in July 2024 and MIRI/LRS in November 2025, which at $z = 9.878$ implies a rest-frame time difference of 45.86 days.

Although the low resolution of the MIRI/LRS observations does not allow for a statistically significant line deblending, we force a two-Gaussian fit to the \Hbeta+\OIIIL[4959,5007] emission, one for the Balmer line and the other for the \OIII emission (see Appendix \ref{ss: hbeta} for more details). If we allow a free fit (i.e. no information from the NIRSpec measurement is used), then we only obtain a total flux for \Hbeta of $6\times 10^{-19}$ erg$\cdot$s$^{-1}\cdot$cm$^{-2}$ from MIRI/LRS, which would be fairly consistent with the Case B and the attenuation magnitude derived from the SED fitting ($A_{V} \sim 0.2$), but comparable to the flux uncertainty of the total feature. If we constrain the fit to preserve the flux measured from NIRSpec, then a bluer tail in the \Hbeta+\OIII emission is expected, although still compatible within the flux uncertainty.

NIRSpec data reductions that extend the wavelength range, and consequently the overlapping region between NIRSpec/PRISM and MIRI/LRS, also yield inconclusive results. Thanks to these new reductions, we can directly measure the independent contribution of \Hbeta, \OIIIL[4959] and \OIIIL[5007], and compare it to the total flux measured in MIRI/LRS. From the reduction provided by the JADES collaboration \citep[][see Appendix \ref{ss: stefano}]{Scholtz_2026} we measure a total flux of $62(\pm2)\times10^{-19}$ erg$\cdot$s$^{-1}\cdot$cm$^{-2}$, which is $\sim 21\%$ lower than that of MIRI/LRS. On the contrary, the reduction provided by the Dawn JWST Archive \citep[DJA,][see Appendix \ref{ss: dja}]{Heintz2024dja, deGraaff2025} gives a total flux of $92(\pm4)\times10^{-19}$ erg$\cdot$s$^{-1}\cdot$cm$^{-2}$, which is $\sim23\%$ higher than the observed flux in MIRI/LRS. As the total flux uncertainty in the extended wavelength range for NIRSpec is $\sim 20\%$ (i.e. the difference between DJA, JADES and MIRI/LRS is below $< 1.5\sigma$), we cannot conclude whether there is actual variation in the total \Hbeta+\OIIIL[4959, 5007] emission.

Based on the \Hbeta emission from NIRSpec/PRISM, the predicted \Halpha emission should be $\sim 28\times10^{-19}$ erg$\cdot$s$^{-1}\cdot$cm$^{-2}$, which implies at least $\sim21\%$ of the \Halpha emission is lost in the MIRI/LRS observations. This decrease is in agreement with the upper limit found for the AGN contamination ($\sim 26\%$). Moreover, translated into luminosity for the 2-10 keV X-ray band \citep{Jin_2012}, this would imply $L_{2-10\ \mathrm{keV}} \sim 3.6\cdot 10^{42}$ erg$\cdot$s$^{-1}$, compatible with the upper limit found (see Appendix \ref{s:xray}). If we assume that line variability is only present in Balmer lines, and it is driven by the Broad Line Region (BLR) of an AGN hosted by the system, we can also infer some constraints on its properties. The rest-frame time of 45.86 days would imply an upper limit of $R_{BLR} \leq 0.04$ pc. We use the most recent calibration between the $R_{BLR}$ and the $\lambda L_{5100}$ luminosity \citep{Gravity_2024} and the bolometric corrections provided by \citet{Netzer_2019} to derive a total X-ray luminosity in the 2-10 keV band from this BLR upper limit, which is $5.83\cdot 10^{43}$ erg$\cdot$s$^{-1}$, fully consistent with the upper limit of $< 7\cdot10^{43}$ erg$\cdot$s$^{-1}$, corrected from magnification, found for GHZ1 (see Appendix \ref{s:xray}).

Overall, line variability is a possibility that, with the current observations, cannot be ruled out. The X-ray upper limit found for the source is consistent with the picture of line variability driven by AGN contamination \citep[in particular considering the intrinsic X-ray weakness found in high-$z$ AGN, e.g.][]{Tortosa_2026}. However, the low spectral resolution of MIRI/LRS observations prevents further testing of other important implications (line kinematic profiles, deblending of \Hbeta) which would significantly increase the constraints on the feasibility for this scenario.

\subsection{Physical and chemical properties of the ISM}\label{ss: phys_chem}
The simultaneous detection of \OIIIL[4363] and \OIIIL[4959,5007] allows us to place a direct constraint on $T_{e}$ based on their emissivities \citep[\textsc{pyneb}][]{pyneb}, although the lack of spectral resolution forces us to make an assumption on the density conditions. Under the standard assumption of $n_{e} = 10^{3}$ cm$^{-3}$ \citep{Abdurrouf_2024}, we obtain $T_{e} = 18,600^{+5100}_{-2500}K$.

\begin{figure*}
\centering
\includegraphics[width=0.8\linewidth]{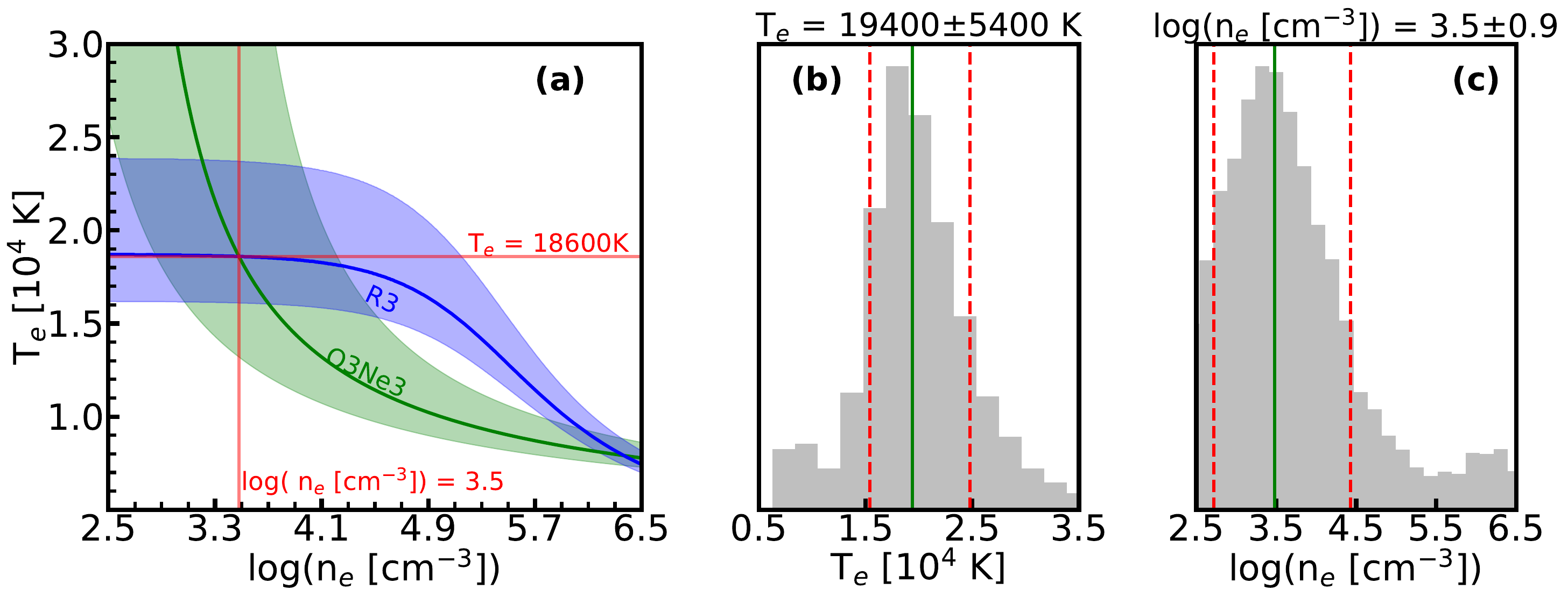}
\caption{Temperature and density conditions in GHZ1. Panel (a) shows the $T_{e}$ vs $n_{e}$ diagram for the two observed emission line ratios sensitive to both properties: R3 and O3Ne3. The solid line represents the observed values and the shaded areas their uncertainties. Panel (b) and (c) shows the posterior distributions from the MCMC fit.}
\label{fig_physics}
\end{figure*}

A recent approach to derive $T_{e}$ is the O3Ne3 ($\equiv$\OIIIL[4959,5007]/\NeIIIL[3868]) ratio proposed by \citet{Perez-Diaz_2026}, which is based on \NeIIIL[3868], \OIIIL[4959,5007] and \OIIL[3727,3729], but it also depends on the density conditions. Hence, we can simultaneously use both approaches to create a diagnostic diagram for $n_{e}$ and $T_{e}$ (see Fig. \ref{fig_physics} (a)). Complementarily, we also fit both ratios within an MCMC technique in which we explore $T_{e}$ and $n_{e}$ (see Figs. \ref{fig_physics} (b) and (c)). Overall, we find that the previous assumptions are consistent with our preliminary estimates, although we remark that there is a potential solution involving high density conditions (log($n_{e}$ [cm$^{-3}$]) $>$ 6) and very low temperatures ($T_{e} <$ 7,000 K).

While the tentative high density, low temperature solution is feasible from the sole point of view of the R3 and O3Ne3 ratios, the simultaneous coverage of UV and optical lines allows us to rule out this scenario. Such high density conditions will significantly drop the emissivity of several of the optical lines (e.g. \OIIIL[4959,5007], \OIIL[3726,3729]) as their critical densities would be well below such values. On the contrary, UV lines (e.g. \sOIIIL, \CIIIL[1907,1909]) should remain bright, as their critical densities ($n_{c} \sim 10^{7}-10^{9}$ cm$^{-3}$) are much higher. The immediate implications would be much brighter UV lines than those currently observed in GHZ1. Hence, we consider that the only feasible conditions for reproducing the observed emission lines are $n_{e} = 10^{3}$cm$^{-3}$ and $T_{e} = 18,600^{+5100}_{-2500}K$.

Having determined the physical conditions of the ISM, we can obtain robust constraints on several chemical abundance ratios. First, we estimate the oxygen abundance from \OIIIL[4959, 5007] and \OIIL[3727,3729] using \textsc{pyneb} \citep{pyneb}. From the former, we infer an ionic abundance of 12+log(O$^{++}$/H$^{+}$) = 7.46$^{+0.22}_{-0.30}$, and from the latter\footnote{We use the non-parametric gas-phase temperature estimation code \textsc{genesis-metallicity} \citep{Langeroodi_2026} to estimate the temperature of O$^{+}$ from the available set of emission lines, yielding $T_{e} ( O^{+}) = 14,400\pm800$ K, which is consistent with the predictions from our photoionization models.} we infer 12+log(O$^{+}$/H$^{+}$) = 7.05$^{+0.14}_{-0.15}$. Overall, we estimate that the total\footnote{These estimations have been performed from the reddening-corrected emission lines, and we highlight that assuming no dust-attenuation yields consistent results within the errors.} oxygen abundance is12+log(O/H)=7.60$^{+0.20}_{-0.25}$ (i.e. $Z = 0.08^{+0.05}_{-0.04}Z_{\odot}$), where we have not applied any ionization correction factor (ICF) as the upper limit on \HeII suggests that its impact is negligible.

We also perform additional tests on the O abundance. Particularly, by means of the $\chi^{2}$ minimization, we find which particular set of parameters (namely 12+log(O/H) and log(U)) best reproduces sensitive emission line ratios (R23, R3 O3Ne3 and O3O2). Additionally, we compute the 12+log(O/H) from the calibration proposed by \citet{Sanders_2024}. We provide in Table \ref{OH_abundances} the results of all different methods.

\begin{table}[h!]
	\caption{Chemical abundances and ionization parameter (log(U)) derived from different techniques.}
	\label{OH_abundances}
	\centering
	\begin{tabular}{c | c  c  c} 
		\hline\hline
		  & \textbf{Direct} & \textbf{Models} & \textbf{Sanders+24 (R23)}\\ 
		   & \textbf{(1)} & \textbf{(2)} & \textbf{(3)}\\ \hline
        12+log(O$^{+}$/H$^{+}$) & 7.05$^{+0.14}_{-0.015}$ & 7.08$^{+0.18}_{-0.17}$ & - \\
        12+log(O$^{++}$/H$^{+}$) & 7.46$^{+0.22}_{-0.30}$ & 7.68$^{+0.13}_{-0.12}$ & - \\ 
        12+log(O/H) & 7.60$^{+0.20}_{-0.25}$ & 7.81$^{+0.12}_{-0.11}$ & 7.92$^{+0.02}_{-0.02}$ \\ \hline 
        log(C$^{++}$/O$^{++}$) & -1.06$^{+0.55}_{-0.13}$ & -0.62$^{+0.13}_{-0.12}$ & - \\
        log(C/O) & -1.06$^{+0.55}_{-0.13}$ & -0.66$^{+0.13}_{-0.13}$ & - \\ \hline
        log(N$^{++}$/O$^{++}$) & $<-0.56$ & $<-1.05$ & - \\
        log(N/O) & $<-0.56$ & $<-1.04$ & - \\ \hline
        log(U) & - & -2.45$^{+0.07}_{-0.07}$ & - \\
        \end{tabular}
\end{table}

Similarly, we derive log(C/O) and an upper limit for log(N/O) abundance ratios, based on the physical conditions previously derived for $T_{e}$ and $n_{e}$ \citep{pyneb}. In this case, we only have a few species to actually assess the total abundance, so we assume that the ionic abundance ratio derived from the direct method is representative for the total abundance. In the model-based determination, based on the previous estimates on 12+log(O/H) and log(U), we just search for the best abundance ratio that reproduces the corresponding emission line ratio: C3O3 for log(C/O) and N3O3 ($\equiv$ \NIIIL[1750]/\OIIIL[4959,5007]) for log(N/O). While for the log(C/O) abundance ratio both estimates are compatible within the errors, the situation is intrinsically more complicated for log(N/O) as we are dealing with an upper limit. To estimate the total log(N/O) from the photoionization models, we have also imposed as a constraint that the predicted N3O3 ratio cannot be above the observed upper limit. This last constraint explains why we found a much tighter upper limit for log(N/O) from the model-based methodology.

For consistency, we will assume the chemical abundances derived from photoionization models: 12+log(O/H)$=7.81_{-0.11}^{+0.12}$ ($Z_{O}=0.13_{-0.03}^{+0.04}Z_{\odot}$), log(C/O)$=-0.66^{+0.13}_{-0.13}$ ($Z_{C}=0.05_{-0.02}^{+0.03}Z_{\odot}$), log(N/O) $<-1.04$ ($Z_{N}<0.08Z_{\odot}$). Whereas the results among different methodologies are compatible within the errors (or upper limits), by assuming these derivations, we are already accounting for any ICF correction factors, which would be largely unconstrained in the case of C and N abundances due to the lack of further emission lines.
\subsection{Chemical enrichment: the evolution of C, N and O}
The constraints on the 12+log(O/H) and log(C/O) abundance ratios, together with the upper limit on log(N/O) allow us to inspect the differential chemical enrichment in GHZ1. For reference, we consider the chemical enrichment trends reported by \citet{Nicholls_2017} for nearby stars, HII regions and galaxies. 

At lower abundances (12+log(O/H) $\lesssim$ 8.4), C and N are mainly produced by the same massive stars that produced O, so a flat log(C/O) and log(N/O) abundance is expected (primary production). Beyond that abundance, and given that there is enough time ($\sim$1 Gyr) for post-asymptotic giant branch (pAGB) stars to produce and release more C and N by means of their CNO cycles, an increase is expected with the presence of more and more O that fuels such production (secondary production).

\begin{figure}
\centering
\includegraphics[width=1\linewidth]{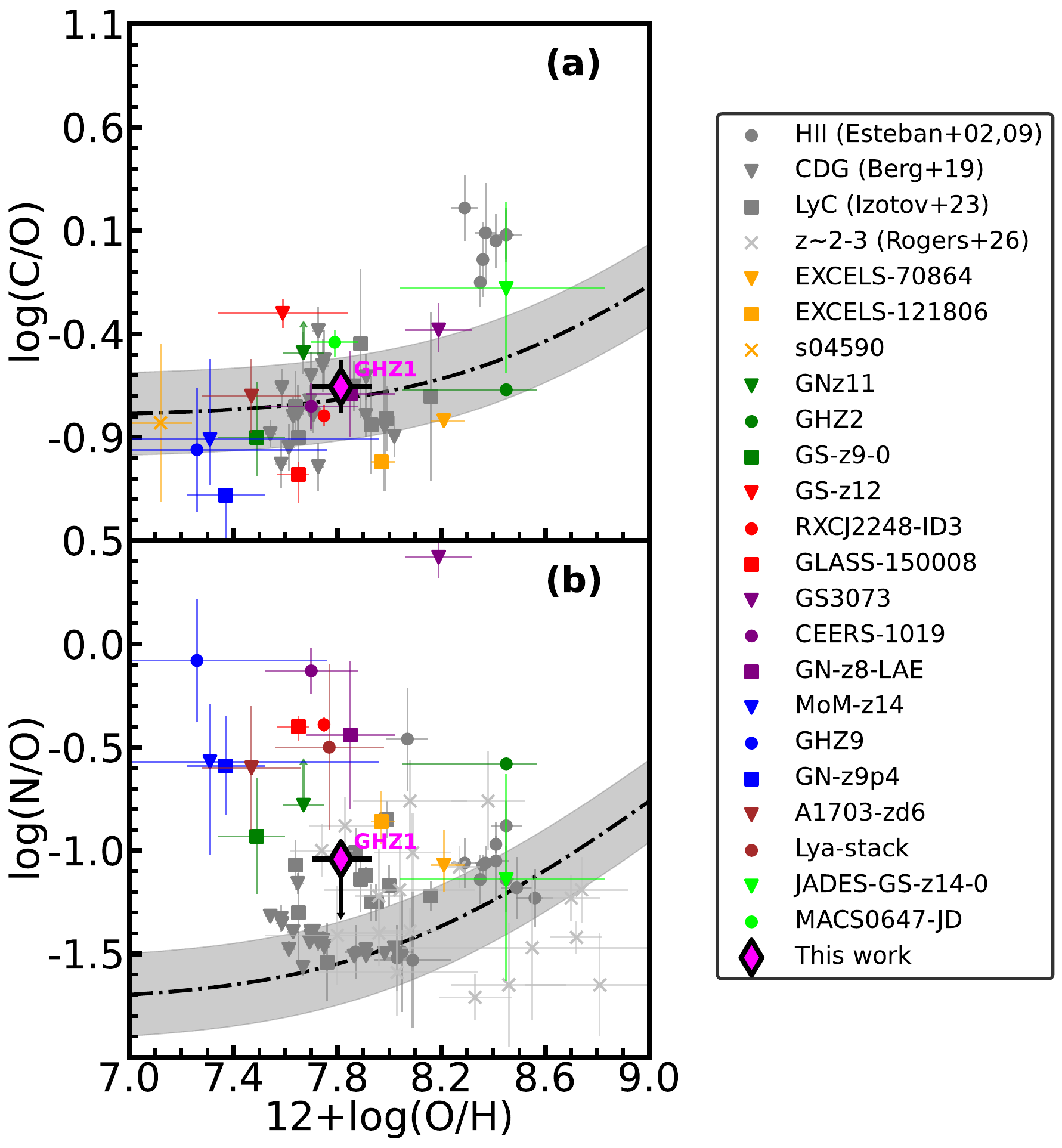}
\caption{Chemical enrichment patterns for local and high-$z$ sources for log(C/O) (a) and log(N/O) (b). In both plots, the gray dashed-dotted line shows the reported relation by \citet{Nicholls_2017} and the shaded area corresponds to an uncertainty of 0.3 dex. List of references can be found in Table \ref{Tab_gal_ref}.}
\label{fig_secondary}
\end{figure}

As we show in Fig. \ref{fig_secondary} (a), GHZ1 follows the expected trend for C primary production \citep{Nicholls_2017}, also in agreement with many of the high-$z$ sources analyzed in the literature. In contrast to the N-enhancement reported in the literature, the upper limit on the log(N/O) puts GHZ1 on the expected trend for N primary production (see Fig. \ref{fig_secondary} (b)), considering the usual scatter at low metallicities \citep[e.g.][]{Perez-Montero_2009}. This differential behavior is not surprising, as most of the high-$z$ sources that report anomalously high log(N/O) ratios are either characterized by AGN activity (e.g. GHZ9, \citealt{Napolitano_2025b}), extreme density conditions (e.g. GHZ2, \citealt{Castellano_2025}), VMS in dense clumps (e.g. CEERS-1019, \citealt{Marques-Chaves_2026b}) or WR signatures (e.g. RXCJ2248-ID3, \citealt{Berg_2026}).

We conclude that GHZ1 shows the expected chemical enrichment for a source that is dominated by star-forming events, from which only massive stars have had enough time to produce metals. In Sec. \ref{ss: compactness}, we further discuss these findings in the context of high-$z$ literature.

\subsection{An early look onto the mass-metallicity relation}
\begin{figure}
\centering
\includegraphics[width=1\linewidth]{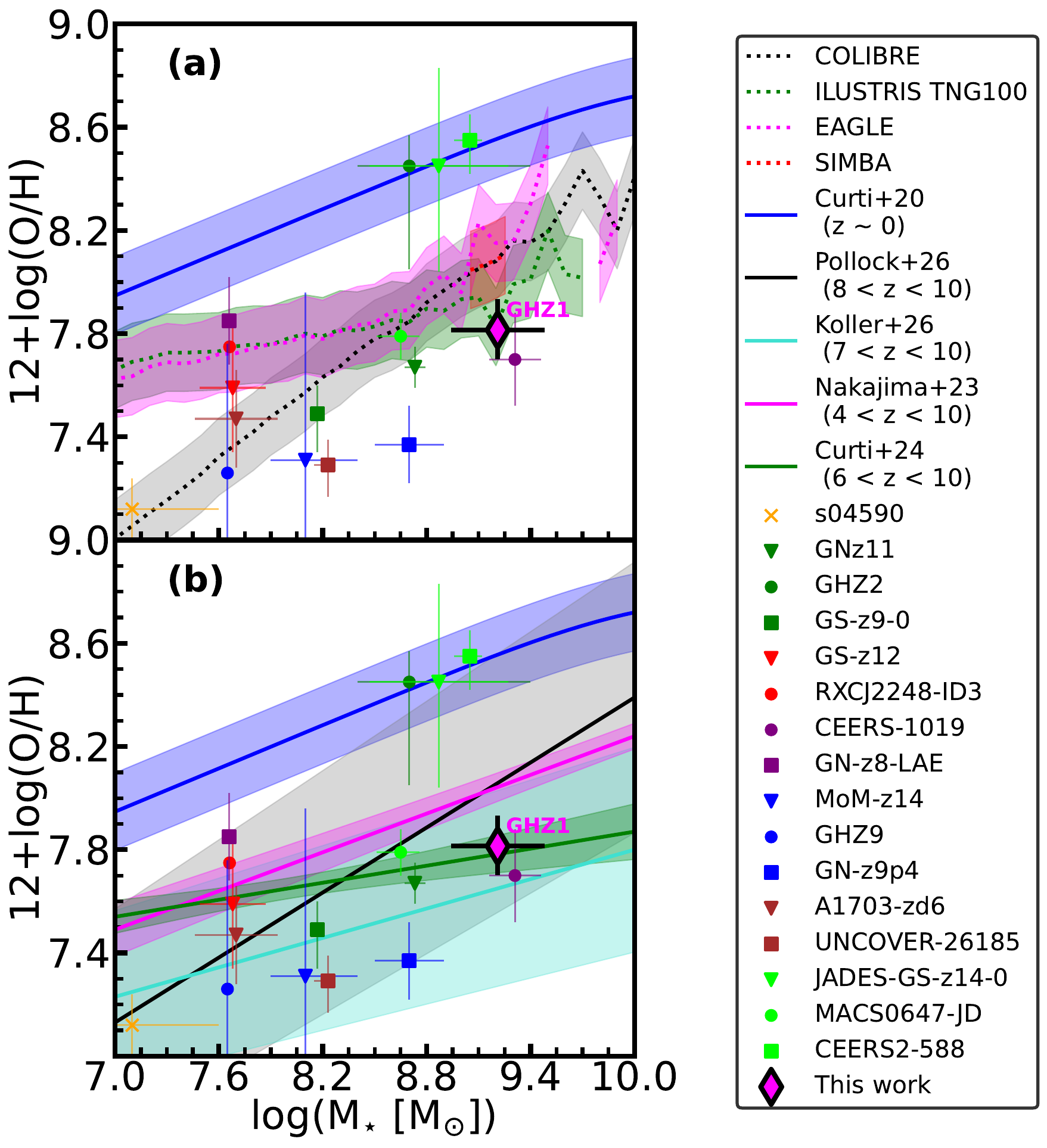}
\caption{The mass-metallicity relation for high-$z$ sources. Panel (a) shows the trends predicted from cosmological simulations. Panel (b) shows the reported trends based on observational data. References can be found in Tables \ref{Tab_gal_ref} and \ref{Tab_mzr_ref}. In both plots, we show the derived relation by \citet{Curti_2020} for nearby galaxies as reference.}
\label{fig_mzr}
\end{figure}

The build-up of heavy elements is tightly linked to the mass growth in galaxies \citep{Maiolino_2019, Curti_2020, Sharda_2026}. Although the main mechanisms that drive such a tight correlation are still under debate, the examination of the mass-metallicity relation (MZR) offers a unique opportunity to test such evolution. Whereas the MZR relation has been carefully revisited at low-$z$ for many decades, there is still a dependence on the methodology followed and the samples considered \citep[e.g.][]{Kewley_2019, Perez-Diaz_2024}. Such biases are more accentuated at high $z$, where both the statistics and methodologies are more scarce \citep{Langeroodi_2023, Nakajima_2023, Curti_2024, Koller_2026, Pollock_2026}. Hence, we consider simultaneously the predictions from cosmological simulations (Fig. \ref{fig_mzr} (a)) and observational trends beyond $z > 4$ (Fig. \ref{fig_mzr} (b)).

Together with CEERS-1019 \citep[$z\sim8.7$,][]{Marques-Chaves_2024} and CEERS2-588 \citep[$z\sim11.04$,][]{Harikane_2026}, GHZ1 stands out for being among the most massive systems at high-$z$, allowing us to further probe the upper end of the MZR. From Fig. \ref{fig_mzr} (a), when compared to cosmological simulations, only Illustris TNG100 \citep{Nelson_2019} predicts a trend that matches the observed position for GHZ1, while the rest of simulations overpredict the gas-phase abundance.

From an observational perspective, as there is a large scatter in the trends observed due to the poor statistics, GHZ1 remains compatible with the reported relations. Whereas the whole sample of objects presented in Fig. \ref{fig_mzr} (b) indicates a flattening of the curve (in agreement with the result found by \citealt{Curti_2024}), it must be noted that such a conclusion requires a larger number of objects and homogeneous analysis, which are beyond the scope of this work. This and other properties will be the subject of a forthcoming paper (\'{A}lvarez-M\'{a}rquez et al. in prep.) that will combine all PRISMS sources with additional galaxies covering $z \sim$ 9-14 and with available NIRSpec$+$MIRI spectroscopy.

\section{Discussion}\label{s: discussion}
\subsection{The galaxy evolution tale of GHZ1}\label{ss: evolution}
GHZ1 shows the properties of a regular star-forming system. The spectro-photometric SED fitting (see Sec. \ref{ss: SED_fitting}) shows that the bulk ($\gtrsim80\%$) of its stellar mass (log(M$_{\star}$ [M$_{\odot}$]) $\sim$ 9.2) was formed $\sim$50-100 Myr ago, and it is now experiencing a second but much more moderate burst of star formation (log(SFR [M$_{\odot}\cdot$yr$^{-1}$])$\sim 0.9$). This SFH matches the average fit reported by \citet{Roberts-Borsani_2026} for extended ($R_{e} > 400$ pc) systems at $z \gtrsim 10$, but clearly contrasts with the cosmological simulations that favor a more stable increasing SFH \citep[see the predictions for CEERS2-588 based on the COLIBRE simulation,][]{Chaikin_2026b}.

The derived sSFR ($\sim 10^{-2.2}$ Myr$^{-1}$) also matches the expectations for a star forming dominated system at $z \gtrsim 10$ ($> 10^{-3.4}$ Myr$^{-1}$ \citep{Cole_2025, Chaikin_2026}. Although there arere not enough statistically significant samples, GHZ1 is consistent with the study presented by \citep{Cole_2025}, which shows that for objects with log(M$_{\star}$ [M$_{\odot}$]) $\in$ [8.7, 9.3], at redshift $z \sim 10$ the expected sSFR for main sequence systems is $\sim 10^{-2}$ Myr$^{-1}$. When compared to the predictions for the main sequence (SFR vs M$_{\star}$) at $z \gtrsim 10$, GHZ1 follows the reported trend by \citet{Chaikin_2026} based on the COLIBRE cosmological simulation.

The analysis of the emission line spectra anchors the picture of regular star formation. We did not retrieve any highly ionized emission lines (e.g. \HeIIL[1640] or \NeVL) nor a detection in X-rays (see Sec. \ref{ss: ionization} for further details). Additionally, the physical properties of the ionized gas ($T_{e} \sim 18,500$ K, $n_{e} \sim 1,000$ cm$^{-3}$ and $\log(U) \sim -2.5$) rule out any extreme ionizing scenario, consistent with the lack of strong UV lines. Puzzlingly enough, the low flux of \Halpha is hard to reconcile with the mentioned properties (see Sec. \ref{ss: missing_ha} for more details).

The chemical enrichment picture is consistent with the expectations from chemical evolution models. GHZ1 is building up metals by means of massive stars, yielding a 12+log(O/H)$\sim 7.8$ and log(C/O)$\sim$-0.66, which are consistent with cosmological simulations \citep[e.g. ILLUSTRIS TNG100][]{Nelson_2019} and with a flat mass-metallicity relation \citep{Koller_2026, Curti_2024}. Due to the faintness of \NIII and \NII lines, the log(N/O) abundance cannot be properly constrained, but the derived upper limit based on photoionization models is consistent with the picture of primary production by massive stars \citep[e.g.][]{Henry_2000}.

In summary, although being among the most massive systems at $z\sim$10, GHZ1 behaves like a regular star-forming dominated system, whose chemical abundances show the expected primary enrichment from massive stars. The ionizing conditions from CELs reveal a coherent picture of massive stars being the predominant source of ionization, although the anomalous \Halpha/\Hbeta ratio found points towards a more complex scenario for its ISM.

\subsection{Compact versus extended sources}\label{ss: compactness}
Our findings of standard chemical evolution conditions differ within high-$z$ literature not only because GHZ1 follows the expected trends for primary production of C, N, and O, but also because GHZ1 stands out as a massive, extended source. Hence, it is essential not only to motivate the chemical enrichment trends (see Fig. \ref{fig_secondary}) but also its relationship to other properties like stellar mass (see Fig. \ref{fig_mzr}) and size.

\begin{figure*}
\centering
\includegraphics[width=1\linewidth]{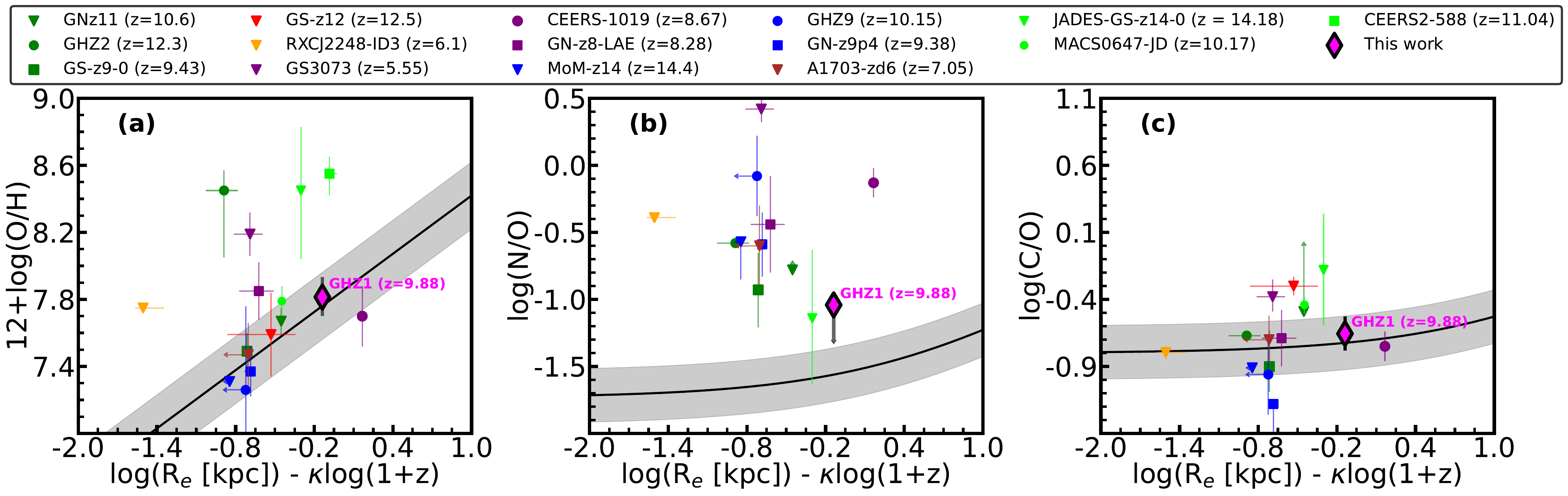}
\caption{The size-metallicity relation for high-$z$ galaxies. For each plot, we represent the predictions from observational trends as derived in Appendix \ref{ap: r-oh}. References can be found in Table \ref{Tab_gal_ref}.}
\label{fig_sizemet}
\end{figure*}

By assuming that the effective-light radius and the stellar mass are correlated \citep[with a non-negligible dependence on redshift, e.g.][]{Morishita_2024b}, we can compare the expected chemical enrichment for the stellar mass and the observed system sizes (see Appendix \ref{ap: r-oh} for further details in the derivation). In this case, we assume the value of $\kappa = -0.24$ for the redshift scaling (see Eq. \ref{size-metallicity}) found by \citet{Morishita_2024b} for extended and compact sources. The predictions (black lines) are shown in Fig. \ref{fig_sizemet} with a 0.3 dex uncertainty (gray-shaded area) to account for differences in the methodology to derive chemical abundances.

These results show that, if we compare by size rather than stellar mass, then in terms of 12+log(O/H) and log(C/O) most of the systems accurately match the predictions as shown in Fig. \ref{fig_sizemet} (a) and (c). The major differences arise, as expected, in the log(N/O) ratio, for which the most compact sources show a higher excess of N than GHZ1.

This picture is consistent with the ongoing discussion about the dichotomy at high-$z$ between compact N-enhanced sources and extended sources that barely show signatures for the excess \citep[e.g.][]{Isobe_2023, Charbonnel_2023, Marques-Chaves_2024}. The physical and chemical properties of GHZ1 rule out extreme scenarios, and favor a chemical enrichment pathway through massive stars. If there had been an extreme event of VMS pollution in the past \citep[e.g.][]{Vink_2023} or selective outflows \citep[e.g.][]{Rizzuti_2025}, those signatures have been diluted within the chemical footprint observed in GHZ1.

\subsection{Where are the \Halpha photons?}\label{ss: missing_ha}
While the emission line spectra based on CELs provides a fully consistent picture of galaxy evolution, the anomalous Balmer line ratios observed in GHZ1 point towards a higher level of complexity. Anomalous Balmer line ratios have already been reported in the literature prior to \jwst \citep[e.g.][]{Atek_2009, Yang_2017}, attributing such deviations to either hidden absorptions not properly corrected or extreme ionizing conditions (O3O2 $\sim$ 10, large Lyman escape fractions). However, with the advent of \jwst a particular effort has been made in other to fully understand the physical origin for Case B deviations \citep{Scarlata_2024, Yanagisawa_2024, McClymont_2025, Nikopoulos_2026}.

\citet{McClymont_2025} proposed a scenario in which matter-bounded conditions successfully reproduced the low \Halpha/\Hbeta ratios reported in JADES. We explore this scenario in Sec. \ref{ss: matter-bounded} and we conclude that it fails to reproduce other key emission line ratios (e.g. O3O2). Additionally, based on the observed UV properties \citep{Chisholm_2022}, the corresponding LyC escape fraction would be in the order of $\sim 2\%$, well below that expected for these conditions. 

\citet{Yanagisawa_2024} proposed that the deviations from Case B originated from optical opacity in Balmer lines, scenario already explored in Sec. \ref{ss: depths}.  Although it provides a simultaneous explanation for many of the observed properties, the associated physical interpretation is more puzzling: the coexistence of a moderately highly ionized cloud(s) and the presence of an optically thick neutral H gas reservoir that has been excited from the ground state. 

\citet{Scarlata_2024} proposed that such abnormal Balmer ratios might be a consequence of a non-spherical symmetry that favors \Halpha scattering from the line-of-sight. Distinguishing between this scenario and that proposed by \citet{Yanagisawa_2024} requires a larger set of H recombination lines (e.g. the full Balmer sequence, Paschen lines,...) that is not accessible from the available data. 

Another scenario that might be explored to explain the Balmer ratios involves line variability. Between NIRSpec and MIRI observations, a total rest-frame time of $\sim 45$ days has passed. In the nearby Universe it has been reported that the variability in the AGN contribution affects the most broad inner emission \citep[see][for a review]{Ricci_2023}. Whereas for typical masses of the massive black hole (SMBH, $~ 10^{6}-10^{8}$ M$_{\odot}$) the broad component is expected to be $\sigma_{broad} \gtrsim 430$ km$\cdot$s$^{-1}$ \citep{Harikane_2023}, less massive black holes (i.e. Intermediate Massive Black Holes, IMBH) might show narrower components that can fall within the instrumental broadening of NIRSpec/PRISM and MIRI/LRS \citep[e.g.][]{Fei_2026}. Although much less abundant than their massive counterparts, there is a clear example of a low massive SMBH in the nearby Universe to which we can compare this scenario: NGC4395 \citep{Filippenko_1989, Lira_1999, Moran_2005, Peterson_2005, Desroches_2006, Laor_2006}. NGC4395 hosts an active BH of $\sim 3\cdot10^{5}$ M$_{\odot}$ \citep{Peterson_2005}, and displays variability of months that has been detected in its X-ray, UV and optical emission \citep{Desroches_2006, ONeill_2006} but also in the brightness of its Balmer broad lines \citep{Lira_1999}, with a measured X-ray emission in the 2-10 keV band of $\lesssim 10^{40}$ erg$\cdot$s$^{-1}$ \citep{Moran_2005, ONeill_2006}. Under these conditions, if we are witnessing variability from the broad line clouds associated to an IMBH, which intrinsically displays a shorter variability than SMBHs, then the effects are expected to be shown in the \Halpha emission, but also in the \Hbeta component, although they are expected to be more prominent in the \Halpha components \citep[e.g.][]{Lira_1999}.

Lastly, another possibility which cannot be explored with the current data, is the presence of multiple Balmer components driven by galactic winds. A clear example of this scenario is the system RXCJ2248-ID \citep[$z\sim6$][]{Topping_2024, Crespo-Gomez_2025, Berg_2026}. The initial NIRSpec-medium grating observations of this system reveal a small Balmer decrement \Halpha/\Hbeta$=2.54\pm0.02$ whereas the other Balmer ratios were above Case B predictions \citep{Topping_2024, Yanagisawa_2024}. The follow-up analysis presented by \citet{Crespo-Gomez_2025} based on integral field spectroscopy with NIRSpec and high resolution gratings, revealed that both \Hbeta and \Halpha were characterized by multiple broad components associated to dusty outflows. However, it must be noted RXCJ2248-ID shows a clear enhancement in its log(N/O) abundance, which might be produced by these outflows \citep[e.g.][]{Rizzuti_2025}, that is not observed in GHZ1.

The combined NIRSpec and MIRI observations have allowed us to rule out some of the discussed scenarios (e.g. density-bounded conditions), and we have extensively analyzed the major systematics that might affect the ratio (see Appendices \ref{ss: sarah}, \ref{ss: stefano}, \ref{ss: dja},  and \ref{ss: 2d_spec}). To further discriminate between these scenarios, multi-wavelength observations as well as follow-ups with higher resolution spectroscopy are necessary to properly unveil the hidden mechanisms beyond the anomalous Balmer decrements.

\section{Conclusions}\label{s: conclusions}
This work presents the results from the combined NIRSpec/PRISM and MIRI/LRS observations of GHZ1, an extended ($R_{e} \simeq 410$ pc) and massive (log(M [M$_{\odot}$]) $\simeq$ 9.2) system when the Universe was $\simeq$ 500 Myr old, in the GLASS-JWST (Abell2744) field. We present a detailed analysis of its continuum and line properties from the UV and optical emission. We find that:
\begin{enumerate}
\item GHZ1 shows all the characteristics of a star-forming dominated system. The inferred SFR from the spectro-photometric SED fitting (log(SFR [M$_{\odot}\cdot$yr$^{-1}$])$\simeq 0.9$ over the last 10 Myr) is consistent with that of other star-forming sources at similar redshifts. The sSFR ($\simeq 10^{-2.2}$ Myr$^{-1}$) shows that GHZ1 is efficiently forming stars, without reaching the threshold for strong feedback events. This ongoing event of star formation is more moderate than a previous ($\sim$ 50-100 Myr) burst that characterizes its old stellar population. The lack of highly-ionized emission lines and X-ray emission shows a coherent picture in which the main ionizing mechanism is provided by massive stars formed in a recent burst of star formation.
\item The chemical enrichment of GHZ1 is consistent with a source that is undergoing a primary production of metals by massive stars. The oxygen abundance (12+log(O/H) $\simeq7.8$) is consistent with the predicted values for its stellar mass based on observed and simulated MZRs. The carbon-to-oxygen abundance ratio (log(C/O) $\simeq-0.7$) is consistent with the expectations for primary production and shows a similar behavior to that observed in other high-$z$ sources. The upper limit derived for the nitrogen-to-oxygen abundance ratio (log(N/O) $\lesssim-1.0$) is also consistent with this picture, implying that if any of the proposed scenarios for N-enhanced galaxies are ubiquitous, GHZ1 has not experienced them yet or has been efficient in erasing those chemical footprints.
\item The observed slope in the UV continuum ($\beta_{UV} \simeq -1.9$) and the inferred stellar attenuation magnitude ($A_{V} \simeq -0.2$ mag) point towards the presence of a moderate amount of dust. This result contrasts with the observed Balmer line ratios, in particular \Halpha/\Hbeta and \Halpha/\Hgamma, for which we report values below the Case B predictions. We have analyzed any potential systematics in the measured \Halpha flux, and conclude that there is an intrinsic lack of \Halpha photons that drives Balmer ratios below the theoretical expectations.
\item We have investigated different ionizing scenarios that would simultaneously reproduce the redder UV slope and the observed Balmer ratios. With the current observations, only three scenarios would yield a coherent picture. The first scenario explored is that Balmer lines might be optically thick due to the presence of a mass of excited hydrogen. The second scenario involves variability linked to a hidden IMBH. The third scenario consists of complex kinematic profiles in the brightest Balmer lines, \Halpha and \Hbeta, driven by galactic outflows.
\end{enumerate}

Although actively forming stars and building up heavy elements in a regular pathway, the anomalous Balmer emission in the second epoch of observations from MIRI/LRS points towards a complex picture that cannot be fully explored with the current observations. The first scenario explored would imply that even star forming dominated systems at high-$z$ can still harbor moderate amounts of gas during bursty events, allowing future events of star formation. The second scenario would imply a link between ongoing star formation and the onset of central engines. The last scenario would have major implications for a correct interpretation of the emission line spectrum of this system, and to understand how the most massive and extended systems at high-$z$ sustain such feedback. Disentangling these scenarios requires not only higher resolution data, but also spatially resolved spectroscopy that would reveal whether this is a global or a localized process. GHZ1 shows an excellent example for the necessity of high-resolution and spatially resolved spectroscopy for a full census of UV-bright objects at Cosmic Dawn  to understand what is driving their UV brightness, how long the N-enrichment phase can last, signatures of high-ionized emission or how the mass-metallicity relation is built.

\begin{acknowledgements}
B.P.-D., M.C., L.N., A.F and P.S. acknowledge support from: INAF RF2024 Large Grant "UNDUST: UNveiling the Dawn of the Universe with JWST"; INAF RF2024 GO Grant ”Revealing the nature of bright galaxies at cosmic dawn with deep JWST spectroscopy”. J.A.-M., C.P.-J., B.R.P., A.V-O acknowledge support by grant PID2024-158856NA-I00, J.A.-M., L.C., and C.P.-J. acknowledge support by grant PIB2021-127718NB-I00, P.G.P.-G acknowledges support from grant PID2022-139567NB-I00, and L.C. and P.G.P.-G acknowledge support from grant PID2025-169195NB-I00 from the Spanish Ministry of Science and Innovation/State Agency of Research MCIN/AEI/10.13039/501100011033 and by “ERDF A way of making Europe”. J.A.-M., L.C., C.P.-J., B.R.P., P.G.P.-G. acknowledge support by grant CSIC/BILATERALES2025/BIJSP25022. A.B. and G.Ö. acknowledges support from the Swedish National Space Administration (SNSA). R.F.A. and A.V.O. are supported by the European Research Council (ERC) under the European Union’s Horizon 2020 research and innovation programme (DistantDust, Grant agreement No. 101117541). J.M.H. acknowledges support from the Evolving Universe Fellowship, which is made possible by a generous donation from Dr. Keiko Miwa Ross; J.M.H. also acknowledges support from JWST Program 8544. D.L. acknowledges support from the Carlsberg Foundation, grant CF25-0662. 
\end{acknowledgements}
\vspace{-0.2in}
\bibliography{main}{}

\begin{thebibliography}{158}
\expandafter\ifx\csname natexlab\endcsname\relax\def\natexlab#1{#1}\fi

\bibitem[{{Abdurro'uf} {et~al.}(2024){Abdurro'uf}, {Larson}, {Coe}, {Hsiao},
  {{\'A}lvarez-M{\'a}rquez}, {G{\'o}mez}, {Adamo}, {Bhatawdekar}, {Bik},
  {Bradley}, {Conselice}, {Dayal}, {Diego}, {Fujimoto}, {Furtak}, {Hutchison},
  {Jung}, {Killi}, {Kokorev}, {Mingozzi}, {Norman}, {Resseguier}, {Ricotti},
  {Rigby}, {Vanzella}, {Welch}, {Windhorst}, {Xu}, \&
  {Zitrin}}]{Abdurrouf_2024}
{Abdurro'uf}, {Larson}, R.~L., {Coe}, D., {et~al.} 2024, \apj, 973, 47

\bibitem[{{Ahn} {et~al.}(2012){Ahn}, {Alexandroff}, {Allende Prieto},
  {Anderson}, {Anderton}, {Andrews}, {Aubourg}, {Bailey}, {Balbinot}, {Barnes},
  {Bautista}, {Beers}, {Beifiori}, {Berlind}, {Bhardwaj}, {Bizyaev}, {Blake},
  {Blanton}, {Blomqvist}, {Bochanski}, {Bolton}, {Borde}, {Bovy}, {Brandt},
  {Brinkmann}, {Brown}, {Brownstein}, {Bundy}, {Busca}, {Carithers}, {Carnero},
  {Carr}, {Casetti-Dinescu}, {Chen}, {Chiappini}, {Comparat}, {Connolly},
  {Crepp}, {Cristiani}, {Croft}, {Cuesta}, {da Costa}, {Davenport}, {Dawson},
  {de Putter}, {De Lee}, {Delubac}, {Dhital}, {Ealet}, {Ebelke}, {Edmondson},
  {Eisenstein}, {Escoffier}, {Esposito}, {Evans}, {Fan}, {Femen{\'\i}a
  Castell{\'a}}, {Fern{\'a}ndez Alvar}, {Ferreira}, {Filiz Ak}, {Finley},
  {Fleming}, {Font-Ribera}, {Frinchaboy}, {Garc{\'\i}a-Hern{\'a}ndez},
  {Garc{\'\i}a P{\'e}rez}, {Ge}, {G{\'e}nova-Santos}, {Gillespie}, {Girardi},
  {Gonz{\'a}lez Hern{\'a}ndez}, {Grebel}, {Gunn}, {Guo}, {Haggard}, {Hamilton},
  {Harris}, {Hawley}, {Hearty}, {Ho}, {Hogg}, {Holtzman}, {Honscheid},
  {Huehnerhoff}, {Ivans}, {Ivezi{\'c}}, {Jacobson}, {Jiang}, {Johansson},
  {Johnson}, {Kauffmann}, {Kirkby}, {Kirkpatrick}, {Klaene}, {Knapp}, {Kneib},
  {Le Goff}, {Leauthaud}, {Lee}, {Lee}, {Long}, {Loomis}, {Lucatello},
  {Lundgren}, {Lupton}, {Ma}, {Ma}, {MacDonald}, {Mack}, {Mahadevan}, {Maia},
  {Majewski}, {Makler}, {Malanushenko}, {Malanushenko}, {Manchado},
  {Mandelbaum}, {Manera}, {Maraston}, {Margala}, {Martell}, {McBride},
  {McGreer}, {McMahon}, {M{\'e}nard}, {Meszaros}, {Miralda-Escud{\'e}},
  {Montero-Dorta}, {Montesano}, {Morrison}, {Muna}, {Munn}, {Murayama},
  {Myers}, {Neto}, {Nguyen}, {Nichol}, {Nidever}, {Noterdaeme}, {Nuza},
  {Ogando}, {Olmstead}, {Oravetz}, {Owen}, {Padmanabhan},
  {Palanque-Delabrouille}, {Pan}, {Parejko}, {Parihar}, {P{\^a}ris},
  {Pattarakijwanich}, {Pepper}, {Percival}, {P{\'e}rez-Fournon},
  {P{\'e}rez-R{\`a}fols}, {Petitjean}, {Pforr}, {Pieri}, {Pinsonneault}, {Porto
  de Mello}, {Prada}, {Price-Whelan}, {Raddick}, {Rebolo}, {Rich}, {Richards},
  {Robin}, {Rocha-Pinto}, {Rockosi}, {Roe}, {Ross}, {Ross}, {Rossi},
  {Rubi{\~n}o-Martin}, {Samushia}, {Sanchez Almeida}, {S{\'a}nchez},
  {Santiago}, {Sayres}, {Schlegel}, {Schlesinger}, {Schmidt}, {Schneider},
  {Schultheis}, {Schwope}, {Sc{\'o}ccola}, {Seljak}, {Sheldon}, {Shen}, {Shu},
  {Simmerer}, {Simmons}, {Skibba}, {Skrutskie}, {Slosar}, {Sobreira}, {Sobeck},
  {Stassun}, {Steele}, \& {Steinmetz}}]{Ahn_2012}
{Ahn}, C.~P., {Alexandroff}, R., {Allende Prieto}, C., {et~al.} 2012, \apjs,
  203, 21

\bibitem[{{{\'A}lvarez-M{\'a}rquez} {et~al.}(2026){{\'A}lvarez-M{\'a}rquez},
  {Colina}, {Crespo-Gomez}, {Kendrew}, {Zavala}, {Marques-Chaves},
  {Prieto-Jim{\'e}nez}, {Abdurro'uf}, {Blanco-Prieto}, {Boogaard},
  {Castellano}, {Fontana}, {Fudamoto}, {Fujimoto}, {Garc{\'\i}a-Mar{\'\i}n},
  {Harikane}, {Harish}, {Hashimoto}, {Hsiao}, {Iani}, {Inoue}, {Langeroodi},
  {Lin}, {Melinder}, {Napolitano}, {Ostlin}, {P{\'e}rez-Gonz{\'a}lez},
  {Rinaldi}, {Rodr{\'\i}guez Del Pino}, {Santini}, {Sugahara}, {Treu},
  {Varo-O'ferral}, \& {Wright}}]{Alvarez-Marquez_2026}
{{\'A}lvarez-M{\'a}rquez}, J., {Colina}, L., {Crespo-Gomez}, A., {et~al.} 2026,
  arXiv e-prints, arXiv:2602.02323

\bibitem[{{{\'A}lvarez-M{\'a}rquez} {et~al.}(2025){{\'A}lvarez-M{\'a}rquez},
  {Crespo G{\'o}mez}, {Colina}, {Langeroodi}, {Marques-Chaves},
  {Prieto-Jim{\'e}nez}, {Bik}, {Alonso-Herrero}, {Boogaard}, {Costantin},
  {Garc{\'\i}a-Mar{\'\i}n}, {Gillman}, {Hjorth}, {Iani}, {Jermann}, {Labiano},
  {Melinder}, {Meyer}, {{\"O}stlin}, {P{\'e}rez-Gonz{\'a}lez}, {Rinaldi},
  {Walter}, {van der Werf}, \& {Wright}}]{Alvarez-Marquez_2025}
{{\'A}lvarez-M{\'a}rquez}, J., {Crespo G{\'o}mez}, A., {Colina}, L., {et~al.}
  2025, \aap, 695, A250

\bibitem[{{Arellano-C{\'o}rdova} {et~al.}(2022){Arellano-C{\'o}rdova}, {Berg},
  {Chisholm}, {Arrabal Haro}, {Dickinson}, {Finkelstein}, {Leclercq}, {Rogers},
  {Simons}, {Skillman}, {Trump}, \& {Kartaltepe}}]{Arellano-Cordova_2022}
{Arellano-C{\'o}rdova}, K.~Z., {Berg}, D.~A., {Chisholm}, J., {et~al.} 2022,
  \apjl, 940, L23

\bibitem[{{Arellano-C{\'o}rdova} {et~al.}(2025){Arellano-C{\'o}rdova},
  {Cullen}, {Carnall}, {Scholte}, {Stanton}, {Kobayashi}, {Martinez}, {Berg},
  {Barrufet}, {Begley}, {Donnan}, {Dunlop}, {Hamadouche}, {McLeod}, {McLure},
  {Rowlands}, \& {Shapley}}]{Arellano-Cordova_2025}
{Arellano-C{\'o}rdova}, K.~Z., {Cullen}, F., {Carnall}, A.~C., {et~al.} 2025,
  \mnras, 540, 2991

\bibitem[{{Arrabal Haro} {et~al.}(2023){Arrabal Haro}, {Dickinson},
  {Finkelstein}, {Kartaltepe}, {Donnan}, {Burgarella}, {Carnall}, {Cullen},
  {Dunlop}, {Fern{\'a}ndez}, {Fujimoto}, {Jung}, {Krips}, {Larson}, {Papovich},
  {P{\'e}rez-Gonz{\'a}lez}, {Amor{\'\i}n}, {Bagley}, {Buat}, {Casey},
  {Chworowsky}, {Cohen}, {Ferguson}, {Giavalisco}, {Huertas-Company},
  {Hutchison}, {Kocevski}, {Koekemoer}, {Lucas}, {McLeod}, {McLure}, {Pirzkal},
  {Seill{\'e}}, {Trump}, {Weiner}, {Wilkins}, \& {Zavala}}]{Arrabal-Haro_2023}
{Arrabal Haro}, P., {Dickinson}, M., {Finkelstein}, S.~L., {et~al.} 2023, \nat,
  622, 707

\bibitem[{{Asplund} {et~al.}(2009){Asplund}, {Grevesse}, {Sauval}, \&
  {Scott}}]{Asplund_2009}
{Asplund}, M., {Grevesse}, N., {Sauval}, A.~J., \& {Scott}, P. 2009, \araa, 47,
  481

\bibitem[{{Atek} {et~al.}(2009){Atek}, {Kunth}, {Schaerer}, {Hayes},
  {Deharveng}, {{\"O}stlin}, \& {Mas-Hesse}}]{Atek_2009}
{Atek}, H., {Kunth}, D., {Schaerer}, D., {et~al.} 2009, \aap, 506, L1

\bibitem[{{Berg} {et~al.}(2019){Berg}, {Erb}, {Henry}, {Skillman}, \&
  {McQuinn}}]{Berg_2019}
{Berg}, D.~A., {Erb}, D.~K., {Henry}, R. B.~C., {Skillman}, E.~D., \&
  {McQuinn}, K. B.~W. 2019, \apj, 874, 93

\bibitem[{{Berg} {et~al.}(2026){Berg}, {Naidu}, {Chisholm}, {Atek}, {Fujimoto},
  {Kokorev}, {Furtak}, {Kobayashi}, {Schaerer}, {Adamo}, {Fei}, {Korber},
  {Matthee}, {Marques-Chaves}, {Martinez}, {McQuinn}, {Mu{\~n}oz}, {Oesch},
  {Saldana-Lopez}, {Stark}, {Stephenson}, \& {Hsiao}}]{Berg_2026}
{Berg}, D.~A., {Naidu}, R.~P., {Chisholm}, J., {et~al.} 2026, \apj, 1003, 112

\bibitem[{{Bergamini} {et~al.}(2023){Bergamini}, {Acebron}, {Grillo}, {Rosati},
  {Caminha}, {Mercurio}, {Vanzella}, {Mason}, {Treu}, {Angora}, {Brammer},
  {Meneghetti}, {Nonino}, {Boyett}, {Brada{\v{c}}}, {Castellano}, {Fontana},
  {Morishita}, {Paris}, {Prieto-Lyon}, {Roberts-Borsani}, {Roy}, {Santini},
  {Vulcani}, {Wang}, \& {Yang}}]{Bergamini_2023}
{Bergamini}, P., {Acebron}, A., {Grillo}, C., {et~al.} 2023, \apj, 952, 84

\bibitem[{{Burgarella} {et~al.}(2026){Burgarella}, {Buat}, {Inoue}, {Takeuchi},
  {Aurin}, {Bouret}, {Dayal}, {Dewachter}, {Dickinson}, {Kobayashi},
  {Nikopoulos}, \& {Somerville}}]{Burgarella_2026}
{Burgarella}, D., {Buat}, V., {Inoue}, A.~K., {et~al.} 2026, arXiv e-prints,
  arXiv:2605.09829

\bibitem[{{Bushouse} {et~al.}(2025){Bushouse}, {Eisenhamer}, {Dencheva},
  {Davies}, {Greenfield}, {Morrison}, {Hodge}, {Simon}, {Grumm}, {Droettboom},
  {Slavich}, {Sosey}, {Pauly}, {Miller}, {Jedrzejewski}, {Hack}, {Davis},
  {Crawford}, {Law}, {Gordon}, {Regan}, {Cara}, {MacDonald}, {Bradley},
  {Shanahan}, {Jamieson}, {Teodoro}, {Williams}, {Pena-Guerrero}, {Graham},
  {Molter}, {Brandt}, {Hayes}, {Cooper}, {Clarke}, \&
  {Filippazzo}}]{Bushouse_2025}
{Bushouse}, H., {Eisenhamer}, J., {Dencheva}, N., {et~al.} 2025, {JWST
  Calibration Pipeline}

\bibitem[{{Calabr{\`o}} {et~al.}(2024){Calabr{\`o}}, {Castellano}, {Zavala},
  {Pentericci}, {Arrabal Haro}, {Bakx}, {Burgarella}, {Casey}, {Dickinson},
  {Finkelstein}, {Fontana}, {Llerena}, {Mascia}, {Merlin}, {Mitsuhashi},
  {Napolitano}, {Paris}, {P{\'e}rez-Gonz{\'a}lez}, {Roberts-Borsani},
  {Santini}, {Treu}, \& {Vanzella}}]{Calabro_2024}
{Calabr{\`o}}, A., {Castellano}, M., {Zavala}, J.~A., {et~al.} 2024, \apj, 975,
  245

\bibitem[{{Calzetti} {et~al.}(2000){Calzetti}, {Armus}, {Bohlin}, {Kinney},
  {Koornneef}, \& {Storchi-Bergmann}}]{Calzetti_2000}
{Calzetti}, D., {Armus}, L., {Bohlin}, R.~C., {et~al.} 2000, \apj, 533, 682

\bibitem[{{Cameron} {et~al.}(2026){Cameron}, {Carreira}, {Simmonds}, {Bunker},
  {Saxena}, {Carniani}, {Charlot}, {Chevallard}, {Curtis-Lake}, {Hainline},
  {Hausen}, {Ji}, {Ji}, {Johnson}, {Rinaldi}, {Robertson}, {Scholtz},
  {Silcock}, {Tacchella}, {Trussler}, {{\"U}bler}, {Williams}, {Willmer},
  {Willott}, \& {Witstok}}]{Cameron_2026}
{Cameron}, A.~J., {Carreira}, C., {Simmonds}, C., {et~al.} 2026, \mnras
  [\eprint[arXiv]{2601.15964}]

\bibitem[{{Cameron} {et~al.}(2023){Cameron}, {Katz}, {Rey}, \&
  {Saxena}}]{Cameron_2023}
{Cameron}, A.~J., {Katz}, H., {Rey}, M.~P., \& {Saxena}, A. 2023, \mnras, 523,
  3516

\bibitem[{{Cardelli} {et~al.}(1989){Cardelli}, {Clayton}, \&
  {Mathis}}]{Cardelli_1989}
{Cardelli}, J.~A., {Clayton}, G.~C., \& {Mathis}, J.~S. 1989, \apj, 345, 245

\bibitem[{{Carnall} {et~al.}(2019){Carnall}, {McLure}, {Dunlop}, {Cullen},
  {McLeod}, {Wild}, {Johnson}, {Appleby}, {Dav{\'e}}, {Amorin}, {Bolzonella},
  {Castellano}, {Cimatti}, {Cucciati}, {Gargiulo}, {Garilli}, {Marchi},
  {Pentericci}, {Pozzetti}, {Schreiber}, {Talia}, \& {Zamorani}}]{Carnall_2019}
{Carnall}, A.~C., {McLure}, R.~J., {Dunlop}, J.~S., {et~al.} 2019, \mnras, 490,
  417

\bibitem[{{Carnall} {et~al.}(2018){Carnall}, {McLure}, {Dunlop}, \&
  {Dav{\'e}}}]{Carnall_2018}
{Carnall}, A.~C., {McLure}, R.~J., {Dunlop}, J.~S., \& {Dav{\'e}}, R. 2018,
  \mnras, 480, 4379

\bibitem[{{Castellano} {et~al.}(2025){Castellano}, {Fontana}, {Merlin},
  {Santini}, {Napolitano}, {Menci}, {P{\'e}rez-Gonz{\'a}lez}, {Calabr{\`o}},
  {Paris}, {Pentericci}, {Zavala}, {Dickinson}, {Finkelstein}, {Treu},
  {Amorin}, {Arrabal Haro}, {Bergamini}, {Bisigello}, {Catone}, {Daddi},
  {Dayal}, {Dekel}, {Ferrara}, {Fortuni}, {Gandolfi}, {Giavalisco}, {Grillo},
  {Guida}, {Hathi}, {Holwerda}, {Koekemoer}, {Kokorev}, {Li}, {Llerena},
  {Lucas}, {Mascia}, {Metha}, {Morishita}, {Nanayakkara}, {Pacucci},
  {Roberts-Borsani}, {Rodighiero}, {Rosati}, {Salazar}, {Schneider},
  {Somerville}, {Taylor}, {Trenti}, {Trinca}, {Wang}, {Watson}, {Yang}, \&
  {Yung}}]{Castellano_2025}
{Castellano}, M., {Fontana}, A., {Merlin}, E., {et~al.} 2025, \aap, 704, A158

\bibitem[{{Castellano} {et~al.}(2022){Castellano}, {Fontana}, {Treu},
  {Santini}, {Merlin}, {Leethochawalit}, {Trenti}, {Vanzella}, {Mestric},
  {Bonchi}, {Belfiori}, {Nonino}, {Paris}, {Polenta}, {Roberts-Borsani},
  {Boyett}, {Brada{\v{c}}}, {Calabr{\`o}}, {Glazebrook}, {Grillo}, {Mascia},
  {Mason}, {Mercurio}, {Morishita}, {Nanayakkara}, {Pentericci}, {Rosati},
  {Vulcani}, {Wang}, \& {Yang}}]{Castellano_2022}
{Castellano}, M., {Fontana}, A., {Treu}, T., {et~al.} 2022, \apjl, 938, L15

\bibitem[{{Castellano} {et~al.}(2026){Castellano}, {Napolitano}, {Moreschini},
  {Calabr{\`o}}, {Christensen}, {Llerena}, {Bakx}, {Belfiore}, {Bevacqua},
  {Dickinson}, {Fontana}, {Gandolfi}, {Gasparetto}, {Marconi}, {Mascia},
  {Merlin}, {Morishita}, {Nanayakkara}, {Paris}, {Pentericci},
  {P{\'e}rez-D{\'\i}az}, {Roberts-Borsani}, {Ruiz}, {Santini}, {Treu},
  {Vanzella}, {Vulcani}, {Wang}, {Yoon}, \& {Zavala}}]{Castellano_2026}
{Castellano}, M., {Napolitano}, L., {Moreschini}, B., {et~al.} 2026, The Open
  Journal of Astrophysics, 9, 60281

\bibitem[{{Chaikin} {et~al.}(2026{\natexlab{a}}){Chaikin}, {Pontzen}, {Frenk},
  {Schaye}, {Lu}, {Crain}, \& {Durrant}}]{Chaikin_2026b}
{Chaikin}, E., {Pontzen}, A., {Frenk}, C.~S., {et~al.} 2026{\natexlab{a}},
  arXiv e-prints, arXiv:2608.19007

\bibitem[{{Chaikin} {et~al.}(2026{\natexlab{b}}){Chaikin}, {Schaye},
  {Schaller}, {Ploeckinger}, {Ben{\'\i}tez-Llambay}, {Frenk}, {Hu{\v{s}}ko},
  {McGibbon}, {Richings}, \& {Trayford}}]{Chaikin_2026}
{Chaikin}, E., {Schaye}, J., {Schaller}, M., {et~al.} 2026{\natexlab{b}},
  \mnras, 548, stag740

\bibitem[{{Charbonnel} {et~al.}(2023){Charbonnel}, {Schaerer}, {Prantzos},
  {Ram{\'\i}rez-Galeano}, {Fragos}, {Kuruvanthodi}, {Marques-Chaves}, \&
  {Gieles}}]{Charbonnel_2023}
{Charbonnel}, C., {Schaerer}, D., {Prantzos}, N., {et~al.} 2023, \aap, 673, L7

\bibitem[{{Chen} {et~al.}(2026){Chen}, {Stark}, {Mason}, {Senchyna}, {Tang},
  {Keerthi Vasan G.}, {Whitler}, {Plat}, \& {Gelli}}]{Chen_2026}
{Chen}, Z., {Stark}, D.~P., {Mason}, C.~A., {et~al.} 2026, arXiv e-prints,
  arXiv:2608.12699

\bibitem[{{Chisholm} {et~al.}(2022){Chisholm}, {Saldana-Lopez}, {Flury},
  {Schaerer}, {Jaskot}, {Amor{\'\i}n}, {Atek}, {Finkelstein}, {Fleming},
  {Ferguson}, {Fern{\'a}ndez}, {Giavalisco}, {Hayes}, {Heckman}, {Henry}, {Ji},
  {Marques-Chaves}, {Mauerhofer}, {McCandliss}, {Oey}, {{\"O}stlin},
  {Rutkowski}, {Scarlata}, {Thuan}, {Trebitsch}, {Wang}, {Worseck}, \&
  {Xu}}]{Chisholm_2022}
{Chisholm}, J., {Saldana-Lopez}, A., {Flury}, S., {et~al.} 2022, \mnras, 517,
  5104

\bibitem[{{Cole} {et~al.}(2025){Cole}, {Papovich}, {Finkelstein}, {Bagley},
  {Dickinson}, {Iyer}, {Yung}, {Ciesla}, {Amor{\'\i}n}, {Arrabal Haro},
  {Bhatawdekar}, {Calabr{\`o}}, {Cleri}, {de la Vega}, {Dekel}, {Endsley},
  {Gawiser}, {Giavalisco}, {Hathi}, {Hirschmann}, {Holwerda}, {Kartaltepe},
  {Koekemoer}, {Lucas}, {Mascia}, {Mobasher}, {P{\'e}rez-Gonz{\'a}lez},
  {Rodighiero}, {Ronayne}, {Tacchella}, {Weiner}, \& {Wilkins}}]{Cole_2025}
{Cole}, J.~W., {Papovich}, C., {Finkelstein}, S.~L., {et~al.} 2025, \apj, 979,
  193

\bibitem[{{Crespo G{\'o}mez} {et~al.}(2025){Crespo G{\'o}mez}, {Tamura},
  {Colina}, {{\'A}lvarez-M{\'a}rquez}, {Hashimoto}, {Marques-Chaves},
  {Nakazato}, {Blanco-Prieto}, {Sunaga}, {Costantin}, {Inoue}, {Hamada},
  {Arribas}, {Ceverino}, {Hagimoto}, {Mawatari}, {Osone}, {Sugahara},
  {Harikane}, {Lee}, {Taniguchi}, \& {Umehata}}]{Crespo-Gomez_2025}
{Crespo G{\'o}mez}, A., {Tamura}, Y., {Colina}, L., {et~al.} 2025, arXiv
  e-prints, arXiv:2511.14658

\bibitem[{{Curti} {et~al.}(2023){Curti}, {D'Eugenio}, {Carniani}, {Maiolino},
  {Sandles}, {Witstok}, {Baker}, {Bennett}, {Piotrowska}, {Tacchella},
  {Charlot}, {Nakajima}, {Maheson}, {Mannucci}, {Amiri}, {Arribas}, {Belfiore},
  {Bonaventura}, {Bunker}, {Chevallard}, {Cresci}, {Curtis-Lake},
  {Hayden-Pawson}, {Jones}, {Kumari}, {Laseter}, {Looser}, {Marconi}, {Maseda},
  {Scholtz}, {Smit}, {{\"U}bler}, \& {Wallace}}]{Curti_2023}
{Curti}, M., {D'Eugenio}, F., {Carniani}, S., {et~al.} 2023, \mnras, 518, 425

\bibitem[{{Curti} {et~al.}(2024){Curti}, {Maiolino}, {Curtis-Lake},
  {Chevallard}, {Carniani}, {D'Eugenio}, {Looser}, {Scholtz}, {Charlot},
  {Cameron}, {{\"U}bler}, {Witstok}, {Boyett}, {Laseter}, {Sandles}, {Arribas},
  {Bunker}, {Giardino}, {Maseda}, {Rawle}, {Rodr{\'\i}guez Del Pino}, {Smit},
  {Willott}, {Eisenstein}, {Hausen}, {Johnson}, {Rieke}, {Robertson},
  {Tacchella}, {Williams}, {Willmer}, {Baker}, {Bhatawdekar}, {Egami},
  {Helton}, {Ji}, {Kumari}, {Perna}, {Shivaei}, \& {Sun}}]{Curti_2024}
{Curti}, M., {Maiolino}, R., {Curtis-Lake}, E., {et~al.} 2024, \aap, 684, A75

\bibitem[{{Curti} {et~al.}(2020){Curti}, {Mannucci}, {Cresci}, \&
  {Maiolino}}]{Curti_2020}
{Curti}, M., {Mannucci}, F., {Cresci}, G., \& {Maiolino}, R. 2020, \mnras, 491,
  944

\bibitem[{{Curti} {et~al.}(2025){Curti}, {Witstok}, {Jakobsen}, {Kobayashi},
  {Curtis-Lake}, {Hainline}, {Ji}, {D'Eugenio}, {Chevallard}, {Maiolino},
  {Scholtz}, {Carniani}, {Arribas}, {Baker}, {Bhatawdekar}, {Boyett}, {Bunker},
  {Cameron}, {Cargile}, {Charlot}, {Eisenstein}, {Ji}, {Johnson}, {Kumari},
  {Maseda}, {Robertson}, {Silcock}, {Tacchella}, {{\"U}bler}, {Venturi},
  {Williams}, {Willmer}, \& {Willott}}]{Curti_2025}
{Curti}, M., {Witstok}, J., {Jakobsen}, P., {et~al.} 2025, \aap, 697, A89

\bibitem[{{de Graaff} {et~al.}(2025){de Graaff}, {Brammer}, {Weibel}, {Lewis},
  {Maseda}, {Oesch}, {Bezanson}, {Boogaard}, {Cleri}, {Cooper}, {Gottumukkala},
  {Greene}, {Hirschmann}, {Hviding}, {Katz}, {Labb{\'e}}, {Leja}, {Matthee},
  {McConachie}, {Miller}, {Naidu}, {Price}, {Rix}, {Setton}, {Suess}, {Wang},
  {Whitaker}, \& {Williams}}]{deGraaff2025}
{de Graaff}, A., {Brammer}, G., {Weibel}, A., {et~al.} 2025, \aap, 697, A189

\bibitem[{{Desroches} {et~al.}(2006){Desroches}, {Filippenko}, {Kaspi}, {Laor},
  {Maoz}, {Ganeshalingam}, {Li}, {Moran}, {Swift}, {Bentz}, {Ho}, {Nandra},
  {O'Neill}, \& {Peterson}}]{Desroches_2006}
{Desroches}, L.-B., {Filippenko}, A.~V., {Kaspi}, S., {et~al.} 2006, \apj, 650,
  88

\bibitem[{{D'Eugenio} {et~al.}(2024){D'Eugenio}, {Maiolino}, {Carniani},
  {Chevallard}, {Curtis-Lake}, {Witstok}, {Charlot}, {Baker}, {Arribas},
  {Boyett}, {Bunker}, {Curti}, {Eisenstein}, {Hainline}, {Ji}, {Johnson},
  {Kumari}, {Looser}, {Nakajima}, {Nelson}, {Rieke}, {Robertson}, {Scholtz},
  {Smit}, {Sun}, {Venturi}, {Tacchella}, {{\"U}bler}, {Willmer}, \&
  {Willott}}]{DEugenio_2024}
{D'Eugenio}, F., {Maiolino}, R., {Carniani}, S., {et~al.} 2024, \aap, 689, A152

\bibitem[{{Donnan} {et~al.}(2025){Donnan}, {Dickinson}, {Taylor}, {Arrabal
  Haro}, {Finkelstein}, {Stanton}, {Jung}, {Papovich}, {Akins}, {Koekemoer},
  {McLeod}, {Napolitano}, {Amor{\'\i}n}, {Begley}, {Burgarella}, {Carnall},
  {Casey}, {Calabr{\`o}}, {Cullen}, {Dunlop}, {Ellis}, {Fern{\'a}ndez},
  {Giavalisco}, {Hirschmann}, {Hu}, {Illingworth}, {Kartaltepe}, {Kocevski},
  {Kokorev}, {Leung}, {Lucas}, {Morales}, {McLure}, {Pentericci},
  {P{\'e}rez-Gonz{\'a}lez}, {Somerville}, {Stevenson}, {Trump}, {Yung}, \&
  {Zavala}}]{Donnan_2025}
{Donnan}, C.~T., {Dickinson}, M., {Taylor}, A.~J., {et~al.} 2025, \apj, 993,
  224

\bibitem[{{Donnan} {et~al.}(2024){Donnan}, {McLure}, {Dunlop}, {McLeod},
  {Magee}, {Arellano-C{\'o}rdova}, {Barrufet}, {Begley}, {Bowler}, {Carnall},
  {Cullen}, {Ellis}, {Fontana}, {Illingworth}, {Grogin}, {Hamadouche},
  {Koekemoer}, {Liu}, {Mason}, {Santini}, \& {Stanton}}]{Donnan_2024}
{Donnan}, C.~T., {McLure}, R.~J., {Dunlop}, J.~S., {et~al.} 2024, \mnras, 533,
  3222

\bibitem[{{Dressler} {et~al.}(2024){Dressler}, {Rieke}, {Eisenstein}, {Stark},
  {Burns}, {Bhatawdekar}, {Bonaventura}, {Boyett}, {Bunker}, {Carniani},
  {Charlot}, {Hausen}, {Misselt}, {Tacchella}, \& {Willmer}}]{Dressler_2024}
{Dressler}, A., {Rieke}, M., {Eisenstein}, D., {et~al.} 2024, \apj, 964, 150

\bibitem[{{Eldridge} {et~al.}(2017){Eldridge}, {Stanway}, {Xiao}, {McClelland},
  {Taylor}, {Ng}, {Greis}, \& {Bray}}]{Eldrige_2017}
{Eldridge}, J.~J., {Stanway}, E.~R., {Xiao}, L., {et~al.} 2017, \pasa, 34, e058

\bibitem[{{Esteban} {et~al.}(2009){Esteban}, {Bresolin}, {Peimbert},
  {Garc{\'\i}a-Rojas}, {Peimbert}, \& {Mesa-Delgado}}]{Esteban_2009}
{Esteban}, C., {Bresolin}, F., {Peimbert}, M., {et~al.} 2009, \apj, 700, 654

\bibitem[{{Esteban} \& {Peimbert}(1995)}]{Esteban_1995}
{Esteban}, C. \& {Peimbert}, M. 1995, \aap, 300, 78

\bibitem[{{Esteban} {et~al.}(2002){Esteban}, {Peimbert}, {Torres-Peimbert}, \&
  {Rodr{\'\i}guez}}]{Esteban_2002}
{Esteban}, C., {Peimbert}, M., {Torres-Peimbert}, S., \& {Rodr{\'\i}guez}, M.
  2002, \apj, 581, 241

\bibitem[{{Evans} {et~al.}(2024){Evans}, {Evans}, {Mart{\'\i}nez-Galarza},
  {Miller}, {Primini}, {Azadi}, {Burke}, {Civano}, {D'Abrusco}, {Fabbiano},
  {Graessle}, {Grier}, {Houck}, {Lauer}, {McCollough}, {Nowak}, {Plummer},
  {Rots}, {Siemiginowska}, \& {Tibbetts}}]{Evans_2024}
{Evans}, I.~N., {Evans}, J.~D., {Mart{\'\i}nez-Galarza}, J.~R., {et~al.} 2024,
  \apjs, 274, 22

\bibitem[{{Fei} {et~al.}(2026){Fei}, {Fujimoto}, {Naidu}, {Chisholm}, {Atek},
  {Brammer}, {Asada}, {Berg}, {Bromm}, {Furtak}, {Greene}, {Hsiao}, {Jeon},
  {Kokorev}, {Matthee}, {Natarajan}, {Pan}, {Richard}, {Saldana-Lopez},
  {Schaerer}, {Volonteri}, \& {Zitrin}}]{Fei_2026}
{Fei}, Q., {Fujimoto}, S., {Naidu}, R.~P., {et~al.} 2026, \apj, 1003, 244

\bibitem[{{Ferrara}(2024)}]{Ferrara_2024}
{Ferrara}, A. 2024, \aap, 684, A207

\bibitem[{{Ferrara} {et~al.}(2025){Ferrara}, {Pallottini}, \&
  {Sommovigo}}]{Ferrara_2025}
{Ferrara}, A., {Pallottini}, A., \& {Sommovigo}, L. 2025, \aap, 694, A286

\bibitem[{{Filippenko} \& {Sargent}(1989)}]{Filippenko_1989}
{Filippenko}, A.~V. \& {Sargent}, W. L.~W. 1989, \apjl, 342, L11

\bibitem[{{Foreman-Mackey} {et~al.}(2013){Foreman-Mackey}, {Hogg}, {Lang}, \&
  {Goodman}}]{emcee}
{Foreman-Mackey}, D., {Hogg}, D.~W., {Lang}, D., \& {Goodman}, J. 2013, \pasp,
  125, 306

\bibitem[{{Froese Fischer} \& {Tachiev}(2004)}]{FFT_2004}
{Froese Fischer}, C. \& {Tachiev}, G. 2004, Atomic Data and Nuclear Data
  Tables, 87, 1

\bibitem[{{Galavis} {et~al.}(1997){Galavis}, {Mendoza}, \&
  {Zeippen}}]{GMZ_1997}
{Galavis}, M.~E., {Mendoza}, C., \& {Zeippen}, C.~J. 1997, \aaps, 123, 159

\bibitem[{{Gandolfi} {et~al.}(2026){Gandolfi}, {Rodighiero}, {Bisigello},
  {Grazian}, {Finkelstein}, {Dickinson}, {Castellano}, {Merlin}, {Calabr{\`o}},
  {Papovich}, {Bianchetti}, {Ba{\~n}ados}, {Benotto}, {Catone}, {Buitrago},
  {Daddi}, {Girardi}, {Giulietti}, {Hirschmann}, {Holwerda}, {Arrabal Haro},
  {Lapi}, {Lucas}, {Lyu}, {Massardi}, {Pacucci}, {P{\'e}rez-Gonz{\'a}lez},
  {Ronconi}, {Tarrasse}, {Wilkins}, {Vulcani}, {Yung}, {Zavala}, {Backhaus},
  {Bagley}, {Buat}, {Burgarella}, {Kartaltepe}, {Khusanova}, {Kirkpatrick},
  {Kocevski}, {Koekemoer}, {Lambrides}, {Pirzkal}, \& {Yang}}]{Gandolfi_2026}
{Gandolfi}, G., {Rodighiero}, G., {Bisigello}, L., {et~al.} 2026, \aap, 708,
  A195

\bibitem[{{Garcia} {et~al.}(2025){Garcia}, {Torrey}, {Ellison}, {Grasha},
  {Chen}, {Hemler}, {Zimmerman}, {Wright}, {Zovaro}, {Nelson}, {Sanders},
  {Kewley}, \& {Hernquist}}]{Garcia_2025}
{Garcia}, A.~M., {Torrey}, P., {Ellison}, S.~L., {et~al.} 2025, \mnras, 536,
  119

\bibitem[{{GRAVITY Collaboration} {et~al.}(2024){GRAVITY Collaboration},
  {Amorim}, {Bourdarot}, {Brandner}, {Cao}, {Cl{\'e}net}, {Davies}, {de Zeeuw},
  {Dexter}, {Drescher}, {Eckart}, {Eisenhauer}, {Fabricius}, {Feuchtgruber},
  {F{\"o}rster Schreiber}, {Garcia}, {Genzel}, {Gillessen}, {Gratadour},
  {H{\"o}nig}, {Kishimoto}, {Lacour}, {Lutz}, {Millour}, {Netzer}, {Ott},
  {Paumard}, {Perraut}, {Perrin}, {Peterson}, {Petrucci}, {Pfuhl}, {Prieto},
  {Rabien}, {Rouan}, {Santos}, {Shangguan}, {Shimizu}, {Sternberg},
  {Straubmeier}, {Sturm}, {Tacconi}, {Tristram}, {Widmann}, \&
  {Woillez}}]{Gravity_2024}
{GRAVITY Collaboration}, {Amorim}, A., {Bourdarot}, G., {et~al.} 2024, \aap,
  684, A167

\bibitem[{{Gunasekera} {et~al.}(2025){Gunasekera}, {van Hoof}, {Dehghanian},
  {Chakraborty}, {Shaw}, {Bianchi}, {Chatzikos}, {Tsujimoto}, \&
  {Ferland}}]{cloudy_v25}
{Gunasekera}, C.~M., {van Hoof}, P.~A.~M., {Dehghanian}, M., {et~al.} 2025,
  \rmxaa, 61, 120

\bibitem[{{Harikane} {et~al.}(2026){Harikane}, {Perez-Gonzalez},
  {Alvarez-Marquez}, {Ouchi}, {Nakazato}, {Ono}, {Nakajima}, {Umeda}, {Isobe},
  {Xu}, \& {Zhang}}]{Harikane_2026}
{Harikane}, Y., {Perez-Gonzalez}, P.~G., {Alvarez-Marquez}, J., {et~al.} 2026,
  arXiv e-prints, arXiv:2601.21833

\bibitem[{{Harikane} {et~al.}(2025){Harikane}, {Sanders}, {Ellis}, {Jones},
  {Ouchi}, {Laporte}, {Roberts-Borsani}, {Katz}, {Nakajima}, {Ono}, \&
  {Gupta}}]{Hariakane_2025}
{Harikane}, Y., {Sanders}, R.~L., {Ellis}, R., {et~al.} 2025, \apj, 993, 204

\bibitem[{{Harikane} {et~al.}(2023){Harikane}, {Zhang}, {Nakajima}, {Ouchi},
  {Isobe}, {Ono}, {Hatano}, {Xu}, \& {Umeda}}]{Harikane_2023}
{Harikane}, Y., {Zhang}, Y., {Nakajima}, K., {et~al.} 2023, \apj, 959, 39

\bibitem[{{Heintz} {et~al.}(2025){Heintz}, {Brammer}, {Watson}, {Oesch},
  {Keating}, {Hayes}, {Abdurro'uf}, {Arellano-C{\'o}rdova}, {Carnall},
  {Christiansen}, {Cullen}, {Dav{\'e}}, {Dayal}, {Ferrara}, {Finlator},
  {Fynbo}, {Flury}, {Gelli}, {Gillman}, {Gottumukkala}, {Gould}, {Greve},
  {Hardin}, {Hsiao}, {Hutter}, {Jakobsson}, {Killi}, {Khosravaninezhad},
  {Laursen}, {Lee}, {Magdis}, {Matthee}, {Naidu}, {Narayanan}, {Pollock},
  {Prescott}, {Rusakov}, {Shuntov}, {Sneppen}, {Smit}, {Tanvir}, {Terp},
  {Toft}, {Valentino}, {Vijayan}, {Weaver}, {Wise}, \& {Witstok}}]{Heintz_2025}
{Heintz}, K.~E., {Brammer}, G.~B., {Watson}, D., {et~al.} 2025, \aap, 693, A60

\bibitem[{{Heintz} {et~al.}(2024){Heintz}, {Watson}, {Brammer}, {Vejlgaard},
  {Hutter}, {Strait}, {Matthee}, {Oesch}, {Jakobsson}, {Tanvir}, {Laursen},
  {Naidu}, {Mason}, {Killi}, {Jung}, {Hsiao}, {Abdurro'uf}, {Coe}, {Arrabal
  Haro}, {Finkelstein}, \& {Toft}}]{Heintz2024dja}
{Heintz}, K.~E., {Watson}, D., {Brammer}, G., {et~al.} 2024, Science, 384, 890

\bibitem[{{Helton} {et~al.}(2026){Helton}, {Morrison}, {Hainline}, {D'Eugenio},
  {Rieke}, {Alberts}, {Carniani}, {Leja}, {Li}, {Rinaldi}, {Scholtz}, {Stone},
  {Willmer}, {Wu}, {Baker}, {Bunker}, {Charlot}, {Chevallard}, {Cleri},
  {Curti}, {Curtis-Lake}, {Egami}, {Eisenstein}, {Jakobsen}, {Ji}, {Johnson},
  {Kumari}, {Lin}, {Lyu}, {Maiolino}, {Maseda}, {P{\'e}rez-Gonz{\'a}lez},
  {Rieke}, {Robertson}, {Saxena}, {Sun}, {Tacchella}, {{\"U}bler}, {Venturi},
  {Williams}, {Willott}, {Witstok}, \& {Zhu}}]{Helton_2026}
{Helton}, J.~M., {Morrison}, J.~E., {Hainline}, K.~N., {et~al.} 2026, \apjl,
  1004, L30

\bibitem[{{Henry} {et~al.}(2000){Henry}, {Edmunds}, \&
  {K{\"o}ppen}}]{Henry_2000}
{Henry}, R.~B.~C., {Edmunds}, M.~G., \& {K{\"o}ppen}, J. 2000, \apj, 541, 660

\bibitem[{{HI4PI Collaboration} {et~al.}(2016){HI4PI Collaboration}, {Ben
  Bekhti}, {Fl{\"o}er}, {Keller}, {Kerp}, {Lenz}, {Winkel}, {Bailin},
  {Calabretta}, {Dedes}, {Ford}, {Gibson}, {Haud}, {Janowiecki}, {Kalberla},
  {Lockman}, {McClure-Griffiths}, {Murphy}, {Nakanishi}, {Pisano}, \&
  {Staveley-Smith}}]{HI4PI_2016}
{HI4PI Collaboration}, {Ben Bekhti}, N., {Fl{\"o}er}, L., {et~al.} 2016, \aap,
  594, A116

\bibitem[{{Hsiao} {et~al.}(2024){Hsiao}, {{\'A}lvarez-M{\'a}rquez}, {Coe},
  {Crespo G{\'o}mez}, {Abdurro'uf}, {Dayal}, {Larson}, {Bik}, {Blanco-Prieto},
  {Colina}, {P{\'e}rez-Gonz{\'a}lez}, {Costantin}, {Prieto-Jim{\'e}nez},
  {Adamo}, {Bradley}, {Conselice}, {Fujimoto}, {Furtak}, {Hutchison}, {James},
  {Jim{\'e}nez-Teja}, {Jung}, {Kokorev}, {Mingozzi}, {Norman}, {Ricotti},
  {Rigby}, {Sharon}, {Vanzella}, {Welch}, {Xu}, {Zackrisson}, \&
  {Zitrin}}]{Hsiao_2024}
{Hsiao}, T. Y.-Y., {{\'A}lvarez-M{\'a}rquez}, J., {Coe}, D., {et~al.} 2024,
  \apj, 973, 81

\bibitem[{{Hsiao} {et~al.}(2025){Hsiao}, {Topping}, {Coe}, {Chisholm}, {Berg},
  {Abdurro'uf}, {{\'A}lvarez-M{\'a}rquez}, {Maiolino}, {Dayal}, \&
  {Furtak}}]{Tiger_2025}
{Hsiao}, T. Y.-Y., {Topping}, M.~W., {Coe}, D., {et~al.} 2025, \apj, 993, 70

\bibitem[{{Isobe} {et~al.}(2023){Isobe}, {Ouchi}, {Tominaga}, {Watanabe},
  {Nakajima}, {Umeda}, {Yajima}, {Harikane}, {Fukushima}, {Xu}, {Ono}, \&
  {Zhang}}]{Isobe_2023}
{Isobe}, Y., {Ouchi}, M., {Tominaga}, N., {et~al.} 2023, \apj, 959, 100

\bibitem[{{Iyer} {et~al.}(2019){Iyer}, {Gawiser}, {Faber}, {Ferguson},
  {Kartaltepe}, {Koekemoer}, {Pacifici}, \& {Somerville}}]{Iyer_2019}
{Iyer}, K.~G., {Gawiser}, E., {Faber}, S.~M., {et~al.} 2019, \apj, 879, 116

\bibitem[{{Izotov} {et~al.}(2023){Izotov}, {Schaerer}, {Worseck}, {Berg},
  {Chisholm}, {Ravindranath}, \& {Thuan}}]{Izotov_2023}
{Izotov}, Y.~I., {Schaerer}, D., {Worseck}, G., {et~al.} 2023, \mnras, 522,
  1228

\bibitem[{{Ji} {et~al.}(2024){Ji}, {{\"U}bler}, {Maiolino}, {D'Eugenio},
  {Arribas}, {Bunker}, {Charlot}, {Perna}, {Rodr{\'\i}guez Del Pino},
  {B{\"o}ker}, {Cresci}, {Curti}, {Kumari}, \& {Lamperti}}]{Ji_2024}
{Ji}, X., {{\"U}bler}, H., {Maiolino}, R., {et~al.} 2024, \mnras, 535, 881

\bibitem[{{Jin} {et~al.}(2012){Jin}, {Ward}, \& {Done}}]{Jin_2012}
{Jin}, C., {Ward}, M., \& {Done}, C. 2012, \mnras, 422, 3268

\bibitem[{{Katz} {et~al.}(2025){Katz}, {Cameron}, {Saxena}, {Barrufet},
  {Choustikov}, {Cleri}, {de Graff}, {Ellis}, {Fosbury}, {Heintz}, {Maseda},
  {Matthee}, {McConachie}, \& {Oesch}}]{Katz_2025}
{Katz}, H., {Cameron}, A.~J., {Saxena}, A., {et~al.} 2025, The Open Journal of
  Astrophysics, 8, 104

\bibitem[{{Kendrew} {et~al.}(2015){Kendrew}, {Scheithauer}, {Bouchet},
  {Amiaux}, {Azzollini}, {Bouwman}, {Chen}, {Dubreuil}, {Fischer}, {Glasse},
  {Greene}, {Lagage}, {Lahuis}, {Ronayette}, {Wright}, \&
  {Wright}}]{Kendrew_2015}
{Kendrew}, S., {Scheithauer}, S., {Bouchet}, P., {et~al.} 2015, \pasp, 127, 623

\bibitem[{{Kewley} {et~al.}(2019){Kewley}, {Nicholls}, \&
  {Sutherland}}]{Kewley_2019}
{Kewley}, L.~J., {Nicholls}, D.~C., \& {Sutherland}, R.~S. 2019, \araa, 57, 511

\bibitem[{{Kisielius} {et~al.}(2009){Kisielius}, {Storey}, {Ferland}, \&
  {Keenan}}]{Kisielius_2009}
{Kisielius}, R., {Storey}, P.~J., {Ferland}, G.~J., \& {Keenan}, F.~P. 2009,
  \mnras, 397, 903

\bibitem[{{Kokorev} {et~al.}(2025){Kokorev}, {Ch{\'a}vez Ortiz}, {Taylor},
  {Finkelstein}, {Arrabal Haro}, {Dickinson}, {Chisholm}, {Fujimoto}, {noz},
  {Endsley}, {Hu}, {Napolitano}, {Wilkins}, {Akins}, {Amori{\'\i}n}, {Casey},
  {Cheng}, {Cleri}, {Cole}, {Cullen}, {Daddi}, {Davis}, {Donnan}, {Dunlop},
  {Fern{\'a}ndez}, {Giavalisco}, {Grogin}, {Hathi}, {Hirschmann}, {Kartaltepe},
  {Koekemoer}, {Leung}, {Lucas}, {McLeod}, {Papovich}, {Pentericci},
  {P{\'e}rez-Gonz{\'a}lez}, {Somerville}, {Wang}, {Yung}, \&
  {Zavala}}]{Kokorev_2025}
{Kokorev}, V., {Ch{\'a}vez Ortiz}, {\'O}.~A., {Taylor}, A.~J., {et~al.} 2025,
  \apjl, 988, L10

\bibitem[{{Koller} {et~al.}(2026){Koller}, {Maiolino}, {{\"U}bler}, {Duan},
  {Scholtz}, {Arribas}, {Baker}, {Carniani}, {Charlot}, {Curti}, {Graziani},
  {Jones}, {McClymont}, {Perna}, {Del Pino}, {Tacchella}, {Venditti},
  {Venturi}, \& {Witstok}}]{Koller_2026}
{Koller}, M., {Maiolino}, R., {{\"U}bler}, H., {et~al.} 2026, \mnras

\bibitem[{{Kov{\'a}cs} {et~al.}(2024){Kov{\'a}cs}, {Bogd{\'a}n}, {Natarajan},
  {Werner}, {Azadi}, {Volonteri}, {Tremblay}, {Chadayammuri}, {Forman},
  {Jones}, \& {Kraft}}]{Kovacs_2024}
{Kov{\'a}cs}, O.~E., {Bogd{\'a}n}, {\'A}., {Natarajan}, P., {et~al.} 2024,
  \apjl, 965, L21

\bibitem[{{Kroupa} {et~al.}(1993){Kroupa}, {Tout}, \& {Gilmore}}]{Kroupa_1993}
{Kroupa}, P., {Tout}, C.~A., \& {Gilmore}, G. 1993, \mnras, 262, 545

\bibitem[{{Langeroodi} \& {Hjorth}(2026)}]{Langeroodi_2026}
{Langeroodi}, D. \& {Hjorth}, J. 2026, \apjl, 997, L30

\bibitem[{{Langeroodi} {et~al.}(2023){Langeroodi}, {Hjorth}, {Chen}, {Kelly},
  {Williams}, {Lin}, {Scarlata}, {Zitrin}, {Broadhurst}, {Diego}, {Huang},
  {Filippenko}, {Foley}, {Jha}, {Koekemoer}, {Oguri}, {Perez-Fournon},
  {Pierel}, {Poidevin}, \& {Strolger}}]{Langeroodi_2023}
{Langeroodi}, D., {Hjorth}, J., {Chen}, W., {et~al.} 2023, \apj, 957, 39

\bibitem[{{Laor}(2006)}]{Laor_2006}
{Laor}, A. 2006, \apj, 643, 112

\bibitem[{{Lira} {et~al.}(1999){Lira}, {Lawrence}, {O'Brien}, {Johnson},
  {Terlevich}, \& {Bannister}}]{Lira_1999}
{Lira}, P., {Lawrence}, A., {O'Brien}, P., {et~al.} 1999, \mnras, 305, 109

\bibitem[{{Luridiana} {et~al.}(2015){Luridiana}, {Morisset}, \& {Shaw}}]{pyneb}
{Luridiana}, V., {Morisset}, C., \& {Shaw}, R.~A. 2015, \aap, 573, A42

\bibitem[{{Maiolino} \& {Mannucci}(2019)}]{Maiolino_2019}
{Maiolino}, R. \& {Mannucci}, F. 2019, \aapr, 27, 3

\bibitem[{{Maiolino} {et~al.}(2024){Maiolino}, {Scholtz}, {Witstok},
  {Carniani}, {D'Eugenio}, {de Graaff}, {{\"U}bler}, {Tacchella},
  {Curtis-Lake}, {Arribas}, {Bunker}, {Charlot}, {Chevallard}, {Curti},
  {Looser}, {Maseda}, {Rawle}, {Rodr{\'\i}guez del Pino}, {Willott}, {Egami},
  {Eisenstein}, {Hainline}, {Robertson}, {Williams}, {Willmer}, {Baker},
  {Boyett}, {DeCoursey}, {Fabian}, {Helton}, {Ji}, {Jones}, {Kumari},
  {Laporte}, {Nelson}, {Perna}, {Sandles}, {Shivaei}, \& {Sun}}]{Maiolino_2024}
{Maiolino}, R., {Scholtz}, J., {Witstok}, J., {et~al.} 2024, \nat, 627, 59

\bibitem[{{Marques-Chaves} {et~al.}(2026{\natexlab{a}}){Marques-Chaves},
  {{\'A}lvarez-M{\'a}rquez}, {Colina}, {Kendrew}, {Abdurro'uf},
  {Blanco-Prieto}, {Boogaard}, {Castellano}, {Caputi}, {Crespo-G{\'o}mez},
  {Fontana}, {Fudamoto}, {Fujimoto}, {Garc{\'\i}a-Mar{\'\i}n}, {Harikane},
  {Harish}, {Hashimoto}, {Hsiao}, {Iani}, {Inoue}, {Langeroodi}, {Lin},
  {Melinder}, {Napolitano}, {{\"O}stlin}, {P{\'e}rez-Gonz{\'a}lez},
  {Prieto-Jim{\'e}nez}, {Rinaldi}, {Rodr{\'\i}guez Del Pino}, {Santini},
  {Sugahara}, {Varo-O'Ferrall}, {Wright}, \& {Zavala}}]{Marques-Chaves_2026}
{Marques-Chaves}, R., {{\'A}lvarez-M{\'a}rquez}, J., {Colina}, L., {et~al.}
  2026{\natexlab{a}}, \aap, 711, A301

\bibitem[{{Marques-Chaves} {et~al.}(2026{\natexlab{b}}){Marques-Chaves},
  {Martins}, {Schaerer}, {Dessauges-Zavadsky}, \&
  {Palacios}}]{Marques-Chaves_2026b}
{Marques-Chaves}, R., {Martins}, F., {Schaerer}, D., {Dessauges-Zavadsky}, M.,
  \& {Palacios}, A. 2026{\natexlab{b}}, \aap, 711, L3

\bibitem[{{Marques-Chaves} {et~al.}(2024){Marques-Chaves}, {Schaerer},
  {Kuruvanthodi}, {Korber}, {Prantzos}, {Charbonnel}, {Weibel}, {Izotov},
  {Messa}, {Brammer}, {Dessauges-Zavadsky}, \& {Oesch}}]{Marques-Chaves_2024}
{Marques-Chaves}, R., {Schaerer}, D., {Kuruvanthodi}, A., {et~al.} 2024, \aap,
  681, A30

\bibitem[{{Mauerhofer} {et~al.}(2025){Mauerhofer}, {Dayal}, {Haehnelt}, {Kimm},
  {Rosdahl}, \& {Teyssier}}]{Mauerhofer_2025}
{Mauerhofer}, V., {Dayal}, P., {Haehnelt}, M.~G., {et~al.} 2025, \aap, 696,
  A157

\bibitem[{{Mazzolari} {et~al.}(2024){Mazzolari}, {{\"U}bler}, {Maiolino}, {Ji},
  {Nakajima}, {Feltre}, {Scholtz}, {D'Eugenio}, {Curti}, {Mignoli}, \&
  {Marconi}}]{Mazzolari_2024}
{Mazzolari}, G., {{\"U}bler}, H., {Maiolino}, R., {et~al.} 2024, \aap, 691,
  A345

\bibitem[{{McClymont} {et~al.}(2025){McClymont}, {Tacchella}, {D'Eugenio},
  {Witten}, {Ji}, {Smith}, {Maiolino}, {Arribas}, {Scholtz}, {Simmonds}, \&
  {Witstok}}]{McClymont_2025}
{McClymont}, W., {Tacchella}, S., {D'Eugenio}, F., {et~al.} 2025, \mnras, 540,
  190

\bibitem[{{McClymont} {et~al.}(2026){McClymont}, {Tacchella}, {Smith},
  {Kannan}, {Garaldi}, {Puchwein}, {Isobe}, {Ji}, {Shen}, {Wang}, {Belokurov},
  {Borrow}, {D'Eugenio}, {Keating}, {Maiolino}, {Monty}, {Vogelsberger}, \&
  {Zier}}]{McClymon_2026}
{McClymont}, W., {Tacchella}, S., {Smith}, A., {et~al.} 2026, \mnras, 548,
  stag016

\bibitem[{{McGaugh} {et~al.}(2024){McGaugh}, {Schombert}, {Lelli}, \&
  {Franck}}]{McGaugh_2024}
{McGaugh}, S.~S., {Schombert}, J.~M., {Lelli}, F., \& {Franck}, J. 2024, \apj,
  976, 13

\bibitem[{{McLaughlin} {et~al.}(2011){McLaughlin}, {Lee}, {Ludlow}, {Landi},
  {Loch}, {Pindzola}, \& {Ballance}}]{ML_2011}
{McLaughlin}, B.~M., {Lee}, T.-G., {Ludlow}, J.~A., {et~al.} 2011, Journal of
  Physics B Atomic Molecular Physics, 44, 175206

\bibitem[{{Menci} {et~al.}(2024){Menci}, {Sen}, \& {Castellano}}]{Menci_2024}
{Menci}, N., {Sen}, A.~A., \& {Castellano}, M. 2024, \apj, 976, 227

\bibitem[{{Merlin} {et~al.}(2024){Merlin}, {Santini}, {Paris}, {Castellano},
  {Fontana}, {Treu}, {Finkelstein}, {Dunlop}, {Arrabal Haro}, {Bagley},
  {Boyett}, {Calabr{\`o}}, {Correnti}, {Davis}, {Dickinson}, {Donnan},
  {Ferguson}, {Fortuni}, {Giavalisco}, {Glazebrook}, {Grazian}, {Grogin},
  {Hathi}, {Hirschmann}, {Kartaltepe}, {Kewley}, {Kirkpatrick}, {Kocevski},
  {Koekemoer}, {Leung}, {Lotz}, {Lucas}, {Magee}, {Marchesini}, {Mascia},
  {McLeod}, {McLure}, {Nanayakkara}, {Napolitano}, {Nonino}, {Papovich},
  {Pentericci}, {P{\'e}rez-Gonz{\'a}lez}, {Pirzkal}, {Ravindranath},
  {Roberts-Borsani}, {Somerville}, {Trenti}, {Trump}, {Vulcani}, {Wang},
  {Watson}, {Wilkins}, {Yang}, \& {Yung}}]{Merlin_2024}
{Merlin}, E., {Santini}, P., {Paris}, D., {et~al.} 2024, \aap, 691, A240

\bibitem[{{Moran} {et~al.}(2005){Moran}, {Eracleous}, {Leighly}, {Chartas},
  {Filippenko}, {Ho}, \& {Blanco}}]{Moran_2005}
{Moran}, E.~C., {Eracleous}, M., {Leighly}, K.~M., {et~al.} 2005, \aj, 129,
  2108

\bibitem[{{Morishita} {et~al.}(2024{\natexlab{a}}){Morishita}, {Stiavelli},
  {Chary}, {Trenti}, {Bergamini}, {Chiaberge}, {Leethochawalit},
  {Roberts-Borsani}, {Shen}, \& {Treu}}]{Morishita_2024b}
{Morishita}, T., {Stiavelli}, M., {Chary}, R.-R., {et~al.} 2024{\natexlab{a}},
  \apj, 963, 9

\bibitem[{{Morishita} {et~al.}(2024{\natexlab{b}}){Morishita}, {Stiavelli},
  {Grillo}, {Rosati}, {Schuldt}, {Trenti}, {Bergamini}, {Boyett}, {Chary},
  {Leethochawalit}, {Roberts-Borsani}, {Treu}, \& {Vanzella}}]{Morishita_2024}
{Morishita}, T., {Stiavelli}, M., {Grillo}, C., {et~al.} 2024{\natexlab{b}},
  \apj, 971, 43

\bibitem[{{Naidu} {et~al.}(2026){Naidu}, {Oesch}, {Brammer}, {Weibel}, {Li},
  {Matthee}, {Chisolm}, {Pollock}, {Heintz}, {Johnson}, {Shen}, {Hviding},
  {Leja}, {Tacchella}, {Ganguly}, {Witten}, {Atek}, {Belli}, {Bose}, {Bouwens},
  {Dayal}, {Decarli}, {de Graaff}, {Fudamoto}, {Giovinazzo}, {Greene},
  {Illingworth}, {Inoue}, {Kane}, {Labbe}, {Leonova}, {Marques-Chaves},
  {Meyer}, {Nelson}, {Roberts-Borsani}, {Schaerer}, {Simcoe}, {Stefanon},
  {Sugahara}, {Toft}, {van der Wel}, {van Dokkum}, {Walter}, {Watson},
  {Weaver}, \& {Whitaker}}]{Naidu_2026}
{Naidu}, R.~P., {Oesch}, P.~A., {Brammer}, G., {et~al.} 2026, The Open Journal
  of Astrophysics, 9, 56033

\bibitem[{{Naidu} {et~al.}(2022){Naidu}, {Oesch}, {van Dokkum}, {Nelson},
  {Suess}, {Brammer}, {Whitaker}, {Illingworth}, {Bouwens}, {Tacchella},
  {Matthee}, {Allen}, {Bezanson}, {Conroy}, {Labbe}, {Leja}, {Leonova},
  {Magee}, {Price}, {Setton}, {Strait}, {Stefanon}, {Toft}, {Weaver}, \&
  {Weibel}}]{Naidu_2022}
{Naidu}, R.~P., {Oesch}, P.~A., {van Dokkum}, P., {et~al.} 2022, \apjl, 940,
  L14

\bibitem[{{Nakajima} {et~al.}(2023){Nakajima}, {Ouchi}, {Isobe}, {Harikane},
  {Zhang}, {Ono}, {Umeda}, \& {Oguri}}]{Nakajima_2023}
{Nakajima}, K., {Ouchi}, M., {Isobe}, Y., {et~al.} 2023, \apjs, 269, 33

\bibitem[{{Nakane} \& {Ouchi}(2026)}]{Nakane_2026}
{Nakane}, M. \& {Ouchi}, M. 2026, arXiv e-prints, arXiv:2608.12466

\bibitem[{{Napolitano} {et~al.}(2025{\natexlab{a}}){Napolitano}, {Castellano},
  {Pentericci}, {Arrabal Haro}, {Fontana}, {Treu}, {Bergamini}, {Calabr{\`o}},
  {Mascia}, {Morishita}, {Roberts-Borsani}, {Santini}, {Vanzella}, {Vulcani},
  {Zakharova}, {Bakx}, {Dickinson}, {Grillo}, {Leethochawalit}, {Llerena},
  {Merlin}, {Paris}, {Rojas-Ruiz}, {Rosati}, {Wang}, {Yoon}, \&
  {Zavala}}]{Napolitano_2025a}
{Napolitano}, L., {Castellano}, M., {Pentericci}, L., {et~al.}
  2025{\natexlab{a}}, \aap, 693, A50

\bibitem[{{Napolitano} {et~al.}(2025{\natexlab{b}}){Napolitano}, {Castellano},
  {Pentericci}, {Vignali}, {Gilli}, {Fontana}, {Santini}, {Treu},
  {Calabr{\`o}}, {Llerena}, {Piconcelli}, {Zappacosta}, {Mascia}, {Tripodi},
  {Arrabal Haro}, {Bergamini}, {Bakx}, {Dickinson}, {Glazebrook}, {Henry},
  {Leethochawalit}, {Mazzolari}, {Merlin}, {Morishita}, {Nanayakkara}, {Paris},
  {Puccetti}, {Roberts-Borsani}, {Rojas Ruiz}, {Rosati}, {Vanzella}, {Vito},
  {Vulcani}, {Wang}, {Yoon}, \& {Zavala}}]{Napolitano_2025b}
{Napolitano}, L., {Castellano}, M., {Pentericci}, L., {et~al.}
  2025{\natexlab{b}}, \apj, 989, 75

\bibitem[{{Navarro-Carrera} {et~al.}(2025){Navarro-Carrera}, {Caputi}, {Iani},
  {Rinaldi}, {Kokorev}, \& {Kerutt}}]{Navarro-Carrera_2025}
{Navarro-Carrera}, R., {Caputi}, K.~I., {Iani}, E., {et~al.} 2025, \apj, 993,
  194

\bibitem[{{Nelson} {et~al.}(2019){Nelson}, {Springel}, {Pillepich},
  {Rodriguez-Gomez}, {Torrey}, {Genel}, {Vogelsberger}, {Pakmor}, {Marinacci},
  {Weinberger}, {Kelley}, {Lovell}, {Diemer}, \& {Hernquist}}]{Nelson_2019}
{Nelson}, D., {Springel}, V., {Pillepich}, A., {et~al.} 2019, Computational
  Astrophysics and Cosmology, 6, 2

\bibitem[{{Netzer}(2019)}]{Netzer_2019}
{Netzer}, H. 2019, \mnras, 488, 5185

\bibitem[{{Nicholls} {et~al.}(2017){Nicholls}, {Sutherland}, {Dopita},
  {Kewley}, \& {Groves}}]{Nicholls_2017}
{Nicholls}, D.~C., {Sutherland}, R.~S., {Dopita}, M.~A., {Kewley}, L.~J., \&
  {Groves}, B.~A. 2017, \mnras, 466, 4403

\bibitem[{{Nikopoulos} {et~al.}(2026){Nikopoulos}, {Watson}, {Sneppen},
  {Rusakov}, {Heintz}, {Witstok}, \& {Brammer}}]{Nikopoulos_2026}
{Nikopoulos}, G.~P., {Watson}, D., {Sneppen}, A., {et~al.} 2026, \aap, 710,
  A136

\bibitem[{{Oke} \& {Gunn}(1983)}]{Oke_1983}
{Oke}, J.~B. \& {Gunn}, J.~E. 1983, \apj, 266, 713

\bibitem[{{O'Neill} {et~al.}(2006){O'Neill}, {Kaspi}, {Laor}, {Nandra},
  {Moran}, {Peterson}, {Desroches}, {Filippenko}, {Ho}, \&
  {Maoz}}]{ONeill_2006}
{O'Neill}, P.~M., {Kaspi}, S., {Laor}, A., {et~al.} 2006, \apj, 645, 160

\bibitem[{{P{\'e}rez-D{\'\i}az} {et~al.}(2024){P{\'e}rez-D{\'\i}az},
  {P{\'e}rez-Montero}, {Fern{\'a}ndez-Ontiveros}, {V{\'\i}lchez}, \&
  {Amor{\'\i}n}}]{Perez-Diaz_2024}
{P{\'e}rez-D{\'\i}az}, B., {P{\'e}rez-Montero}, E., {Fern{\'a}ndez-Ontiveros},
  J.~A., {V{\'\i}lchez}, J.~M., \& {Amor{\'\i}n}, R. 2024, Nature Astronomy, 8,
  368

\bibitem[{{P{\'e}rez-D{\'\i}az} {et~al.}(2026){P{\'e}rez-D{\'\i}az},
  {V{\'\i}lchez}, {Castellano}, {Amor{\'\i}n}, {Bevacqua}, {Fontana},
  {Gandolfi}, {Gim{\'e}nez-Alc{\'a}zar}, {Pentericci}, {P{\'e}rez-Montero},
  {Santini}, \& {Tripodi}}]{Perez-Diaz_2026}
{P{\'e}rez-D{\'\i}az}, B., {V{\'\i}lchez}, J.~M., {Castellano}, M., {et~al.}
  2026, \aap, 710, L28

\bibitem[{{P{\'e}rez-Gonz{\'a}lez} {et~al.}(2023){P{\'e}rez-Gonz{\'a}lez},
  {Costantin}, {Langeroodi}, {Rinaldi}, {Annunziatella}, {Ilbert}, {Colina},
  {N{\o}rgaard-Nielsen}, {Greve}, {{\"O}stlin}, {Wright}, {Alonso-Herrero},
  {{\'A}lvarez-M{\'a}rquez}, {Caputi}, {Eckart}, {Le F{\`e}vre}, {Labiano},
  {Garc{\'\i}a-Mar{\'\i}n}, {Hjorth}, {Kendrew}, {Pye}, {Tikkanen}, {van der
  Werf}, {Walter}, {Ward}, {Bik}, {Boogaard}, {Bosman}, {G{\'o}mez}, {Gillman},
  {Iani}, {Jermann}, {Melinder}, {Meyer}, {Moutard}, {van Dishoek}, {Henning},
  {Lagage}, {Guedel}, {Peissker}, {Ray}, {Vandenbussche},
  {Garc{\'\i}a-Argum{\'a}nez}, \& {Mar{\'\i}a
  M{\'e}rida}}]{Perez-Gonzalez_2023}
{P{\'e}rez-Gonz{\'a}lez}, P.~G., {Costantin}, L., {Langeroodi}, D., {et~al.}
  2023, \apjl, 951, L1

\bibitem[{{P{\'e}rez-Gonz{\'a}lez} {et~al.}(2025){P{\'e}rez-Gonz{\'a}lez},
  {{\"O}stlin}, {Costantin}, {Melinder}, {Finkelstein}, {Somerville},
  {Annunziatella}, {{\'A}lvarez-M{\'a}rquez}, {Colina}, {Dekel}, {Ferguson},
  {Li}, {Yung}, {Bagley}, {Boogaard}, {Burgarella}, {Calabr{\`o}}, {Caputi},
  {Cheng}, {Dickinson}, {Eckart}, {Giavalisco}, {Gillman}, {Greve}, {Hamed},
  {Hathi}, {Hjorth}, {Huertas-Company}, {Kartaltepe}, {Koekemoer}, {Kokorev},
  {Labiano}, {Langeroodi}, {Leung}, {Natarajan}, {Papovich}, {Peissker},
  {Pentericci}, {Pirzkal}, {Rinaldi}, {van der Werf}, \&
  {Walter}}]{Perez-Gonzalez_2025}
{P{\'e}rez-Gonz{\'a}lez}, P.~G., {{\"O}stlin}, G., {Costantin}, L., {et~al.}
  2025, \apj, 991, 179

\bibitem[{{P{\'e}rez-Montero} \& {Contini}(2009)}]{Perez-Montero_2009}
{P{\'e}rez-Montero}, E. \& {Contini}, T. 2009, \mnras, 398, 949

\bibitem[{{Peterson} {et~al.}(2005){Peterson}, {Bentz}, {Desroches},
  {Filippenko}, {Ho}, {Kaspi}, {Laor}, {Maoz}, {Moran}, {Pogge}, \&
  {Quillen}}]{Peterson_2005}
{Peterson}, B.~M., {Bentz}, M.~C., {Desroches}, L.-B., {et~al.} 2005, \apj,
  632, 799

\bibitem[{{Pirzkal} {et~al.}(2024){Pirzkal}, {Rothberg}, {Papovich}, {Shen},
  {Leung}, {Bagley}, {Finkelstein}, {Vanderhoof}, {Lotz}, {Koekemoer}, {Hathi},
  {Cheng}, {Cleri}, {Grogin}, {Yung}, {Dickinson}, {Ferguson}, {Gardner},
  {Jung}, {Kartaltepe}, {Ryan}, {Simons}, {Ravindranath}, {Berg}, {Backhaus},
  {Casey}, {Castellano}, {Ch{\'a}vez Ortiz}, {Chworowsky}, {Cox}, {Dav{\'e}},
  {Davis}, {Estrada-Carpenter}, {Fontana}, {Fujimoto}, {Giavalisco}, {Grazian},
  {Hutchison}, {Jaskot}, {Kewley}, {Kirkpatrick}, {Kocevski}, {Larson},
  {Matharu}, {Natarajan}, {Pentericci}, {P{\'e}rez-Gonz{\'a}lez}, {Snyder},
  {Somerville}, {Trump}, \& {Wilkins}}]{Pirzkal_2024}
{Pirzkal}, N., {Rothberg}, B., {Papovich}, C., {et~al.} 2024, \apj, 969, 90

\bibitem[{{Planck Collaboration} {et~al.}(2020){Planck Collaboration},
  {Aghanim}, {Akrami}, {Ashdown}, {Aumont}, {Baccigalupi}, {Ballardini},
  {Banday}, {Barreiro}, {Bartolo}, {Basak}, {Battye}, {Benabed}, {Bernard},
  {Bersanelli}, {Bielewicz}, {Bock}, {Bond}, {Borrill}, {Bouchet}, {Boulanger},
  {Bucher}, {Burigana}, {Butler}, {Calabrese}, {Cardoso}, {Carron},
  {Challinor}, {Chiang}, {Chluba}, {Colombo}, {Combet}, {Contreras}, {Crill},
  {Cuttaia}, {de Bernardis}, {de Zotti}, {Delabrouille}, {Delouis}, {Di
  Valentino}, {Diego}, {Dor{\'e}}, {Douspis}, {Ducout}, {Dupac}, {Dusini},
  {Efstathiou}, {Elsner}, {En{\ss}lin}, {Eriksen}, {Fantaye}, {Farhang},
  {Fergusson}, {Fernandez-Cobos}, {Finelli}, {Forastieri}, {Frailis},
  {Fraisse}, {Franceschi}, {Frolov}, {Galeotta}, {Galli}, {Ganga},
  {G{\'e}nova-Santos}, {Gerbino}, {Ghosh}, {Gonz{\'a}lez-Nuevo}, {G{\'o}rski},
  {Gratton}, {Gruppuso}, {Gudmundsson}, {Hamann}, {Handley}, {Hansen},
  {Herranz}, {Hildebrandt}, {Hivon}, {Huang}, {Jaffe}, {Jones}, {Karakci},
  {Keih{\"a}nen}, {Keskitalo}, {Kiiveri}, {Kim}, {Kisner}, {Knox},
  {Krachmalnicoff}, {Kunz}, {Kurki-Suonio}, {Lagache}, {Lamarre}, {Lasenby},
  {Lattanzi}, {Lawrence}, {Le Jeune}, {Lemos}, {Lesgourgues}, {Levrier},
  {Lewis}, {Liguori}, {Lilje}, {Lilley}, {Lindholm}, {L{\'o}pez-Caniego},
  {Lubin}, {Ma}, {Mac{\'\i}as-P{\'e}rez}, {Maggio}, {Maino}, {Mandolesi},
  {Mangilli}, {Marcos-Caballero}, {Maris}, {Martin}, {Martinelli},
  {Mart{\'\i}nez-Gonz{\'a}lez}, {Matarrese}, {Mauri}, {McEwen}, {Meinhold},
  {Melchiorri}, {Mennella}, {Migliaccio}, {Millea}, {Mitra},
  {Miville-Desch{\^e}nes}, {Molinari}, {Montier}, {Morgante}, {Moss}, {Natoli},
  {N{\o}rgaard-Nielsen}, {Pagano}, {Paoletti}, {Partridge}, {Patanchon},
  {Peiris}, {Perrotta}, {Pettorino}, {Piacentini}, {Polastri}, {Polenta},
  {Puget}, {Rachen}, {Reinecke}, {Remazeilles}, {Renzi}, {Rocha}, {Rosset},
  {Roudier}, {Rubi{\~n}o-Mart{\'\i}n}, {Ruiz-Granados}, {Salvati}, {Sandri},
  {Savelainen}, {Scott}, {Shellard}, {Sirignano}, {Sirri}, {Spencer},
  {Sunyaev}, {Suur-Uski}, {Tauber}, {Tavagnacco}, {Tenti}, {Toffolatti},
  {Tomasi}, {Trombetti}, {Valenziano}, {Valiviita}, {Van Tent}, {Vibert},
  {Vielva}, {Villa}, {Vittorio}, {Wandelt}, {Wehus}, {White}, {White},
  {Zacchei}, \& {Zonca}}]{Planck_2020}
{Planck Collaboration}, {Aghanim}, N., {Akrami}, Y., {et~al.} 2020, \aap, 641,
  A6

\bibitem[{{Pollock} {et~al.}(2026){Pollock}, {Gottumukkala}, {Heintz},
  {Brammer}, {Roberts-Borsani}, {Oesch}, {Witstok}, {Arellano-C{\'o}rdova},
  {Cullen}, {Scholte}, {Terp}, {Rowland}, {Sneppen}, {Ito}, {Valentino},
  {Matthee}, {Watson}, \& {Toft}}]{Pollock_2026}
{Pollock}, C.~L., {Gottumukkala}, R., {Heintz}, K.~E., {et~al.} 2026, \aap,
  708, A203

\bibitem[{{Ricci} \& {Trakhtenbrot}(2023)}]{Ricci_2023}
{Ricci}, C. \& {Trakhtenbrot}, B. 2023, Nature Astronomy, 7, 1282

\bibitem[{{Rizzuti} {et~al.}(2025){Rizzuti}, {Matteucci}, {Molaro}, {Cescutti},
  \& {Maiolino}}]{Rizzuti_2025}
{Rizzuti}, F., {Matteucci}, F., {Molaro}, P., {Cescutti}, G., \& {Maiolino}, R.
  2025, \aap, 697, A96

\bibitem[{{Roberts-Borsani} {et~al.}(2026){Roberts-Borsani}, {Oesch}, {Ellis},
  {Weibel}, {Giovinazzo}, {Bouwens}, {Dayal}, {Fontana}, {Heintz}, {Matthee},
  {Meyer}, {Pentericci}, {Shapley}, {Tacchella}, {Treu}, {Walter}, {Atek},
  {Bose}, {Castellano}, {Fudamoto}, {Morishita}, {Naidu}, {Sanders}, \& {van
  der Wel}}]{Roberts-Borsani_2026}
{Roberts-Borsani}, G., {Oesch}, P.~A., {Ellis}, R., {et~al.} 2026, \mnras, 548,
  stag701

\bibitem[{{Rodighiero} {et~al.}(2023){Rodighiero}, {Bisigello}, {Iani},
  {Marasco}, {Grazian}, {Sinigaglia}, {Cassata}, \&
  {Gruppioni}}]{Rodighiero_2023}
{Rodighiero}, G., {Bisigello}, L., {Iani}, E., {et~al.} 2023, \mnras, 518, L19

\bibitem[{{Rodighiero} {et~al.}(2026){Rodighiero}, {Ferrara}, {Catone},
  {Napolitano}, {Cassata}, {Gandolfi}, {Merlin}, {Grazian}, {Renzini},
  {Bisigello}, {Castellano}, {P{\'e}rez-Gonz{\'a}lez}, {P{\'e}rez-D{\'\i}az},
  {Iani}, {Gruppioni}, {Finkelstein}, {Koekemoer}, {Bianchetti}, \&
  {Sinigaglia}}]{Rodighiero_2026}
{Rodighiero}, G., {Ferrara}, A., {Catone}, M., {et~al.} 2026, arXiv e-prints,
  arXiv:2603.15841

\bibitem[{{Rogers} {et~al.}(2026){Rogers}, {Strom}, {Rudie}, {Trainor}, {von
  Raesfeld}, {Raptis}, {Korhonen Cuestas}, {Miller}, {Steidel}, {Maseda},
  {Chen}, \& {Law}}]{Rogers_2026}
{Rogers}, N. S.~J., {Strom}, A.~L., {Rudie}, G.~C., {et~al.} 2026, \apjl, 997,
  L44

\bibitem[{{Salim} {et~al.}(2018){Salim}, {Boquien}, \& {Lee}}]{Salim_2018}
{Salim}, S., {Boquien}, M., \& {Lee}, J.~C. 2018, \apj, 859, 11

\bibitem[{{Sanders} {et~al.}(2024){Sanders}, {Shapley}, {Topping}, {Reddy}, \&
  {Brammer}}]{Sanders_2024}
{Sanders}, R.~L., {Shapley}, A.~E., {Topping}, M.~W., {Reddy}, N.~A., \&
  {Brammer}, G.~B. 2024, \apj, 962, 24

\bibitem[{{Santini} {et~al.}(2026){Santini}, {Castellano}, {Calabr{\`o}},
  {Fontana}, {Merlin}, {Bevacqua}, {Bergamini}, {Boquien}, {Cantarella},
  {Ciesla}, {Ferrara}, {Finkelstein}, {Fortuni}, {Gandolfi}, {Gasparetto},
  {Giallongo}, {Grogin}, {Guida}, {Koekemoer}, {Menci}, {Napolitano}, {Paris},
  {Pentericci}, {Perez-Diaz}, {Stoyanova}, \& {Treu}}]{Santini_2026}
{Santini}, P., {Castellano}, M., {Calabr{\`o}}, A., {et~al.} 2026, \aap, 710,
  A249

\bibitem[{{Santini} {et~al.}(2023){Santini}, {Fontana}, {Castellano},
  {Leethochawalit}, {Trenti}, {Treu}, {Belfiori}, {Birrer}, {Bonchi}, {Merlin},
  {Mason}, {Morishita}, {Nonino}, {Paris}, {Polenta}, {Rosati}, {Yang},
  {Boyett}, {Bradac}, {Calabr{\`o}}, {Dressler}, {Glazebrook}, {Marchesini},
  {Mascia}, {Nanayakkara}, {Pentericci}, {Roberts-Borsani}, {Scarlata},
  {Vulcani}, \& {Wang}}]{Santini_2023}
{Santini}, P., {Fontana}, A., {Castellano}, M., {et~al.} 2023, \apjl, 942, L27

\bibitem[{{Scarlata} {et~al.}(2024){Scarlata}, {Hayes}, {Panagia}, {Mehta},
  {Haardt}, \& {Bagley}}]{Scarlata_2024}
{Scarlata}, C., {Hayes}, M., {Panagia}, N., {et~al.} 2024, arXiv e-prints,
  arXiv:2404.09015

\bibitem[{{Schaerer} {et~al.}(2024){Schaerer}, {Marques-Chaves}, {Xiao}, \&
  {Korber}}]{Schaerer_2024}
{Schaerer}, D., {Marques-Chaves}, R., {Xiao}, M., \& {Korber}, D. 2024, \aap,
  687, L11

\bibitem[{{Schaye} {et~al.}(2015){Schaye}, {Crain}, {Bower}, {Furlong},
  {Schaller}, {Theuns}, {Dalla Vecchia}, {Frenk}, {McCarthy}, {Helly},
  {Jenkins}, {Rosas-Guevara}, {White}, {Baes}, {Booth}, {Camps}, {Navarro},
  {Qu}, {Rahmati}, {Sawala}, {Thomas}, \& {Trayford}}]{Schaye_2015}
{Schaye}, J., {Crain}, R.~A., {Bower}, R.~G., {et~al.} 2015, \mnras, 446, 521

\bibitem[{{Scholtz} {et~al.}(2026){Scholtz}, {Carniani}, {Parlanti},
  {D'Eugenio}, {Curtis-Lake}, {Jakobsen}, {Bunker}, {Cameron}, {Arribas},
  {Baker}, {Charlot}, {Chevellard}, {Circosta}, {Curti}, {Duan}, {Eisenstein},
  {Hainline}, {Ji}, {Johnson}, {Jones}, {Kumari}, {Maiolino}, {Maseda},
  {Perna}, {P{\'e}rez-Gonz{\'a}lez}, {Rawle}, {Rieke}, {Rinaldi}, {Robertson},
  {Saxena}, {Shivaei}, {Silcock}, {Sun}, {Rodr{\'\i}guez Del Pino},
  {Tacchella}, {{\"U}bler}, {Venturi}, {Williams}, {Willmer}, {Willott}, \&
  {Witstok}}]{Scholtz_2026}
{Scholtz}, J., {Carniani}, S., {Parlanti}, E., {et~al.} 2026, \mnras, 549,
  stag939

\bibitem[{{Senchyna} {et~al.}(2024){Senchyna}, {Plat}, {Stark}, {Rudie},
  {Berg}, {Charlot}, {James}, \& {Mingozzi}}]{Senchyna_2024}
{Senchyna}, P., {Plat}, A., {Stark}, D.~P., {et~al.} 2024, \apj, 966, 92

\bibitem[{{Sharda} {et~al.}(2026){Sharda}, {Schaye}, {McGibbon},
  {Ben{\'\i}tez-Llambay}, {Chaikin}, {Frenk}, {Hodge}, {Hu{\v{s}}ko},
  {Ploeckinger}, {Richings}, \& {Schaller}}]{Sharda_2026}
{Sharda}, P., {Schaye}, J., {McGibbon}, R.~J., {et~al.} 2026, arXiv e-prints,
  arXiv:2606.25995

\bibitem[{{Stiavelli} {et~al.}(2025){Stiavelli}, {Morishita}, {Chiaberge},
  {Leethochawalit}, {Norman}, {Ricotti}, {Roberts-Borsani}, {Treu}, {Vanzella},
  {Wyse}, {Zhang}, \& {Boyett}}]{Stiavelli_2025}
{Stiavelli}, M., {Morishita}, T., {Chiaberge}, M., {et~al.} 2025, \apj, 981,
  136

\bibitem[{{Storey} {et~al.}(2014){Storey}, {Sochi}, \& {Badnell}}]{SSB_2014}
{Storey}, P.~J., {Sochi}, T., \& {Badnell}, N.~R. 2014, \mnras, 441, 3028

\bibitem[{{Storey} \& {Zeippen}(2000)}]{SZ_2000}
{Storey}, P.~J. \& {Zeippen}, C.~J. 2000, \mnras, 312, 813

\bibitem[{{Tang} {et~al.}(2026){Tang}, {Stark}, {Mason}, {Gelli}, {Chen}, \&
  {Topping}}]{Tang_2026}
{Tang}, M., {Stark}, D.~P., {Mason}, C.~A., {et~al.} 2026, \apj, 1001, 38

\bibitem[{{Topping} {et~al.}(2025){Topping}, {Stark}, {Senchyna}, {Chen},
  {Zitrin}, {Endsley}, {Charlot}, {Furtak}, {Maseda}, {Plat}, {Smit},
  {Mainali}, {Chevallard}, {Molyneux}, \& {Rigby}}]{Topping_2025}
{Topping}, M.~W., {Stark}, D.~P., {Senchyna}, P., {et~al.} 2025, \apj, 980, 225

\bibitem[{{Topping} {et~al.}(2024){Topping}, {Stark}, {Senchyna}, {Plat},
  {Zitrin}, {Endsley}, {Charlot}, {Furtak}, {Maseda}, {Smit}, {Mainali},
  {Chevallard}, {Molyneux}, \& {Rigby}}]{Topping_2024}
{Topping}, M.~W., {Stark}, D.~P., {Senchyna}, P., {et~al.} 2024, \mnras, 529,
  3301

\bibitem[{{Tortosa} {et~al.}(2026){Tortosa}, {Ricci}, {Du}, {Venturi}, {Ho},
  {Li}, {Wang}, \& {Berton}}]{Tortosa_2026}
{Tortosa}, A., {Ricci}, C., {Du}, P., {et~al.} 2026, \aap, 708, A293

\bibitem[{{Treu} {et~al.}(2022){Treu}, {Roberts-Borsani}, {Bradac}, {Brammer},
  {Fontana}, {Henry}, {Mason}, {Morishita}, {Pentericci}, {Wang}, {Acebron},
  {Bagley}, {Bergamini}, {Belfiori}, {Bonchi}, {Boyett}, {Boutsia},
  {Calabr{\'o}}, {Caminha}, {Castellano}, {Dressler}, {Glazebrook}, {Grillo},
  {Jacobs}, {Jones}, {Kelly}, {Leethochawalit}, {Malkan}, {Marchesini},
  {Mascia}, {Mercurio}, {Merlin}, {Nanayakkara}, {Nonino}, {Paris},
  {Poggianti}, {Rosati}, {Santini}, {Scarlata}, {Shipley}, {Strait}, {Trenti},
  {Tubthong}, {Vanzella}, {Vulcani}, \& {Yang}}]{Treu_2022}
{Treu}, T., {Roberts-Borsani}, G., {Bradac}, M., {et~al.} 2022, \apj, 935, 110

\bibitem[{{Tripodi} {et~al.}(2026){Tripodi}, {Napolitano}, {Pentericci},
  {P{\'e}rez-D{\'\i}az}, {Bhagwat}, {D'Eugenio}, {Arevalo-Gonzalez},
  {Arroyo-Polonio}, {Calabr{\`o}}, {Ciardi}, {Dickinson}, {Ferguson},
  {Gandolfi}, {Hirschmann}, {Hu}, {Koekemoer}, {Llerena}, {Lucas}, {Oey},
  {Papovich}, {Yung}, \& {Wang}}]{Tripodi_2026}
{Tripodi}, R., {Napolitano}, L., {Pentericci}, L., {et~al.} 2026, arXiv
  e-prints, arXiv:2603.06409

\bibitem[{{Vink}(2023)}]{Vink_2023}
{Vink}, J.~S. 2023, \aap, 679, L9

\bibitem[{{Vollmann} \& {Eversberg}(2006)}]{Vollmann_2006}
{Vollmann}, K. \& {Eversberg}, T. 2006, Astronomische Nachrichten, 327, 862

\bibitem[{{Whitler} {et~al.}(2023){Whitler}, {Endsley}, {Stark}, {Topping},
  {Chen}, \& {Charlot}}]{Whitler_2023}
{Whitler}, L., {Endsley}, R., {Stark}, D.~P., {et~al.} 2023, \mnras, 519, 157

\bibitem[{{Yanagisawa} {et~al.}(2024){Yanagisawa}, {Ouchi}, {Nakajima},
  {Yajima}, {Umeda}, {Baba}, {Nakagawa}, {Nakane}, {Matsumoto}, {Ono},
  {Harikane}, {Isobe}, {Xu}, \& {Zhang}}]{Yanagisawa_2024}
{Yanagisawa}, H., {Ouchi}, M., {Nakajima}, K., {et~al.} 2024, \apj, 974, 180

\bibitem[{{Yang} {et~al.}(2017){Yang}, {Malhotra}, {Gronke}, {Rhoads},
  {Leitherer}, {Wofford}, {Jiang}, {Dijkstra}, {Tilvi}, \& {Wang}}]{Yang_2017}
{Yang}, H., {Malhotra}, S., {Gronke}, M., {et~al.} 2017, \apj, 844, 171

\bibitem[{{Yang} {et~al.}(2022){Yang}, {Morishita}, {Leethochawalit},
  {Castellano}, {Calabr{\`o}}, {Treu}, {Bonchi}, {Fontana}, {Mason}, {Merlin},
  {Paris}, {Trenti}, {Roberts-Borsani}, {Bradac}, {Vanzella}, {Vulcani},
  {Marchesini}, {Ding}, {Nanayakkara}, {Birrer}, {Glazebrook}, {Jones},
  {Boyett}, {Santini}, {Strait}, \& {Wang}}]{Yang_2022}
{Yang}, L., {Morishita}, T., {Leethochawalit}, N., {et~al.} 2022, \apjl, 938,
  L17

\bibitem[{{Zavala} {et~al.}(2025){Zavala}, {Castellano}, {Akins}, {Bakx},
  {Burgarella}, {Casey}, {Ch{\'a}vez Ortiz}, {Dickinson}, {Finkelstein},
  {Mitsuhashi}, {Nakajima}, {P{\'e}rez-Gonz{\'a}lez}, {Arrabal Haro},
  {Bergamini}, {Buat}, {Backhaus}, {Calabr{\`o}}, {Cleri},
  {Fern{\'a}ndez-Arenas}, {Fontana}, {Franco}, {Grillo}, {Giavalisco},
  {Grogin}, {Hathi}, {Hirschmann}, {Ikeda}, {Jung}, {Kartaltepe}, {Koekemoer},
  {Larson}, {McKinney}, {Papovich}, {Rosati}, {Saito}, {Santini}, {Terlevich},
  {Terlevich}, {Treu}, \& {Yung}}]{Zavala_2025}
{Zavala}, J.~A., {Castellano}, M., {Akins}, H.~B., {et~al.} 2025, Nature
  Astronomy, 9, 155

\bibitem[{{Zhu} {et~al.}(2025){Zhu}, {Kewley}, {Hsiao}, \&
  {Trussler}}]{Zhu_2025}
{Zhu}, P., {Kewley}, L.~J., {Hsiao}, T. Y.-Y., \& {Trussler}, J. 2025, \apjl,
  994, L29

\bibitem[{{Zou} {et~al.}(2026){Zou}, {Gallo}, {Zuo}, {Hodges-Kluck}, {Nguyen},
  {Roberts-Borsani}, {Madau}, {Pacucci}, {Seth}, \& {Treu}}]{Zou_2026}
{Zou}, F., {Gallo}, E., {Zuo}, Z., {et~al.} 2026, arXiv e-prints,
  arXiv:2603.24893

\bibitem[{{Zou} {et~al.}(2024){Zou}, {Sui}, {Saintonge}, {Scholte},
  {Moustakas}, {Siudek}, {Dey}, {Juneau}, {Guo}, {Canning}, {Aguilar}, {Ahlen},
  {Brooks}, {Claybaugh}, {Dawson}, {de la Macorra}, {Doel}, {Forero-Romero},
  {Gontcho A Gontcho}, {Honscheid}, {Landriau}, {Le Guillou}, {Manera},
  {Meisner}, {Miquel}, {Nie}, {Poppett}, {Rezaie}, {Rossi}, {Sanchez},
  {Schubnell}, {Seo}, {Tarl{\'e}}, {Zhou}, \& {Zou}}]{Zou_2024}
{Zou}, H., {Sui}, J., {Saintonge}, A., {et~al.} 2024, \apj, 961, 173

\end{thebibliography}
\bibliographystyle{Bibtex/aa.bst}

\appendix
\section{Spectral fits}\label{ap: spectral_fits}
\subsection{Fits to emission lines}\label{ap: lines}
\begin{figure*}
\centering
\includegraphics[width=0.48\textwidth]{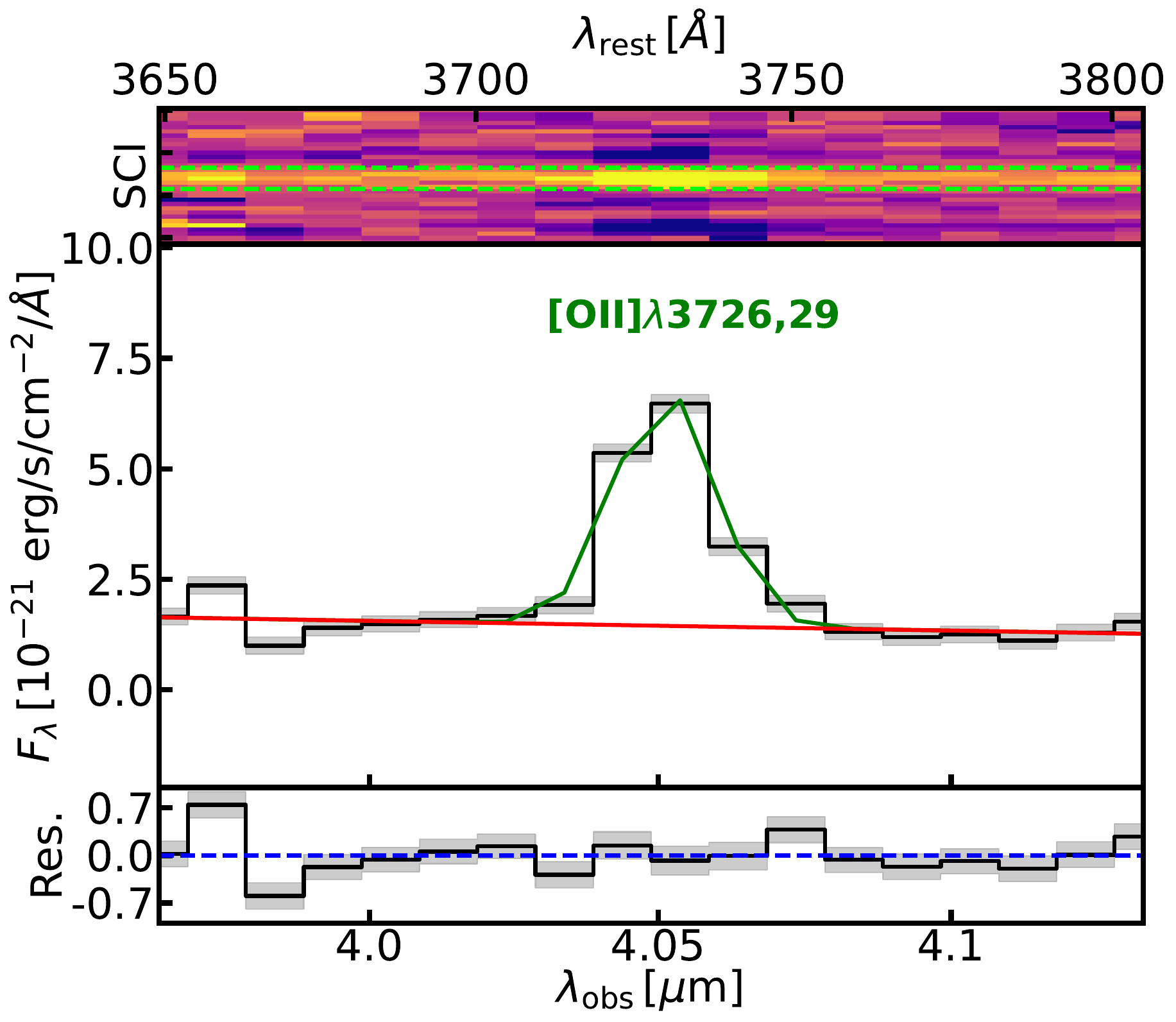}
\includegraphics[width=0.47\textwidth]{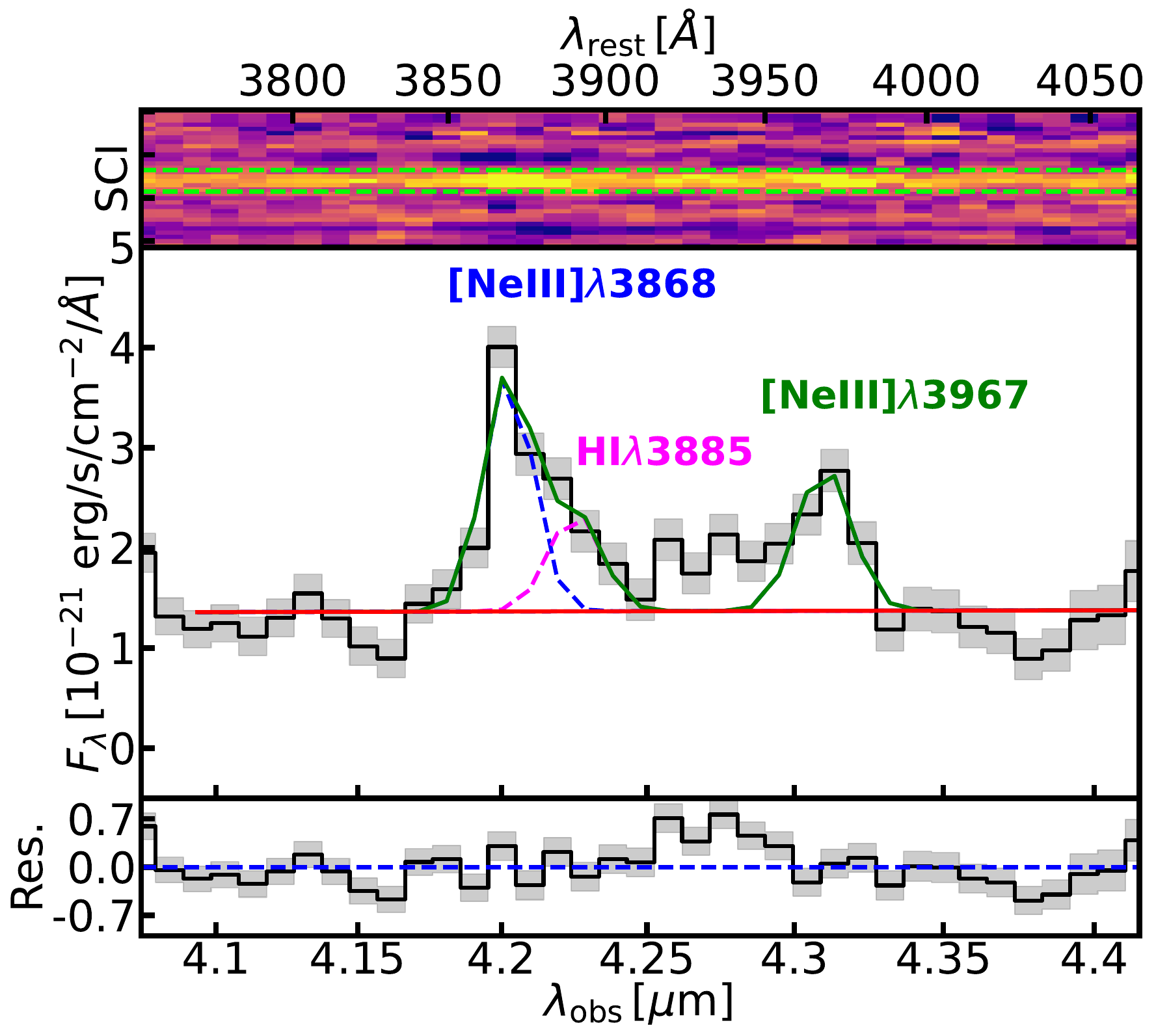}

\includegraphics[width=0.48\textwidth]{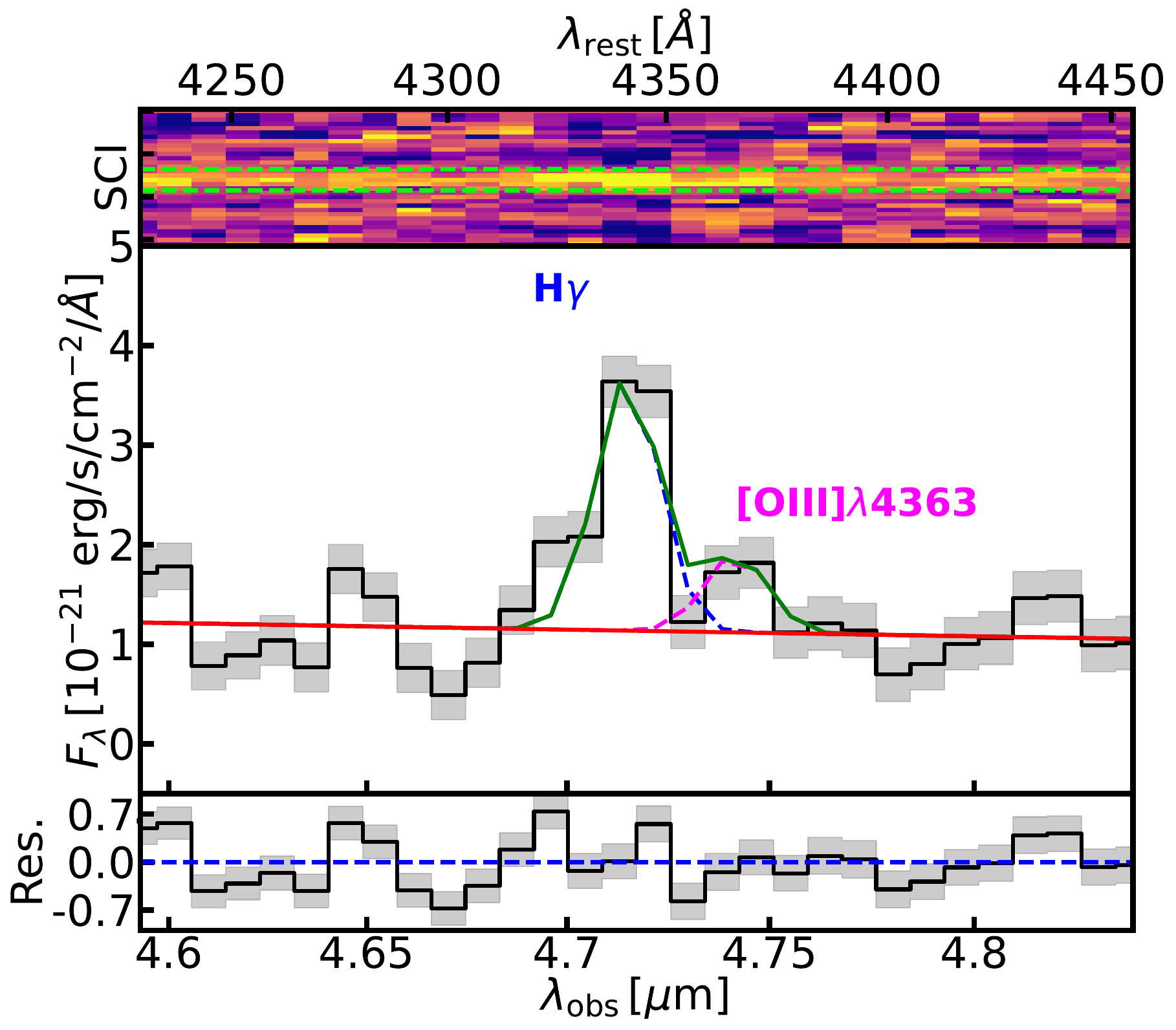}
\includegraphics[width=0.47\textwidth]{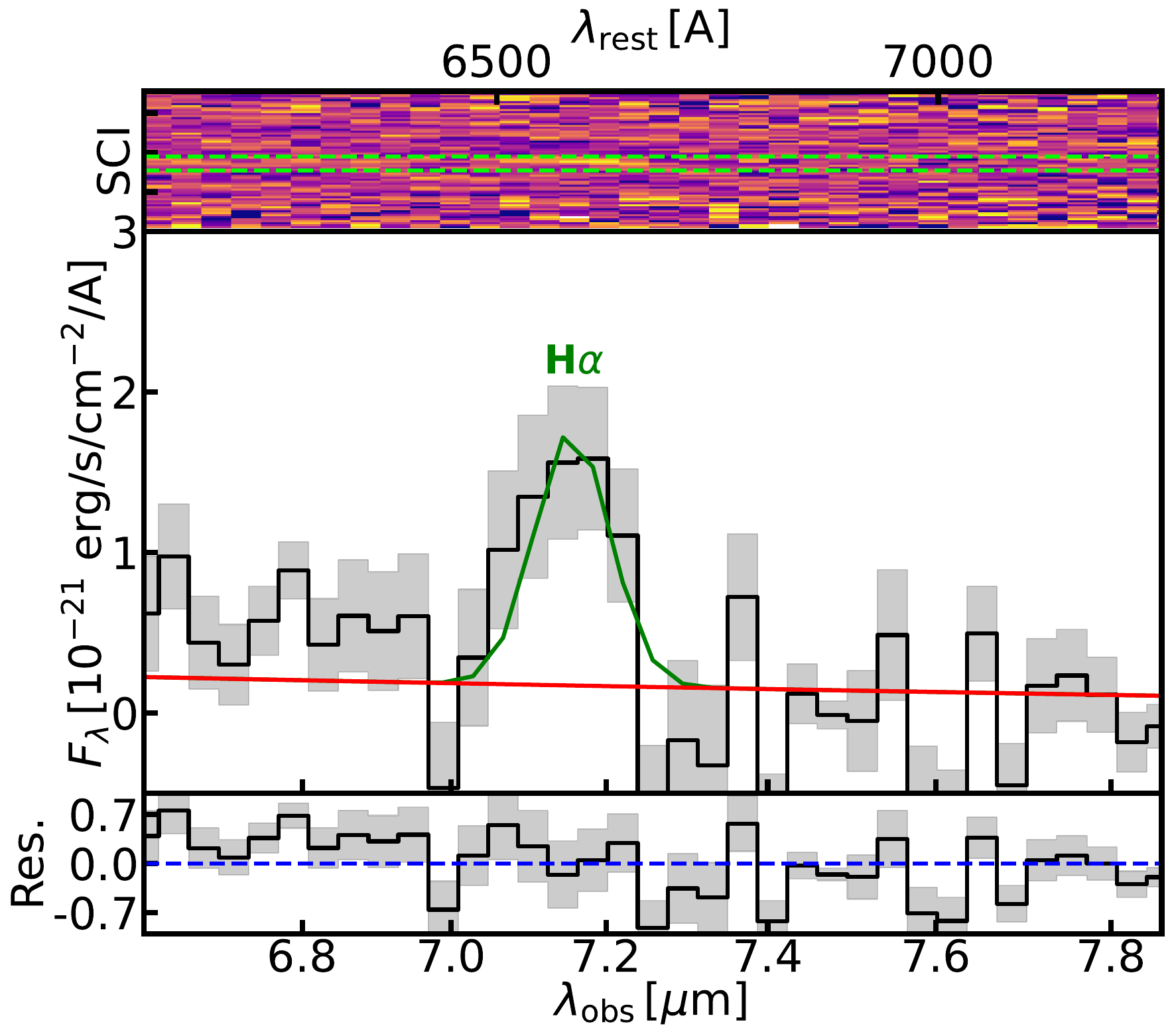}

\includegraphics[width=0.48\textwidth]{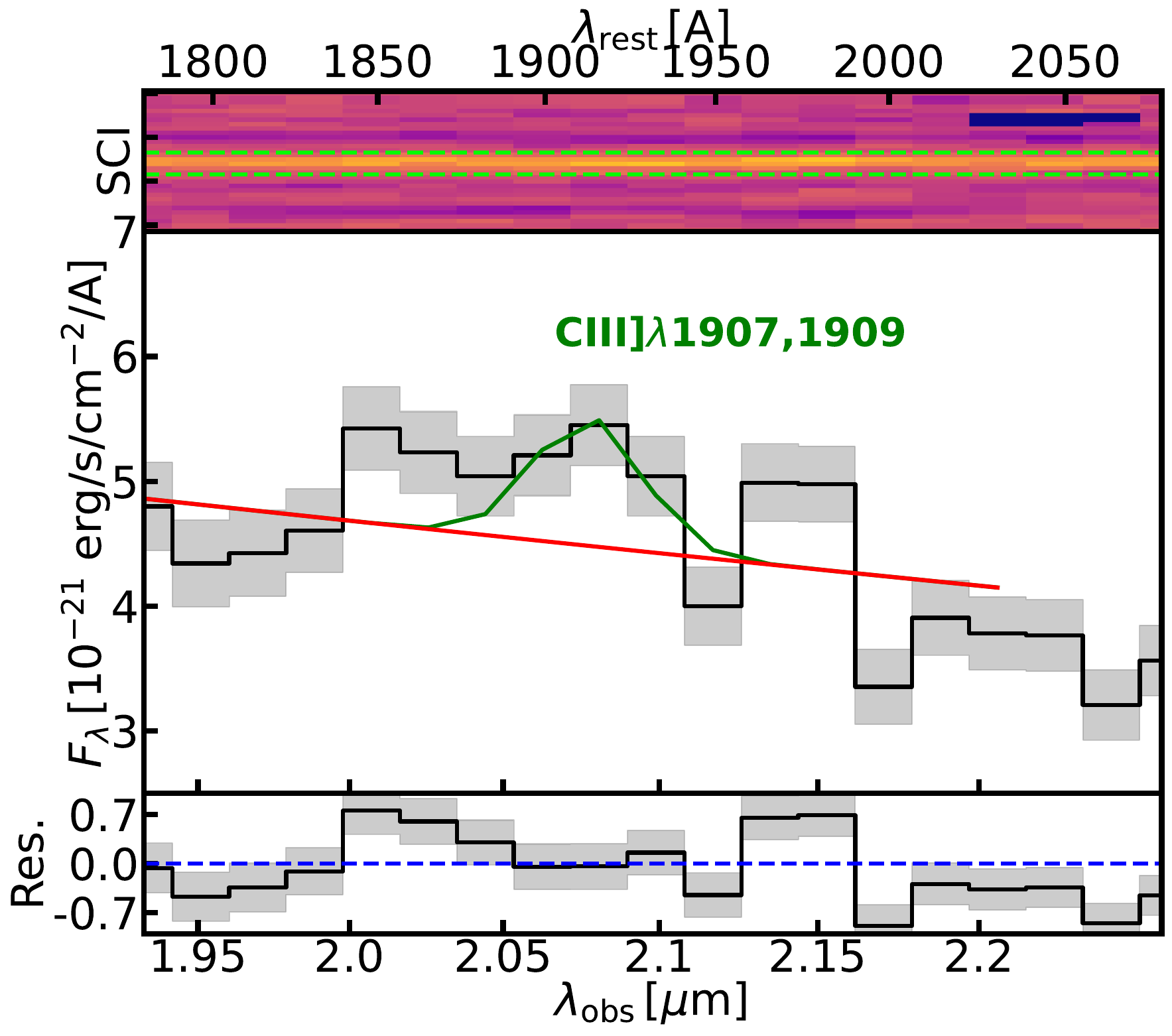}

\caption{Line fits for the brightest emission lines observed in NIRSpec/PRISM and MIRI/LRS. For each emission line (complex), we show: {\it i)} the cutout region in the 2D spectrum (top panel) with the extracted window as lime lines; {\it ii)} the 1D spectrum in the same region, with the continuum (red) and the emission fit (green) over-plotted; and {\it iii)} the residuals from the fit relative to the observed emission. In case more than one emission line is fitted, each component is shown with different colors.}
\label{line_fit_1}
\end{figure*}

We show in Fig. \ref{line_fit_1} the list of fiducial (S/N $>$ 3) emission lines. The fitting procedure is explained in Subsec. \ref{ss: lines}.

Due to the NIRSpec coverage of the \Hbeta emission right in the limit of the NIRSpec wavelength extension, we generate two different spectra from the combination of NIRSpec/PRISM and MIRI/LRS. The first spectrum (hereinafter "GHZ1-spec1"), is the result of concatenating the MIRI/LRS right after the \Hbeta feature in NIRSpec/PRISM, i.e. at 5.3~$\mu$m. The MIRI/LRS spectrum is matched at 5.3$\mu$m after masking the \Hbeta+\OIIIL[4959,5007] feature. The second spectrum (hereinafter "GHZ1-spec2") is the result of matching the MIRI/LRS spectrum to the NIRSpec/PRISM right blueward of the \Hbeta feature (which is masked in NIRSpec/PRISM), i.e. at 5.15~$\mu$m. Hence, we can use GHZ1-spec1 to resolve the \Hbeta component, and subtract it from \Hbeta+\OIIIL[4959,5007] in the GHZ1-spec2 spectrum to confidently measure the contribution of the \OIII lines.

\begin{figure*}
\centering
\includegraphics[width=0.48\textwidth]{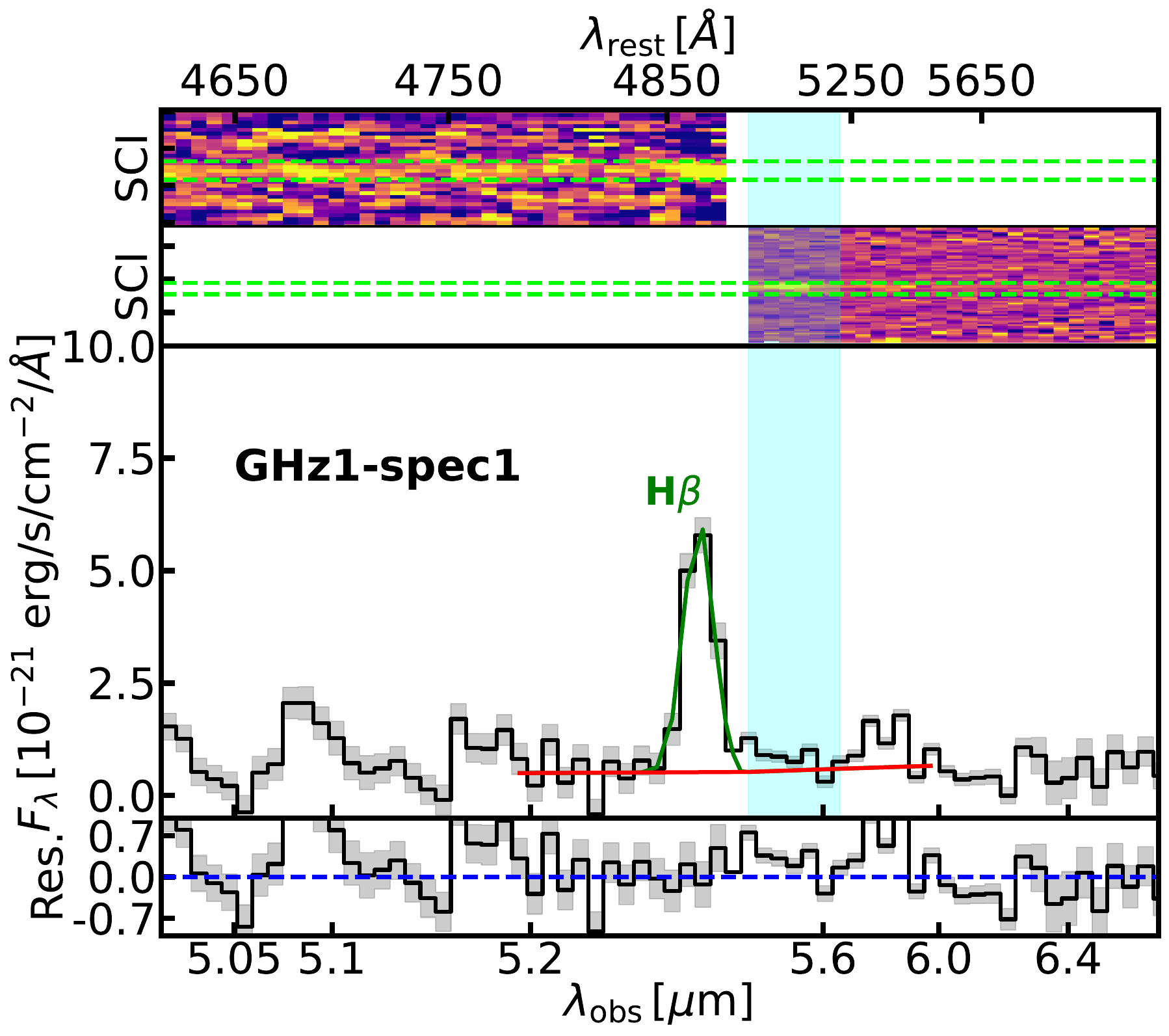}
\includegraphics[width=0.47\textwidth]{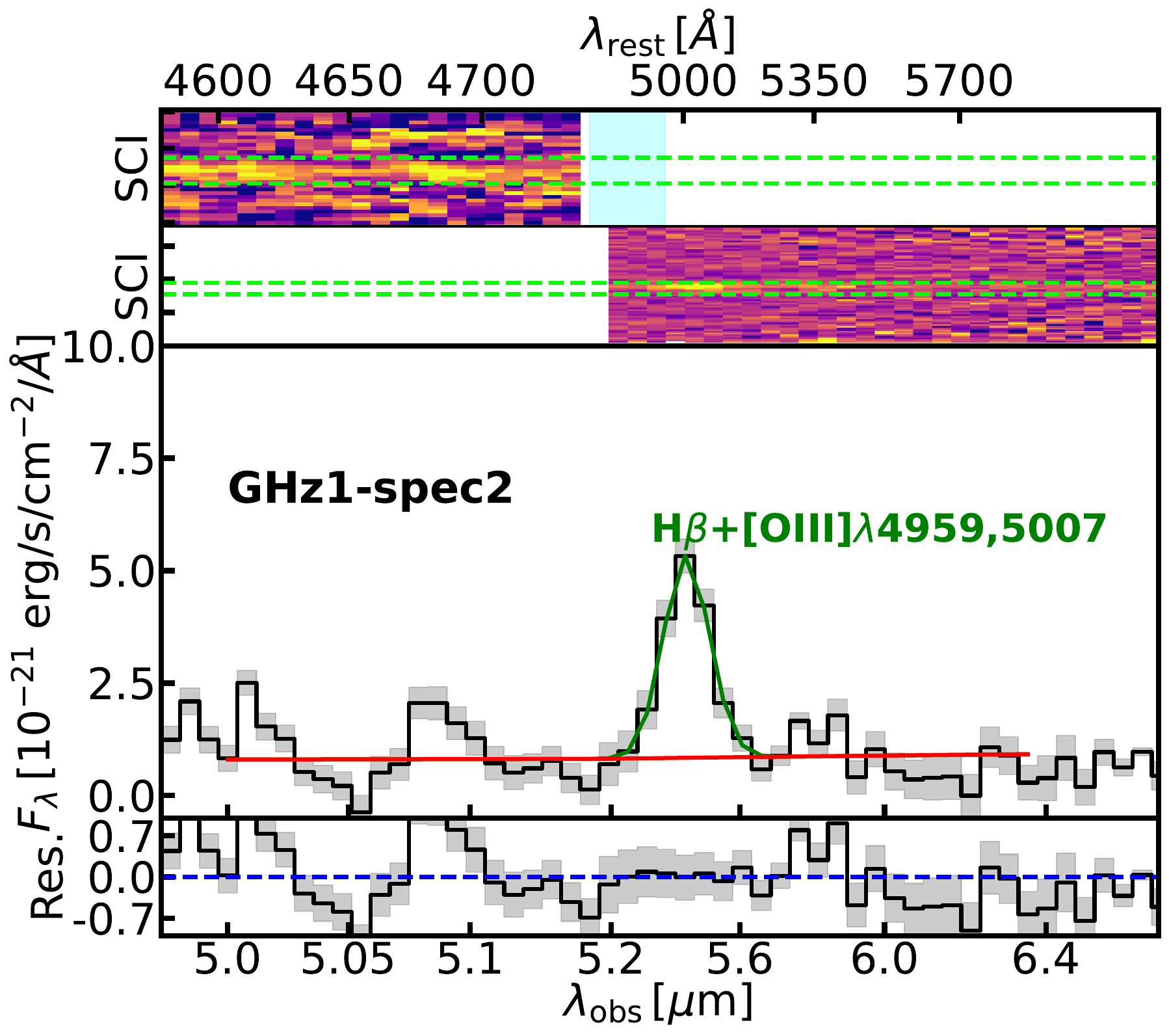}

\caption{Similar to Fig. \ref{line_fit_1}, but in this case focused on the $\Hbeta$ emission. The left panel shows the fit after concatenating the MIRI/LRS spectrum after the \Hbeta line from NIRSpec/PRISM (GHZ1-spec1). The right panel shows the fit from considering the total emission observed in MIRI/LRS (GHZ1-spec2) corresponding to the \Hbeta+\OIII complex. The cyan region represents the masking range for the match between both spectra.}
\label{line_fit_2}
\end{figure*}

The measurements of \Hbeta (GHZ1-spec1) and \Hbeta+\OIIIL[4959,5007] require an independent analysis, as these features are detected before (after) merging the NIRSpec/PRISM and MIRI/LRS data to generate GHZ1-spec1 (-spec2). Contrary to the general procedure outlined in Sec. \ref{ss: lines}, the intervals that allow us to fit the continuum are selected by eye-inspection as: 1) the resolution curve shows a discontinuity due to the merging of two different instruments; and, 2) as part of the considered emission comes from a mask which might not represent the actual shape of the continuum. We show in Fig. \ref{line_fit_2} the fits for \Hbeta (GHZ1-spec1) and \Hbeta+\OIIIL[4959, 5007] (GHZ1-spec2).

\subsection{Fit to the UV continuum}\label{ss: uv_slope}
\begin{figure}
\centering
\includegraphics[width=1\linewidth]{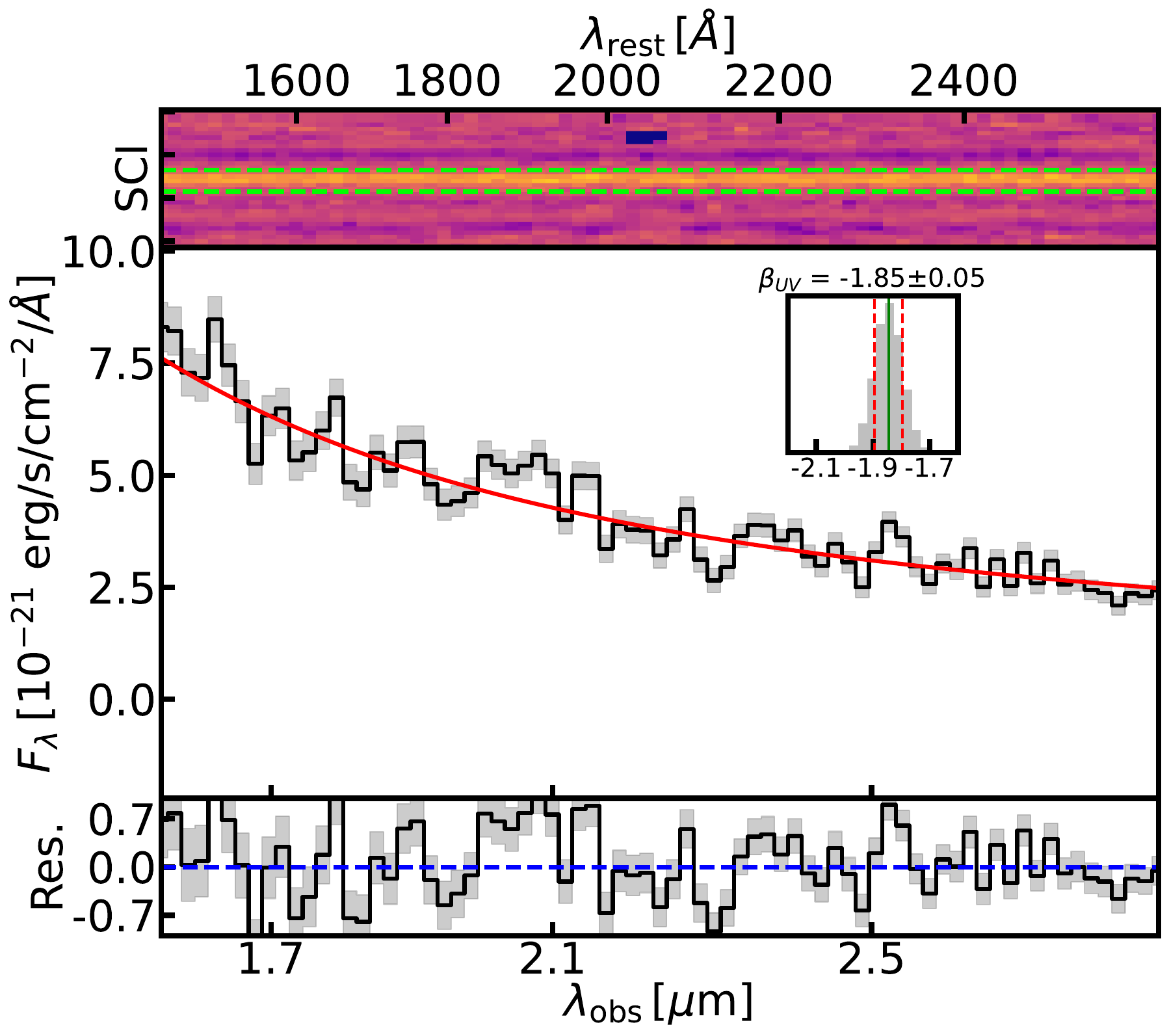}
\caption{Fit to the UV continuum for a power-law of index $\beta_{UV}$. The upper panel shows the 2D spectra whereas the middle panel shows the 1D extracted spectrum. The bottom panel shows the residuals from the fit relative to the observed flux. The histogram shows the distributions for the $\beta_{UV}$ slope fit from the MCMC method.}
\label{fig_slope}
\end{figure}

We assume that the continuum over the rest-frame\footnote{Following the approach proposed by \citet{Heintz_2025}, and further developed by \citet{Napolitano_2025a}, we restrict the range to avoid damping wing contamination in the bluer wavelengths.} range [1400 $\AA$, 2600 $\AA$] follows a power-law (F$_{\lambda}\propto \lambda^{\beta_{UV}}$). We show in Fig. \ref{fig_slope} the best-fitted power-law model to our data, which leads to a robust measurement of $\beta_{UV} = -1.85\pm 0.05$

\subsection{Evaluating the extended profile of emission lines and continuum}\label{ss: 2d_spec}
To investigate the possibility that the extraction windows selected to generate the 1D spectrum from the observations do not introduce biases in the emission line measurements, we evaluate the 2D spectrum from NIRSpec/PRISM and MIRI/LRS. In short, we focus on 4 ranges carefully selected per 2D spectrum that contain the two brightest emission lines and the bluer adjacent continuum. For each range, we stack the 2D spectrum in the wavelength dimension and explore the extended profile across the slits.

\begin{figure*}[p]
\centering
\includegraphics[width=0.85\textwidth]{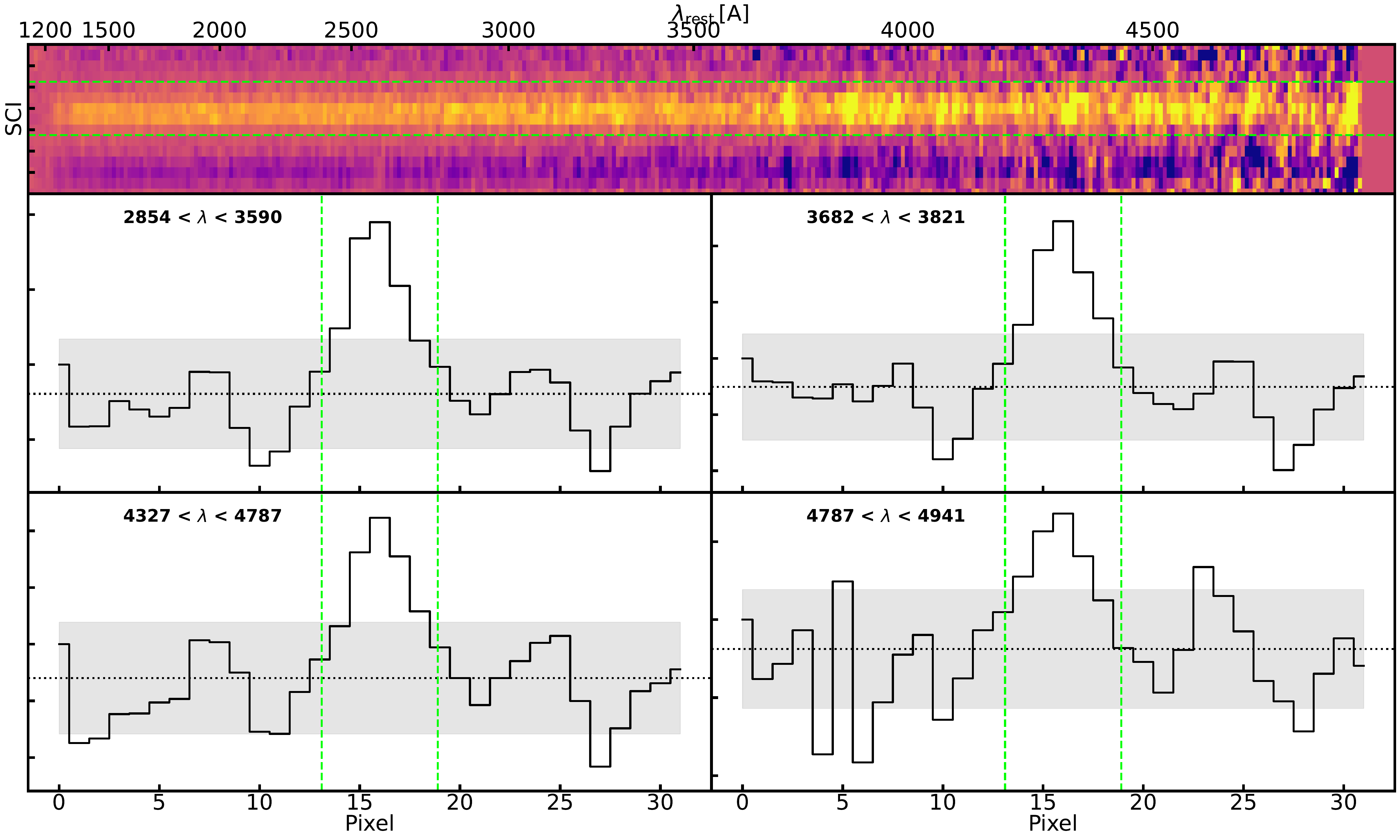}
\caption{Snapshot of the NIRSpec/PRISM 2D spectrum for the observation of GHZ1 and different stacked spectral windows in the slitlet width. The top panel shows a zoom-in view of the full 2D spectrum, with the horizontal lime lines marking the extraction window for the 1D spectrum. From top to bottom, left to right, we show the extended profile of the blue continuum adjacent to \OIIL[3726,3728], the \OIIL[3737,3728] emission, the blue continuum to the \Hbeta emission and the \Hbeta emission. In all cases, the horizontal black line shows the median value and the gray shaded area the average noise across the profile. Vertical lime lines represent the extraction window for the 1D spectrum.}
\label{Fig_nir_2dspec}
\end{figure*}

We show in Fig. \ref{Fig_nir_2dspec} the four ranges and the stacked profiles in NIRSpec/PRISM. Whereas there seems to be a tentative extended emission from the \Hbeta line, the part which is not covered by the extraction window matches the background noise, so any tentative flux loss is accounted by the flux uncertainty.

\begin{figure*}[p]
\centering
\includegraphics[width=0.85\textwidth]{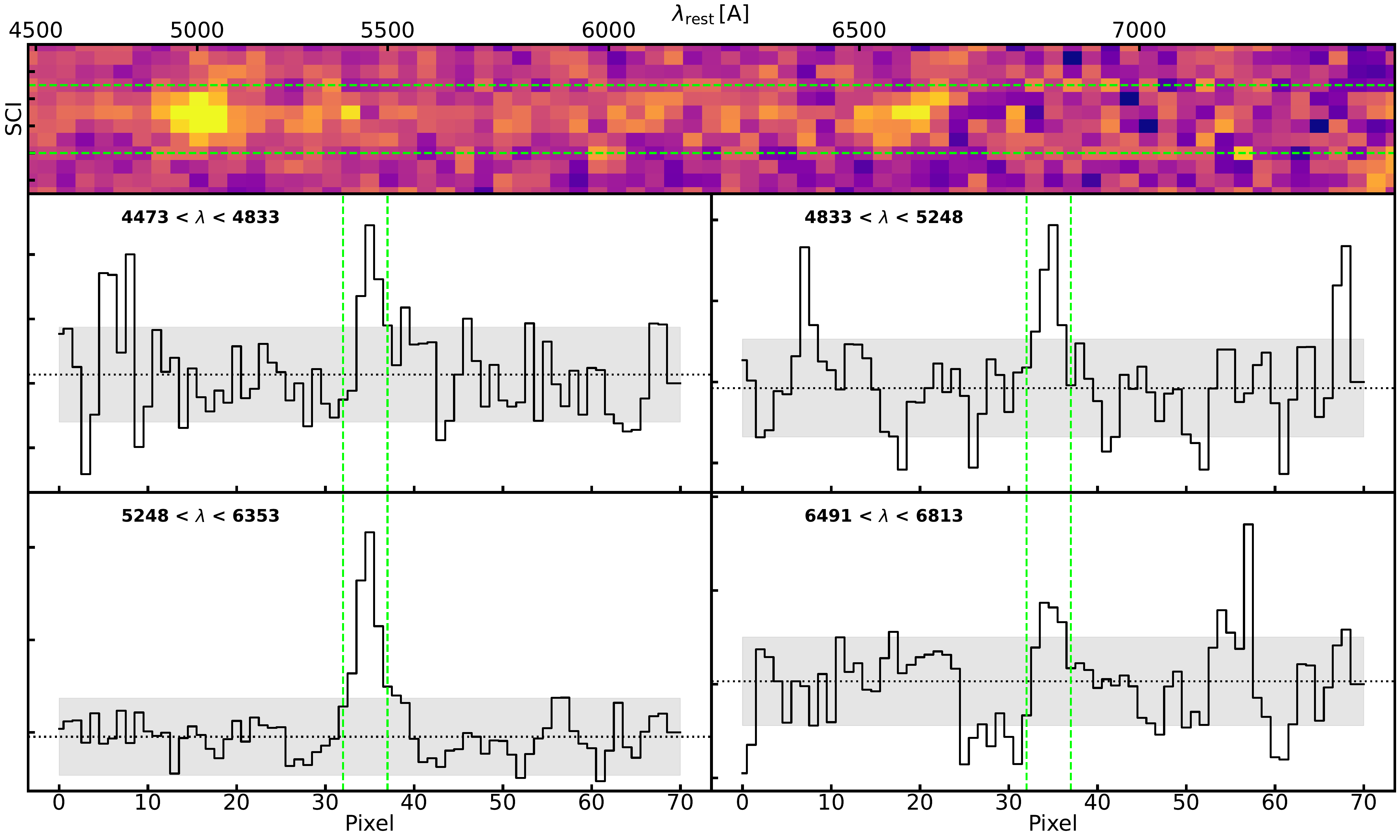}
\caption{Snapshot of the MIRI/LRS 2D spectrum for the observation of GHZ1 and different stacked spectral windows in the slitlet width. The top panel shows a zoom-in view of the full 2D spectrum, with the horizontal lime lines marking the extraction window for the 1D spectrum. From top to bottom, left to right, we show the extended profile of the blue continuum adjacent to the \Hbeta+\OIII complex, the \Hbeta+\OIII emission, the blue continuum to the \Halpha+\NII complex and the \Halpha+\NII emission. In all cases, the horizontal black line shows the median value and the gray shaded area the average noise across the profile. Vertical lime lines represent the extraction window for the 1D spectrum.}
\label{Fig_mir_2dspec}
\end{figure*}

A similar picture is observed from the MIRI/LRS analysis (see Fig. \ref{Fig_mir_2dspec}). The major difference, with respect to the previous analysis, is that the stacked continuum windows reveal a slightly more extended profile than emission lines, but this can be likely explained by the fact that the wavelength range for these windows is much higher, and the background noise is notably higher than in the NIRSpec/PRISM 2D spectrum.

We conclude that the selected extraction windows for the 1D spectra in both NIRSpec/PRISM and MIRI/LRS cannot be accounted for flux losses in the emission lines. While some of the emission lines (e.g. \Hbeta) point towards a slightly more extended emission than the adjacent continuum, we conclude that it is not statistically significant and it is properly accounted for in the error propagation. This result is also consistent with the almost perfect match between the NIRCam broadband photometry and the synthetic photometry generated from the spectrum.
\subsection{\Hbeta emission in MIRI/LRS}\label{ss: hbeta}
\begin{figure*}
\centering
\includegraphics[width=0.48\textwidth]{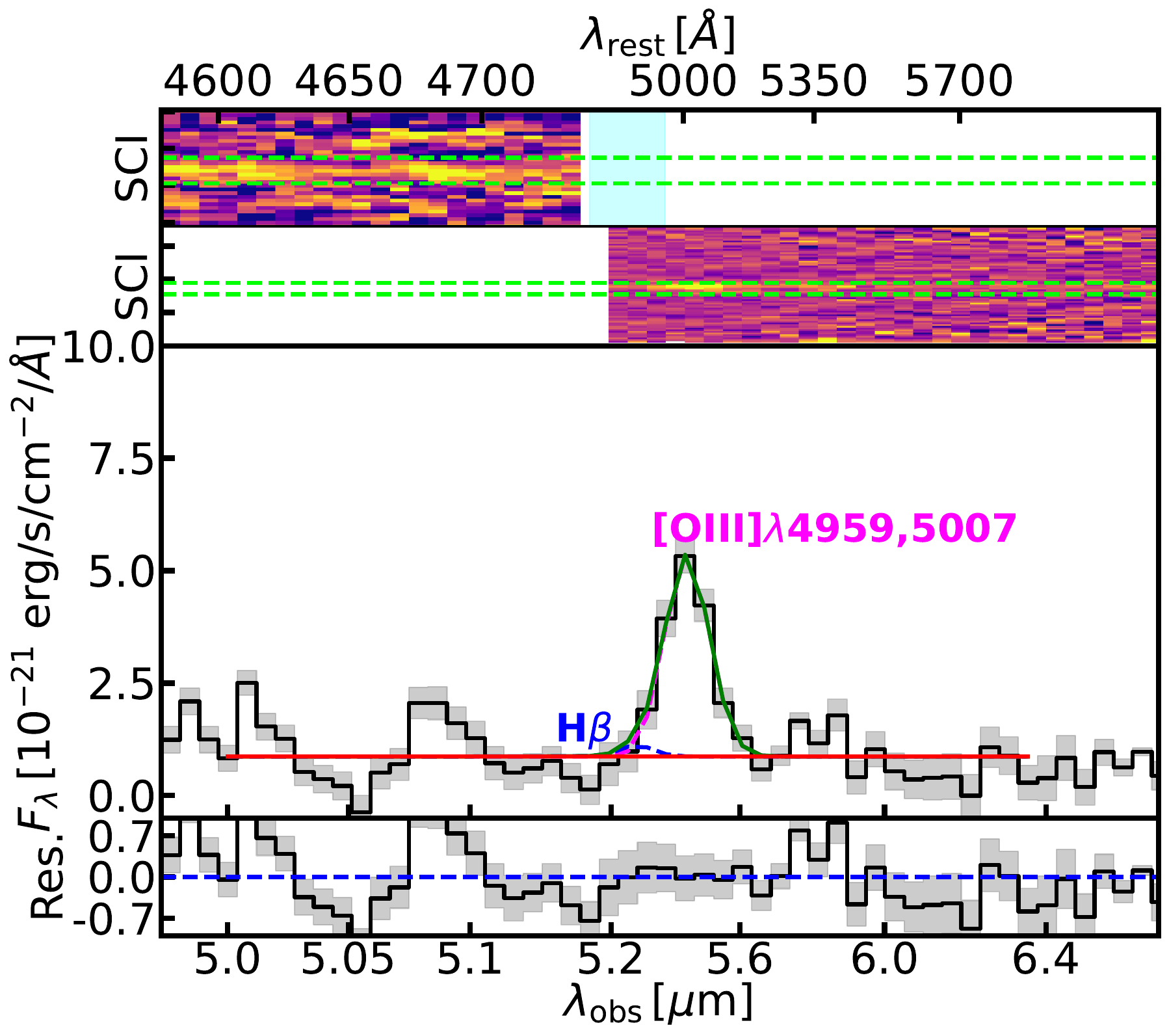}
\includegraphics[width=0.47\textwidth]{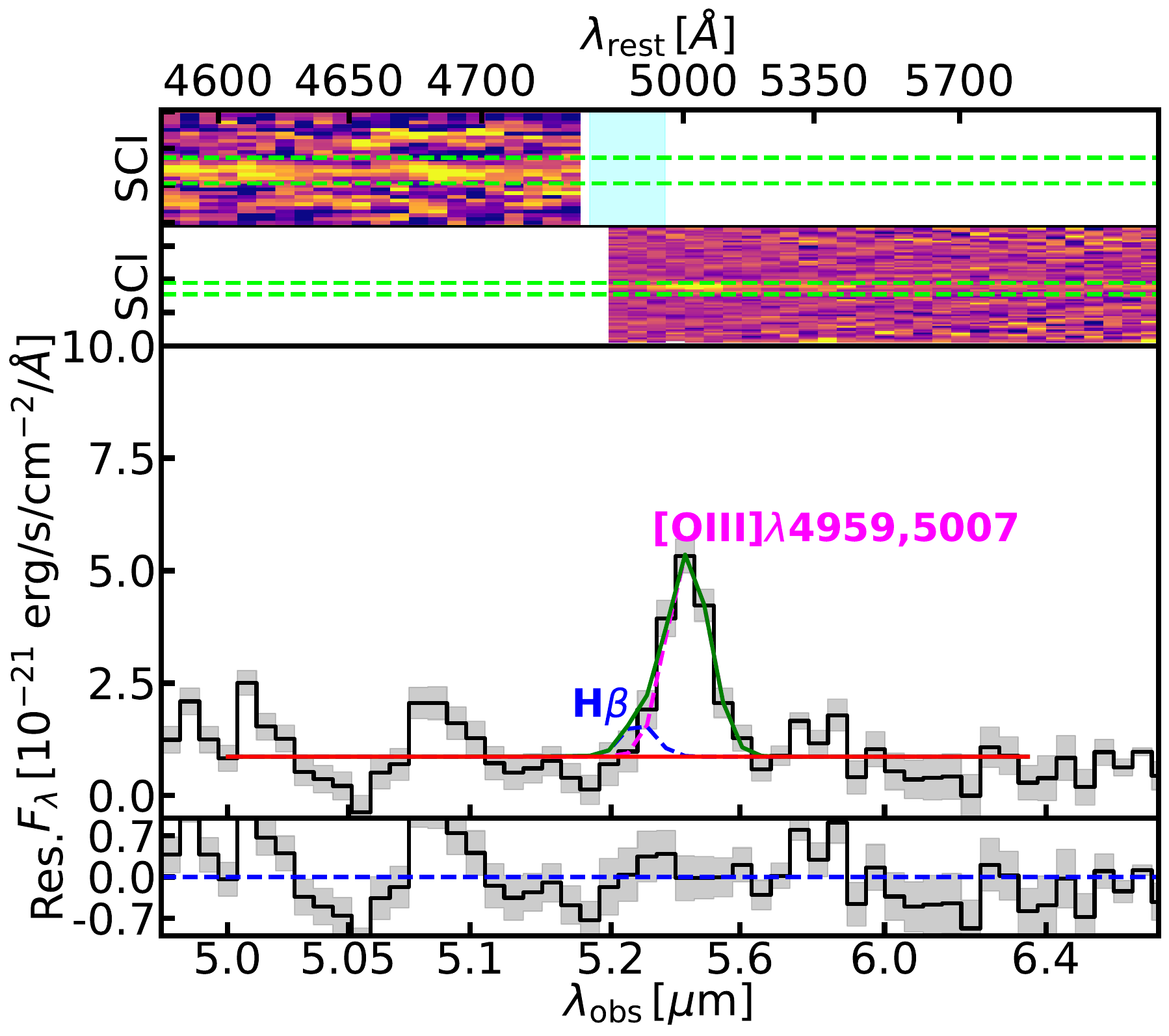}

\caption{Similar to Fig. \ref{line_fit_2}, but in this case focused on the $\Hbeta$ emission from \Hbeta+\OIIIL[4959, 5007] from MIRI/LRS. The left panel shows the fit of two Gaussians, enforcing only \Hbeta to be at the expected wavelength. The right panel shows the same fit if \Hbeta is forced to preserve the flux measured in NIRSpec/PRISM.}
\label{line_fit_hb}
\end{figure*}

Due to the low emission in \Halpha, we have also tried alternative fits to the \Hbeta+\OIIIL[4959,5007] for which we force a two-Gaussian fit: \Hbeta and \OIIIL[4959,5007] separately, in the MIRI/LRS spectrum. The results are shown in Fig. \ref{line_fit_hb}.

If we force a \Hbeta fit which preserves the flux measured from NIRSpec/PRSIM (Fig. \ref{line_fit_hb}, right panel), then a blue tail is expected which is almost compatible within the observed flux uncertainty. If, instead, we allow for a free Gaussian fit, we derived a flux for \Hbeta of F(\Hbeta) $\sim 6\times 10^{-19}$ erg$\cdot$s$^{-1}\cdot$cm$^{-2}$. While, this value would be compatible with Case B recombination and an attenuation magnitude of $A_{V} \sim 0.2$ mag, it must be noted that it is below the $3\sigma$ upper limit expected for such component ($\sim 1.2\times 10^{-18}$ erg$\cdot$s$^{-1}\cdot$cm$^{-2}$).

\section{Alternative data reductions}\label{ap: alt_reductions}
\subsection{Alternative reduction for the MIRI/LRS data}\label{ss: sarah}
In order to analyze the systematics from the MIRI/LRS data reduction, we perform within the collaboration an independent reduction, led by Sarah Kendrew. In short, the major differences from the reduction explained in Sec. \ref{ss: miri} are: {\it i)} this reduction has been done with Pipeline v1.20.1, with the exception of the rate files which were retrieved from v1.19.2; {\it ii)} the path loss step has been skipped; {\it iii)} the background correction followed the official pipeline methodology; and, {\it iv)} the spectral extraction is done on a 6-pixel aperture (rather than the 4-pixel preliminary used) and without the aperture correction.

In this case, we retrieve a higher normalization factor ($\sim$ 1.8) that should be applied to the MIRI/LRS spectrum so the continuum in the 4.9 and 5.2~$\mu$m range is consistent with the NIRSpec/PRISM spectrum. After applying such normalization, and measuring \Halpha with the methodology described in Sec. \ref{ss: lines}, we derive a flux of F(\Halpha) = $19.86(\pm4.17)\times 10^{-19}$ erg$\cdot$s$^{-1}\cdot$cm$^{-2}$, which is consistent with our previous measurement. Hence, we conclude that the initially measured \Halpha flux is consistent with other reductions.

\subsection{Alternative reduction for the NIRSpec/PRISM spectrum: extending the effective wavelength range to 5.5$\mu$m}\label{ss: stefano}
We investigate the systematics involving the NIRSpec/PRISM reduction process by performing an independent data reduction process with the most updated version of the JADES pipeline \citep{Scholtz_2026}. This reduction includes two differential features: \textit{i)} it uses the most recent realization of the data reduction pipeline provided by the NIRSpec GTO team \citep[see][for further details]{Scholtz_2026}, which extends the extracted and calibrated spectral traces on the detector significantly beyond the nominal red-end boundary for each grating/filter configuration; and, \textit{ii)} the extraction window selected encompasses 5 pixels.

We find that the synthetic photometry compared to the NIRCam imaging yields a factor $1.09\pm0.03$, which we apply to the spectrum for consistency. Most of the emission lines retrieved from this reduction and those assumed in the main text (namely \OIIL[3727,3729], \NeIIIL[3868], \NeIIIL[3967], \Hgamma, \Hbeta) yield consistent measurements within the uncertainties, with a pretty close ($~ 8\%$) difference in S/N for bluer wavelengths ($\lambda_{obs} < 4000\AA$), but worse for the redder range ($\sim 16\%$ difference). We do not retrieve any \MgII feature from this reduction.

The continuum over the 4.9-5.2$\mu$m range which overlaps with MIRI/LRS is 1.06($\pm$0.57)$\times10^{-21}$ erg$\cdot$s$^{-1}\cdot$cm$^{-2}$, which yields a normalizing factor of 1.05. The deblended fit of \Hbeta, \OIIIL[4959] and \OIIIL[5007] enabled from this reduction allows us to compare the total \Hbeta+\OIIIL[4959,5007] emission from NIRSpec and MIRI. The former yields a total flux of 61.6($\pm$2.4)$\times10^{-19}$ erg$\cdot$s$^{-1}\cdot$cm$^{-2}$, compared to the measured 79.4($\pm$8.8)$\times10^{-19}$ erg$\cdot$s$^{-1}\cdot$cm$^{-2}$ from MIRI/LRS. This difference is consistent with the absolute flux uncertainty ($\sim$ 20\%) in this extended wavelength range for NIRSpec/PRISM.

The \Hbeta emission line fit gives a total flux of 9.10($\pm$0.51)$\times10^{-19}$ erg$\cdot$s$^{-1}\cdot$cm$^{-2}$ and we report a Balmer decrement for \Halpha/\Hbeta $=2.10\pm0.44$, consistent with the reported value in Table \ref{Balmer_ratios}. While this value is still below the predictions for Case B, the tension is reduced.

\subsection{Alternative reduction for the NIRSpec/PRISM spectrum: DAWN JWST Archive}\label{ss: dja}
We further investigate NIRSpec data reductions by analyzing a third independent reduction. In this case, we use the last reduction offered by the DAWN JWST Archive (DJA\footnote{All data reduced by the DJA collaboration can be found in \url{https://dawn-cph.github.io/dja/}.}) collaboration. Particularly, we retrieve the DJA v4 product which extends the wavelength range calibration for NIRSpec products \citep[e.g.][]{Pollock_2026}. In particular, for GHZ1, this extension allows us to recover the resolved \OIIIL[4959] and \OIIIL[5007] emission lines.

As with our primary NIRSpec/PRISM spectrum, we check for consistency between NIRCam photometry and the NIRSpec/PRISM spectrum from DJA. The observed-to-derived photometry reports a value of 0.72$\pm$0.03, which we apply to the DJA spectrum. In addition, we compare the median continuum emission between NIRSpec/PRISM DJA and the MIRI/LRS. The median flux from NIRSpec/PRISM DJA is 0.92($\pm$0.42)$\times10^{-21}$ erg$\cdot$s$^{-1}\cdot$cm$^{-2}$, leading to a normalization factor of 0.91.

We fit the observed emission lines from the same technique described in Sec. \ref{ss: lines}. In general, the fluxes we report from the DJA spectrum for the lines that are also detected in our reduction are lower (by $17\%$) than those considered in this work, although in all cases they are still compatible within the errors. The S/N on the DJA spectrum is on average $\sim22\%$ below our reduction, originating likely from the different techniques followed to derive the overall flux error from NIRSpec/PRISM. We report that in this case there is no \MgII emission detected in the spectrum.

We report a total flux for the \Hbeta emission line of 8.60($\pm$0.56)$\times10^{-19}$ erg$\cdot$s$^{-1}\cdot$cm$^{-2}$. Particularly, the \Halpha/\Hbeta ratio reported from the NIRSpec/PRISM DJA and MIRI/LRS spectra is 2.03$\pm$0.47, still consistent with the reported value in Table \ref{Balmer_ratios} and below the predicted ratio for Case B. Another important aspect, from the DJA reduction, is that we recover the de-blended \OIIIL[4959] and \OIIIL[5007] lines. The total emission from the emission complex \Hbeta + \OIIIL[4959,5007] would sum up to $92(\pm4)\times10^{-19}$ erg$\cdot$s$^{-1}\cdot$cm$^{-2}$ from NIRSpec/PRISM, while the observed line complex in MIRI/LRS is (after normalization) $75(\pm8)\times10^{-19}$ erg $\cdot$ s $^{-1}\cdot$cm$^{-2}$. 
\\

Considering the different systematics that might affect different reductions in terms of the predicted flux error, the uncertainty in the flux calibration ($\sim$20$\%$) at extended NIRSpec wavelengths and the lack of photometry that would cover wavelengths beyond F444W band, we conclude that the flux uncertainties already considered in this work can explain the induced systematics from different NIRSpec/PRISM reductions, and that the \Hbeta flux reported is consistent within its uncertainties across different reductions.

\section{SED fitting}\label{ap: sed_plot}
\begin{figure*}
\centering
\includegraphics[width=0.8\linewidth]{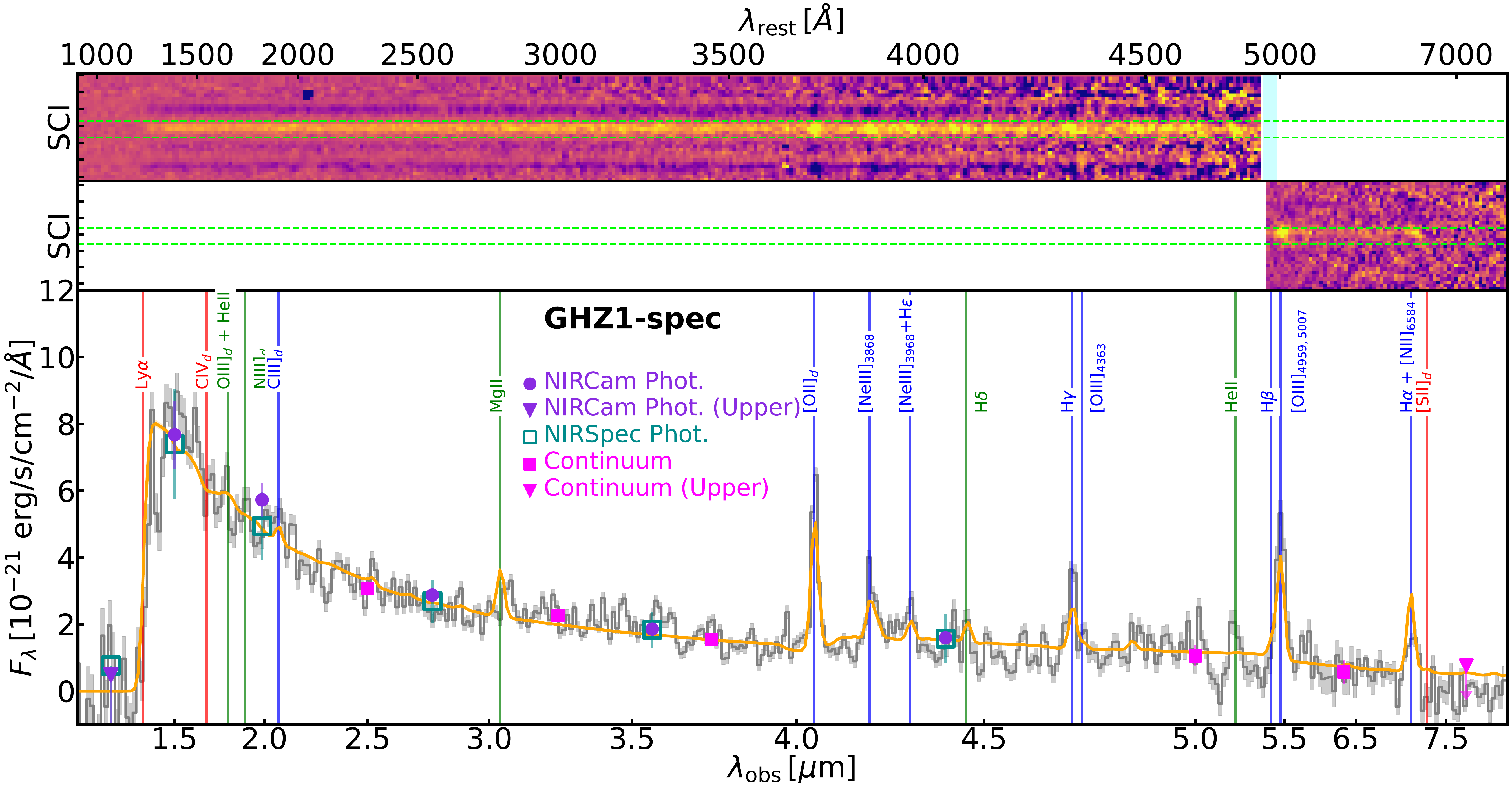}
\caption{The combined NIRSpec+MIRI spectrum \textsc{GHZ1-spec} (black line) and the best-fit template obtained by spectro-photometric fitting with  \textsc{bagpipes} (orange line). Upper panels show the 2D spectra for NIRSpec/PRISM and MIRI/LRS. Photometric and continuum points follow the same convention as in Fig. \ref{Spec_nirmir}.}
\label{sed_fit_plot}
\end{figure*}

We show in Fig.~\ref{sed_fit_plot} the best-fit template for GHZ1 obtained by spectro-photometric fitting with \textsc{bagpipes}. We find a very good agreement between the best-fit and the photometric and continuum bands. The emission lines are also well fitted, albeit with a larger uncertainty, which is not unexpected considering that the simplified nebular modeling implemented in \textsc{bagpipes} struggles at reproducing the emission spectra of high-$z$ sources due to their complex ISM \citep[][]{Castellano_2026}. Most notably, the \Halpha flux is predicted to be $\sim 15\%$ higher than what is actually observed in GHZ1, consistently with our analysis in Sect.~\ref{ss: balmer_anom}.

\section{X-ray emission towards GHZ1}\label{s:xray}
We searched for X-ray emission from GHZ1 using 103 merged \textit{Chandra} ACIS-I observations in the Abell 2744 field. Data reduction followed the procedure described in \citet{Zou_2026} and Wang et al. (in prep.), resulting in a total exposure time of $\sim$ 2.2 Ms. The source lies at an off-axis angle of $\sim$4.2$^{\prime}$, where the 90\% enclosed energy point-spread function (PSF) radius is $\sim$ 2.8$^{\prime\prime}$ \citep{Evans_2024}. Source counts were extracted with the \textsc{CIAO} tool \textsc{srcflux} using a circular aperture of radius 3$^{\prime\prime}$ centered on the optical position, while the background was estimated from a local surrounding annulus with inner and outer radii of 6$^{\prime\prime}$ and 10$^{\prime\prime}$ respectively. 

As the background appears uneven and affected by intra-cluster medium, we further validate the choice of background region. For that purpose, we model the 1$^{\prime}\times1^{\prime}$ region around GHZ1 (after masking any detected point sources together with the GHZ1 aperture), with a 4th-order polynomial in detector coordinates using Poisson statistics. The resulting expected background from this procedure in the 3$^{\prime\prime}$ radius was fully consistent to our preliminary estimate (within 1\% difference level).

The source is not detected ($> 3\sigma$) in any of the 0.5-2, 2-7 and 0.5-7 keV bands, with no marginal signal as the actual net counts are negative. We therefore quote 90\% confidence upper limits on the unabsorbed energy flux from \textsc{srcflux}, assuming a power-law spectrum modified by Galactic absorption ($\Gamma = 2$, $N_{H} = 1.34\times10^{20}$ cm$^{-2}$, \citealt{HI4PI_2016}). The resulting upper limits, without magnification correction, are F$_{X} = 1.6\times10^{-16}$ erg$\cdot$s$^{-1}\cdot$cm$^{-2}$ (0.5-7 keV band) and L$_{X} < 1.2\times10^{44}$ erg/s (2-10 keV band) at $z = 9.878$.

\section{Reference list for additional sources and studies considered in this work}
\subsection{Additional samples}\label{ss: additional_samples}
We present in Table \ref{Tab_gal_ref} the reference list of nearby targets and high-$z$ sources used in this study for comparison with GHZ1.

\begin{table*}[h!]
	\caption{Reference list for nearby and high-$z$ sources used in this work in comparison to GHZ1.}
	\label{Tab_gal_ref}
	\centering
	\begin{tabular}{c c c c c} 
		\hline\hline
		\textbf{Survey/field/name} & \textbf{Number} & \textbf{Redshifts} &  \textbf{Label} & \textbf{Reference} \\ 
		\textbf{(1)} & \textbf{(2)} & \textbf{(3)} & \textbf{(4)} & \textbf{(5)} \\ \hline
        Extragalactic HII regions & 15 & $\sim 0$ & Esteban+02,09 & \citet{Esteban_2002, Esteban_2009} \\
        Bright Compact Dwarf & 20 & $< 0.033$ & Berg+19 & \citet{Berg_2019} \\
		SDSS DR9$^{\dagger}$ & 257 & $<$0.3 & SDSS DR9 & \citet{Ahn_2012} \\
        LyC Leakers & 11 & [0.3, 0.4] & Izotov+23 & \citet{Izotov_2023} \\
        DESI DR1$^{\dagger}$ & 1,423 & $<$0.5 & DESI DR1 & \citet{Zou_2024} \\
        CECILIA & 20 & [2.09, 3.06] & Rogers+26 & \citet{Rogers_2026} \\
        CEERS & 12 & [2.162, 8.679] & Sanders+24 & \citet{Sanders_2024} \\
        MACS J1149.5+2223 & 2 & [4.257, 9.114] & Morishita+24 & \citet{Morishita_2024} \\
        SMACS J0723.3-7327 & 3 & [7.658, 8496] & Curti+23 & \citet{Curti_2023}\\
        Lya-Stack & [Stack] & $> 4$ & - & \citet{Tripodi_2026} \\
        EXCELS-121806 & - & 5.225 & - & \citet{Arellano-Cordova_2025} \\
        EXCELS-70684 & - & 5.255 & - & \citet{Arellano-Cordova_2025} \\
        GS3073 & - & 5.55 & - & \citet{Ji_2024} \\
        RXCJ2248 & - & 6.1 & - & \citet{Berg_2026} \\
        GLASS-150008 & - & 6.23 & - & \citet{Isobe_2023} \\
        A1703-zd6 & - & 7.043 & - & \citet{Topping_2025} \\
        GN-z8-LAE & - & 8.279 & - & \citet{Navarro-Carrera_2025} \\
        s04590 & - & 8.495 & - & \citet{Arellano-Cordova_2022} \\
        CEERS-1019 & - & 8.67 & - & \citet{Marques-Chaves_2024} \\
        GN-z9p4 & - & 9.380 & - & \citet{Schaerer_2024} \\
        GS-z9-0 & - & 9.43 & - & \citet{Curti_2025} \\
        UNCOVER-26185 & - & 10.05 & - & \citet{Alvarez-Marquez_2026} \\
        GHZ9 & - & 10.145 & - & \citet{Napolitano_2025b} \\
        MACS0647-JD & - & 10.17 & - & \citet{Tiger_2025} \\
        GN-z11 & - & 10.60 & - & \citet{Cameron_2023} \\
        CEERS2-588 & - & 11.04 & - & \citet{Harikane_2026} \\
        GHZ2 & - & 12.341 & - & \citet{Castellano_2026} \\
        GS-z12 & - & 12.48 & - & \citet{DEugenio_2024} \\
        JADES-GS-z14-0 & - & 14.18 & - & \citet{Helton_2026} \\
        MoM-z14 & - & 14.44 & - & \citet{Naidu_2026} \\
	\end{tabular}
	\tablefoot{Column (1) shows the name and/or survey. Column (2) shows the total number for the sample only if more than one object is retrieved. Column (3) shows the exact redshift for the source or the redshift range for the sample. Column (4) shows the label used in the plots. Column (5) gives the citation to the work. \\ $^{\dagger}$ The selection criteria are fully explained in \citet{Perez-Diaz_2026}.}
\end{table*}

\subsection{Mass-metallicity fits}\label{ss: additional_samples}
We present in Table \ref{Tab_mzr_ref} the reference list of of all the mass-metallicity relations explored in this study. For cosmological simulations, we have used the homogenized relations provided by \citet{Sharda_2026}.

\begin{table*}[h!]
	\caption{Reference list for mass-metallicity relations used in this study.}
	\label{Tab_mzr_ref}
	\centering
	\begin{tabular}{c c c c} 
		\hline\hline
		\textbf{Label} & \textbf{Type} & \textbf{Redshift} & \textbf{Reference} \\ 
		\textbf{(1)} & \textbf{(2)} & \textbf{(3)} & \textbf{(4)} \\ \hline
        Curti+20 & Observational & $\sim 0$ & \citet{Curti_2020} \\
        Nakajima+23 & Observational & [4, 10] & \citet{Nakajima_2023} \\
        Curti+24 & Observational & [6, 10] & \citet{Curti_2024} \\
        Koller+26 & Observational & [7, 10] & \citet{Koller_2026} \\
        COLIBRE & Simulation & [8, 10] & \citet{Sharda_2026} \\
        ILLUSTRIS & Simulation & [8, 10] & \citet{Nelson_2019} \\
        EAGLE & Simulation & [8, 10] & \citet{Schaye_2015} \\
        SIMBA & Simulation & [8, 10] & \citet{Garcia_2025} \\
        Pollock+26 & Observational & [8, 10] & \citet{Pollock_2026} \\
	\end{tabular}
	\tablefoot{Column (1) gives the label used throughout this work. Column (2) shows the type either observational (i.e. based on observational data) or simulation (based on cosmological simulations). Column (3) gives the redshift range of the analyzed data. Column (4) provides the citation to the corresponding works.}
\end{table*}

\section{Atomic databases}\label{atomic_data}
We use \textsc{PyNeb} \citet{pyneb} to manage the relevant atomic databases (atomic transitions, collisional coefficients and recombination rates) that are involved in the modeling of emission lines. For consistency with the photoionization models computed with \textsc{cloudy} v25 \citet{cloudy_v25}, we try to select the same databases that are used in the modeling of each emission line. Reference papers are listed in Table \ref{data_table}.

\begin{table*}[h!]
	\caption{Sources for the atomic data managed with \textsc{pyneb} \citet{pyneb}.}
	\label{data_table}
	\centering
	\begin{tabular}{c | c  c  } 
		\hline\hline
		\textbf{Ion} & \boldmath{$A_{ij}$} & \boldmath{$\Omega_{ij}$} \\ 
		\textbf{(1)} & \textbf{(2)} & \textbf{(3)} \\ \hline
		O$^{+}$ & \citet{Kisielius_2009} & \citet{FFT_2004} \\ \hline
        O$^{++}$ & \citet{SZ_2000}, & \citet{SSB_2014}  \\
         & \citet{FFT_2004} & \\ \hline
        Ne$^{++}$ & \citet{GMZ_1997} & \citet{ML_2011}  \\  
	\end{tabular}
	\tablefoot{Column (1) shows the considered ion. Column (2) gives the reference for the data containing the atomic transitions and their Einstein coefficients. Column (3) gives the reference for the collisional strengths.}
\end{table*}

\section{Photoionization models}\label{ap: photomodels}
To better explore the physical and ionizing conditions in the ionized gas of GHZ1, we rely on photoionization models to predict and interpret such conditions. We use the \textit{Nebular UltraViolet and Optical emission Lines to Optimize Studies of the iOnized gas} (NUVOLOSO) grid of photoionization models (P\'{e}rez-D\'{\i}az et al. in prep.).

In particular, we used the following grids of photoionization models:
\begin{itemize}
\item The chemical composition of the ionized nebula is governed by the 12+log(O/H) abundance, which varies in the range [6.55, 8.95] in steps of 0.1 dex. The rest of the heavy elements are scaled following the solar proportion with O, with standard depletion recipes depending on the inclusion of dust. Only N, which is independently varied exploring log(N/O) = [-2.0, 1,0] in steps of 0.25 dex, and C, which is assumed to follow log(C/N) = 0.6 for log(N/O) $\leq 0.0$ and log(C/N) = -0.6 for log(N/O) $> 0.0$.
\item The ionizing source is assumed to be a burst of star formation, modeled by means of BPASS v2.3 binary models, for which the stellar metallicity is assumed to be the same as the gas-phase, and we assume different burst ages: 1, 3 or 6 Myr. 
\item The density conditions in the cloud are varied by exploring log(n) = [2, 3, 4, 5, 6, 7, 8] cm$^{-3}$, adopting a density-constant profile. The ionization parameter, which regulates the number of ionizing photons (e.g. the amplitude for the SED), changes in the range [-3.5, -1.0] in steps of 0.5 dex.
\item Two different scenarios for dust are included: i) the Milky Way's dust-to-gas ratio; and, ii) no dust.
\end{itemize}
In total, per density and dust content scenario, we have a grid of 2,230 models. Per model, we retrieved the information of all emission lines, with independence of how faint they are, along with the resolved ionized structure, density and temperature profiles, and the cumulative distribution of the brightest emission lines.

\section{On the derivation of an observationally motivated size-metallicity relation}\label{ap: r-oh}
The work presented by \citet{Morishita_2024b} found that the effective radii of photometrically selected galaxies show a tight dependence with redshift and stellar mass, given by:
\begin{equation}
\log ( R_{e} \left[ \mathrm{kpc} \right] ) = \alpha \log \left( \frac{M_{\star}}{10^{8}M_{\odot}} \right) + \kappa \log \left( 1+z\right) + \ \gamma .
\end{equation}

On the other hand, if we assume the parametrization for the MZR given by \citet{Curti_2024}:
\begin{equation}
12+\log(\mathrm{O/H}) = Z_{0} + \alpha_{Z} \log \left( \frac{M_{\star}}{10^{8}M_{\odot}} \right),
\end{equation}
we can merge both observational trends to obtain a relation between 12+log(O/H), effective radius and redshift given by:
\begin{equation}
\label{size-metallicity} 12+\log(\mathrm{O/H}) = Z_{0} + \frac{\alpha_{Z}}{\alpha} \left[ \log ( R_{e} \left[ \mathrm{kpc} \right] ) - \kappa \log \left( 1+z\right) - \gamma \right],
\end{equation}
where we assume the values derived by \citet{Morishita_2024b} for resolved sources ($\alpha=0.19$, $\kappa = -0.24$, $\gamma = -0.33$) and the parameters of the MZR from \citet{Curti_2024} ($Z_{0} = 7.65$, $\alpha_{Z}=0.11$). Additionally, we can convert Eq. \ref{size-metallicity} into log(C/O) and log(N/O) assuming the reference scaling relations, which in this case we take from \citet{Nicholls_2017}.

\end{document}